Review article

# Nonlocalized Clustering and Evolution of Cluster Structure in Nuclei

Bo Zhou[1,2,*], Yasuro Funaki[3,†], Hisashi Horiuchi[4,‡], Akihiro Tohsaki[4,§]

[1] *Institute for the Advancement of Higher Education, Hokkaido University, Sapporo 060-0817, Japan*
[2] *Department of Physics, Hokkaido University, 060-0810 Sapporo, Japan*
[3] *College of Science and Engineering, Kanto Gakuin University, Yokohama 236-8501, Japan*
[4] *Research Center for Nuclear Physics (RCNP), Osaka University, Osaka 567-0047, Japan*
*Corresponding authors. E-mail: *bo@nucl.sci.hokudai.ac.jp,
[†]funaki@riken.jp,[‡]horiuchi@rcnp.osaka-u.ac.jp,[§]tohsaki@rcnp.osaka-u.ac.jp.*

We explain various facets of the THSR (Tohsaki-Horiuchi-Schuck-Röpke) wave function. We first discuss the THSR wave function as a wave function of cluster-gas state, since the THSR wave function was originally introduced to elucidate the $3\alpha$-condensate-like character of the Hoyle state ($0_2^+$ state) of $^{12}$C. We briefly review the cluster-model studies of the Hoyle state in 1970's in order to explain how there emerged the idea to assign the $\alpha$ condensate character to the Hoyle state. We then explain that the THSR wave function can describe very well also non-gaslike ordinary cluster states with spatial localization of clusters. This fact means that the dynamical motion of clusters is of nonlocalized nature just as in gas-like states of clusters and the localization of clusters is due to the inter-cluster Pauli principle which is against the close approach of two clusters. The nonlocalized cluster dynamics is formulated by the container model of cluster dynamics. The container model describes gas-like state and non-gaslike states as the solutions of the Hill-Wheeler equation with respect to the size parameter of THSR wave function which is just the size parameter of the container. When we notice that fact that the THSR wave function with the smallest value of size parameter is equivalent to the shell-model wave function, we see that the container model describes the evolution of cluster structure from the ground state with shell-model structure up to the gas-like cluster state via ordinary non-gaslike cluster states. For the description of various cluster structure, more generation of THSR wave function have been introduced and we review some typical examples with their actual applications.

# Contents

# 1 Introduction

Nucleons are easy to assemble and disassemble because of the saturation property of binding energy and density, which is one of the important properties that characterize the nuclear many-body system. Nuclear clustering [1–3] is the physics of dynamical assembling and disassembling of nucleons. Formation of clusters is a fundamental aspect of nuclear many-body dynamics together with the formation of mean field.

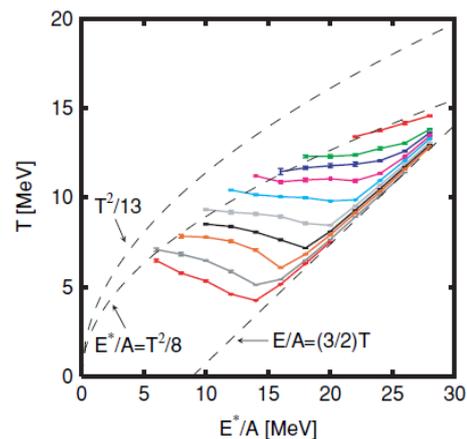

**Fig. 1**  Constant-pressure caloric curve of $^{36}$Ar obtained by AMD. The lines correspond to the pressure $P = 0.02, 0.03, 0.05, 0.07, 0.10, \cdots$ MeV/fm$^3$ from the bottom. Figure was reproduced from Ref. [4].

In the case of liquid-gas transition of infinite nuclear matter, the nuclear matter in the transition region is the mixture of liquid nuclear matter and nucleon-gas. But in the case of finite nuclei, the nucleus in the transition region of liquid-gas transition is composed of not simple mixture of liquid and nucleon-gas but mixture of various clusters, namely hot gas of clusters. Figure 1 shows the constant-pressure caloric curve of the finite nucleus $^{36}$Ar ($N = Z = 18$) obtained by AMD (antisymmetrized molecular dynamics) in Ref. [4]. We can see that the figure shows clearly the existence of nuclear liquid-gas phase transition in finite nucleus. We see the existence of the region of negative heat capacity which is a characteristic feature of finite systems and which corresponds to the plateau region of the constant-pressure caloric curve in the case of infinite nuclear matter. Figure 2 shows fragment mass distribution for the ensembles along the $P = 0.05$ MeV/fm$^3$ line. When the energy is low ($E^*/A = 8$ MeV), the distribution shows $U$-shape with two peaks so that the typical configuration at this energy is a large nucleus coexisting with a few gaseous nucleons. When the energy is increased ($E^*/A = 12 \sim 16$ MeV), the peak at the large fragment becomes smaller and the distribution changes into shoulder-like and power-law-like shapes. Thus complex configurations with many intermediate and light mass fragments are typical at these energies, and the proportion of light fragments increases as the energy increases (from $E^*/A = 12 \sim 16$ MeV).

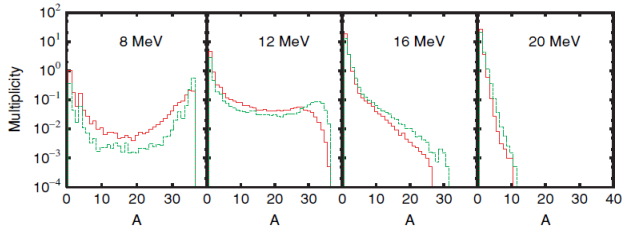

**Fig. 2** Fragment mass distributions along the $P = 0.05$ MeV/fm$^3$ line of the caloric curve of Fig. 1. Full lines are the distributions obtained with $r_{\rm clust} = 2.5$ fm, while dashed lines are those obtained with $r_{\rm clust} = 3.0$ fm. $r_{\rm clust}$ is inter-nucleon distance for identifying clusters. Namely, if the distance between two nucleons is smaller than $r_{\rm clust}$, they are regarded as belonging to the same cluster. Figure was reproduced from Ref. [4].

Recently, study of (cold) cluster gas has started in the field of nuclear structure. Since the breakup thresholds into clusters are much lower than the threshold of nucleon gas, cluster-gas-like structure can be studied spectroscopically. Figure 3 shows excitation energies of $\alpha$-cluster gas-like state and nucleon gas-like state in the case of $^{12}$C. It is now widely accepted that the second $0^+$ state which is known as the Hoyle state is the $3\alpha$-cluster gas-like state having the $3\alpha$-condensate-like structure [5, 6]. The excitation energy 7.65 MeV of this $0_2^+$ state is slightly above the $3\alpha$ and $^8$Be($0_1^+$) + $\alpha$ thresholds.

For the study of the Hoyle state, a new type of cluster-model wave function was introduced [8] which was named THSR (Tohsaki-Horiuchi-Schuck-Röpke) wave function. The THSR wave function in the case of $3\alpha$ system has the form

$$\Phi_{3\alpha}^{\rm THSR}(B) = N(B)\mathcal{A}\Big\{\exp[-\frac{2}{B^2}\sum_{k=1}^{3}(\boldsymbol{X}_k - \boldsymbol{X}_{\rm CM})^2]\prod_{i=1}^{3}\phi(\alpha_i)\Big\} \quad (1)$$

Here $\boldsymbol{X}_k$ and $\boldsymbol{X}_{\rm CM}$ stand for the center-of-mass coordinates of the $\alpha_k$ cluster and the total system, respectively. The intrinsic wave function of the $k$th cluster is repre-sented by the $\phi(\alpha_k)$. For the Hoyle state the size parameter $B$ is much larger than the single-nucleon H.O. (harmonic oscillator) size parameter $b$. It was discovered that the $3\alpha$ cluster model wave functions of the Hoyle state obtained by solving $3\alpha$ RGM (resonating group method) and GCM (generator coordinate method) equations in 1970's are almost 100% equivalent to the single THSR wave functions of the Hoyle state [9]:

$$|\langle \Phi_{3\alpha}^{\rm THSR}|\Phi_{3\alpha}^{\rm RGM/GCM}\rangle|^2 \approx 100\%. \quad (2)$$

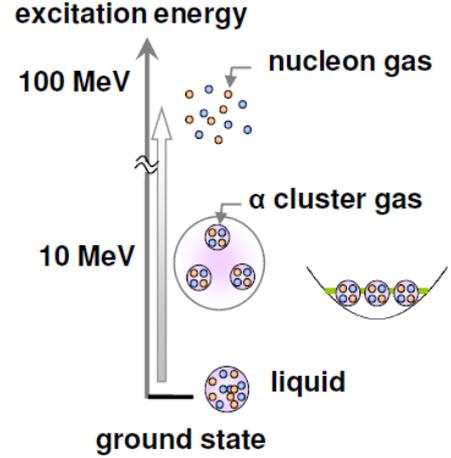

**Fig. 3** Excitation energies of $\alpha$-cluster gas-like state and nucleon gas-like state in the case of $^{12}$C. Figure was reproduced from Ref. [7].

In spite of the fact that the THSR wave function was devised for describing gas-like cluster states, it was found that the THSR wave function can describe very well also non-gaslike ordinary cluster states with spatial localization of clusters [10]. It was reported that the traditional RGM/GCM wave functions of the inversion-doublet band states of $^{20}$Ne are almost 100% equivalent to the $^{16}$O + $\alpha$ single THSR wave functions [10];

$$|\langle \Phi_{^{16}{\rm O}+\alpha}^{\rm THSR}|\Phi_{^{16}{\rm O}+\alpha}^{\rm RGM/GCM}\rangle|^2 \approx 100\%. \quad (3)$$

This discovery clarified that the spatial localization of clusters in two-cluster systems is due to the kinematic effect of inter-cluster Pauli principle [11], which means that the dynamical motion of clusters is of nonlocalized nature. This result teaches us that the cluster dynamics is of nolocalized nature not only in the gas-like cluster states but also in non-gaslike cluster states.

The nonlocalized cluster dynamics can be well formulated by a newly introduced model which is named container model of cluster dynamics [11]. The container model gives us both gas-like state and non-gaslike states as the solutions of the Hill-Wheeler equation with respect to the size parameter of THSR wave function $\Phi^{\rm THSR}(B)$ which is just the size parameter of the container.

$$\sum_B \langle \Phi^{\text{THSR}}(B')|(H-E_k)|\Phi^{\text{THSR}}(B)\rangle f_k(B) = 0. \quad (4)$$

Considering the fact that, in the limit of the smallest value of the size parameter $B$, the THSR wave function is just equivalent to the shell-model wave function, the Hill-Wheeler equation with respect to the size parameter of the container describes the evolution of the cluster structure from the ground state with shell-model like nature to the ordinary cluster states and further to the gas-like states of clusters.

In order to describe variety of cluster states which show up in the process of evolution of cluster states, the THSR wave function should be extended from the simple form with one kind of size parameter to the form of two or more kinds of size parameters. For the sake of the development of the study of cluster dynamics, it is necessary to devise better extension of the THSR wave function.

The content of this review article is as follows. In the next section (Sec. 2), we discuss gas-like states of clusters. Since the Hoyle state is now widely accepted as a gas-like state having the $3\alpha$-condensate-like structure, we first review the research of the Hoyle state. Early in 1970's, the Hoyle state was recognized as being a gaslike state with very large radius composed of $3\alpha$ clusters interacting weakly via $\alpha+\alpha$ relative $S$ waves [12–14]. Then, more than 20 years later, the Hoyle state was reinterpreted as having Bose-condensate-like structure of $3\alpha$ clusters [8]. The THSR wave function was devised for studying $\alpha$ condensate-like structure and it was found the THSR wave function is almost 100% equivalent to the old $3\alpha$ cluster-model wave functions [9]. It is discussed that the recent quantum Monte-Carlo calculation gives the wave function of the Hoyle state [15] whose density distribution is quite similar to that of the $3\alpha$ THSR wave function namely that of the old $3\alpha$ cluster-model wave function [7]. We also review recent studies of the breathing-like excited state [16–18] of the Hoyle state which is assigned to the $0_3^+$ state of $^{12}$C. Next we review the research of the $4\alpha$ condensate-like state in $^{16}$O which is assigned to the observed 6th $0^+$ state at 15.1 MeV. We explain how this assignment of the $4\alpha$ condensatelike state is supported by theoretical calculations and their comparison with experiments. Theoretical calculations are those by $4\alpha$ OCM (orthogonality condition model) [19] and by the GCM with $4\alpha$ THSR wave functions [20] with respect to the size parameter in Eq. (4).

The THSR wave function is a very novel cluster wave function and it has some different features compared with the traditional cluster model. In Sec. 3, the characters of the THSR wave function will be discussed. Firstly, the THSR wave function has the shell-model limit when the size parameter becomes very small while it also has the free-cluster limit when the size parameter becomes very large. Secondly, it was found that the prolate and oblate THSR wave functions are almost equivalent after angular-momentum projection. This is because the rotation-average of a prolate THSR wave function is almost equivalent to an oblate THSR wave function. Thirdly, although the original THSR wave function has the positive-parity character, by introducing another shift parameter, the THSR wave function can also handle the negative-parity cluster states. At last, we also introduce the single-particle property of the THSR wave function of $^{13}$C.

In Sec. 4, we discuss that the THSR wave function can describe very well also non-gaslike ordinary cluster states with spatial localization of clusters. It is explained that the spatial localization of clusters in two-cluster systems is due to the kinematic effect of inter-cluster Pauli principle. Namely the inter-cluster Pauli principle gives rise to the Pauli-forbidden states of the inter-cluster relative motion which are against to the close approach of two clusters. With this understanding of the spatial localization of clusters, it is explained that the dynamical motion of clusters is of nonlocalized nature just like in the case of gas-like states. We review that this understanding was obtained from the study of the inversion doublet bands in $^{20}$Ne [21] by the use of the THSR wave function for $^{16}$O+$\alpha$ clustering [10, 22]. We explain that the inversion doublet bands in $^{20}$Ne have been regarded as a convincing evidence of the spatial localization of clusters since the intrinsic structure with $^{16}$O+$\alpha$ clustering has parity-violating deformation. The equivalence of the RGM/GCM wave function with the THSR wave function shown in Eq. (3) cast a strong doubt on the use of the energy curve by the Brink wave function for determining the spatial localization of clusters. The $^{16}$O+$\alpha$ Brink wave function has the inter-cluster distance parameter $\boldsymbol{D}$ and the optimum value of $|\boldsymbol{D}|$ is obtained from the minimum energy point of the energy curve. It is explained that in Ref. [10] it was shown that this way of determining the inter-cluster distance is misleading and cannot be adopted. Since we prove that the THSR wave function is equivalent to the shell-model wave function in the limit of the smallest value of the size parameter, we see that the THSR wave function can describe, in addition to the gas-like cluster states, the ordinary non-gaslike cluster states and also shell-model-like states.

In Sec. 5, we explain the container model of cluster dynamics. We discuss the solutions of the Hill-Wheeler equation with respect to the single size parameter $B$ in the case of $3\alpha$ and $4\alpha$ systems. The solutions of the Hill-Wheeler equation especially in the case of the $4\alpha$ system show us that, for describing variety of observed cluster states, the THSR wave function should be extended from the simple form with one kind of size parameter to the forms of two or more kinds of size parameters. We review some recent results obtained by using extended THSR wave functions having more than one size parameters. We see that in the case of $^{12}$C the extended THSR calculations predict rich spectra of $3\alpha$ states above the Hoyle state which are in good accordance with experiments when observed data are available. We also report some recent results of the extended THSR calculations in the case of $^{16}$O. The extended THSR calculations are necessary for the study of neutron-rich nuclei because the container for only clusters needs to be different from the containers accommodating valence neutrons. We review the studies of neutron-rich Be isotopes. We also introduce the new development of analytical treatment of the THSR wave function.

In Sec. 6, we discuss the study of evolution of cluster structure by the container model in general. We discuss the duality property of the shell-model ground state which is well described by the THSR wave function as is seen in the fact that the THSR wave function in the limit of the smallest value of the size parameters is equivalent to the shell-model wave function. The important ingredients of the container model is firstly the nonlocalized motion of clusters which is described by the Hill-Wheeler equation of motion with THSR wave functions and secondly the inter-cluster Pauli principle coming from the Fermi statistics of nucleons. We explain that the actions of the inter-cluster Pauli principle are well studied by calculating the Pauli-allowed states of cluster systems, not only two-cluster systems but also many-cluster systems. Based on these studies we discuss the evolution of cluster structure from the ground state with shell-model structure to excited states composed of two clusters and further to excited states composed of many clusters including the gas-like states of clusters. Finally, we give a summary in Sec. 7.

## 2 Hoyle state and gas-like structure of clusters

In this section, we discuss gas-like-structure states with Bose-Einstein condensate character of $\alpha$ clusters. According to the Ikeda threshold rule [23], gas-like $n\alpha$ cluster states in self-conjugate $4n$ nuclei are expected to appear around the $n\alpha$ decaying threshold energy. These energies are in general low enough for the spectroscopic study to be possible, whereas when we consider nucleon gas, its excitation energy is expected to be much higher. The typical example is the Hoyle state in $^{12}$C, whose excitation energy is 7.65 MeV, located at 0.38 MeV above the $3\alpha$ threshold energy. The structure study of the Hoyle state is done by introducing the so-called THSR wave function, and its gas-like and Bose-Einstein condensate characters are investigated.

### 2.1 $3\alpha$ condensate-like structure of the Hoyle state

#### 2.1.1 Reduced $\alpha$-decay width and S-wave dominant structure of the Hoyle state

The Hoyle state, which is predicted by Fred Hoyle, is known to play a crucial role in the helium burning in red giant stars [24, 25]. This state is very closely located at the $3\alpha$ breakup threshold and is well known as one of the mysterious $0^+$ states in light nuclei. The understanding of its structure has been actually one of the most difficult and challenging problems of nuclear physics. Its small excitation energy of 7.65 MeV has been regarded to be difficult to be explained by the shell model. The no-core shell model, which is the most advanced modern shell-model approach, has so far not succeeded to reproduce it [26, 27]. About 50 years ago, Morinaga proposed a famous idea that this state was to be assigned as a linear-chain state of $3\alpha$ clusters [28, 29]. However, a simple theoretical assumption of linear-chain structure turned out to give the reduced $\alpha$-decay width of at most one-third of the Wigner limit value, while from observation it is much larger and exceeds the Wigner limit value [30]. This is because the linear-chain state necessarily includes high partial waves between $^{8}$Be and $\alpha$ clusters, whereas only the $S$ wave contributes to the $\alpha$ decay, due to the small $Q$ value for the Hoyle state.

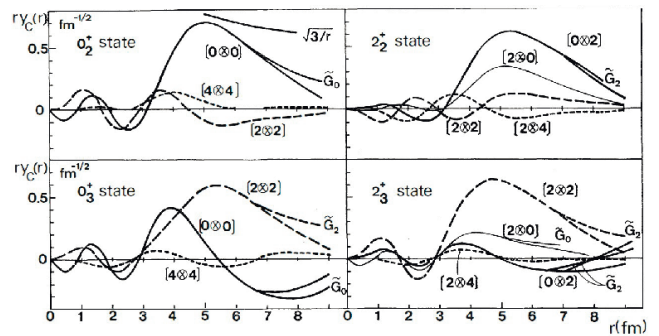

**Fig. 4** The reduced width amplitudes $r\mathcal{Y}_{c=[L,l]_J}(r)$ of $^{8}$Be+$\alpha$ channels. Figure was reproduced from Ref. [2].

The observed large $\alpha$-decay reduced width of the Hoyle state in the $^{8}$Be$(0^+)+\alpha$ channel was successfully reproduced by semi-microscopic three-body calculation [12], called the Orthogonality Condition Model (OCM) [31–33]. This $3\alpha$ OCM calculation gave the $S$-wave dominant structure between $^{8}$Be$(0^+_1)$ and $\alpha$. Since $^{8}$Be$(0^+_1)$ is composed of loosely coupled two $\alpha$ clusters in relative $S$ wave, the Hoyle state was concluded to have a loosely coupled $3\alpha$ structure in relative $S$ waves with large spatial extension, namely a gas-like structure of the $3\alpha$ clusters, which is quite different from the 3$\alpha$ linear-chain structure. This understanding of the structure of the Hoyle state was confirmed, a few years later, by full microscopic 3$\alpha$ calculations, i.e. the 3$\alpha$ RGM calculation [14] and the 3$\alpha$ Brink-GCM calculations [13], where the Brink basis function is used [34]. The 3$\alpha$ Brink-GCM calculation with resonance boundary condition was also performed by Brussels group [35].

**Table 1** Reproduction of the $^{12}$C data by 3$\alpha$ calculations of Refs. [14] and [13]. The numbers of r.m.s. in Ref. [13], $R_{\rm rms}$, are arranged to matter radii, where the proton size effect, $\sqrt{\langle r^2 \rangle_p} = 0.813$ fm, is subtracted.

|  | Exp. | 3$\alpha$ RGM [14] | 3$\alpha$ Brink-GCM [13] |
|---|---|---|---|
| Excitation energy ($0_2^+$) (MeV) | 7.65 | 7.7 | 8.6 |
| Width ($0_2^+$) (eV) | 8.7 ± 2.7 | 7.7 | 7.7 |
| $M(0_2^+ \to 0_1^+)$ (fm$^2$) | 5.4 ± 0,2 | 6.7 | 6.6 |
| $B(E2 : 0_2^+ \to 2_1^+)$ (e$^2$ fm$^4$) | 13 ± 4 | 5.6 | 3.5 |
| $B(E2 : 2_1^+ \to 0_1^+)$ (e$^2$ fm$^4$) | 7.8 | 9.3 | 8.0 |
| $R_{\rm rms}(0_1^+)$ (fm) | 2.43 | 2.40 | 2.40 |
| $R_{\rm rms}(0_2^+)$ (fm) |  | 3.47 | 3.40 |

Both methods, the 3$\alpha$ RGM and 3$\alpha$ Brink-GCM, are to fully solve the relative motions of the $\alpha$ clusters, and are proved to be mutually equivalent [36, 37]. For the former method, the 3$\alpha$ RGM wave function is obtained by solving the following equation of motion,

$$\langle \phi^3(\alpha) | (H - E) | \mathcal{A}\{\chi_{3\alpha}(\boldsymbol{\xi}_1, \boldsymbol{\xi}_2) \phi^3(\alpha)\} \rangle = 0, \quad (5)$$

where $\phi^3(\alpha) \equiv \phi(\alpha_1)\phi(\alpha_2)\phi(\alpha_3)$ and $\phi(\alpha_i)$ is the internal wave function of the $\alpha$ cluster with the four nucleons inside it being constrained to $(0s)^4$ harmonic oscillator configuration, and $\chi_{3\alpha}(\boldsymbol{\xi}_1, \boldsymbol{\xi}_2)$ is the relative wave function between the 3$\alpha$ clusters to be obtained.

These full microscopic calculations nicely reproduced not only the excitation energy of the Hoyle state but also other experimental properties including inelastic electron form factor and $E0$ and $E2$ transition properties. In Table 1 we show that the experimental data available in $^{12}$C are nicely reproduced by the 3$\alpha$ RGM and GCM calculations. The $S$-wave dominant nature of the Hoyle state mentioned above was again clarified from the microscopic point of view. Figure 4 shows the reduced width amplitude (RWA) of $^8$Be + $\alpha$ in $^{12}$C calculated with the 3$\alpha$ GCM, which is defined below,

$$\mathcal{Y}_{c=[L,l]_J}(r) = \sqrt{\frac{12!}{8!4!}} \langle \frac{\delta(r-\xi)}{r^2} [\Phi_{^8{\rm Be}}^L, Y_l(\hat{\boldsymbol{\xi}})]_J \phi(\alpha) | \Phi_{^{12}{\rm C}}^J \rangle. \quad (6)$$

For the Hoyle state ($0_2^+$ state), the $S$-wave component, which is denoted as [0, 0], is shown to be dominant and to have the largest amplitude in outer region at around 5 fm, while the inner oscillation is suppressed, which typically characterizes a cluster structure.

### 2.1.2 Alpha-condensate-like character and the THSR wave function of Hoyle state

The THSR wave function is originally introduced to represent the Bose-Einstein condensate states in self-conjugate 4$n$ nuclei in 2001 [8], and later it was extended so as to include deformation [38], in which the following operator form is assumed:

$$|\Psi_{n\alpha}^{\rm THSR}\rangle = (C_\alpha^\dagger)^n |\text{vac}\rangle, \quad (7)$$

$$C_\alpha^\dagger = \int d\boldsymbol{R} \exp\left[-\sum_{k=x,y,z} \frac{R_k^2}{\beta_k^2}\right] \int d\boldsymbol{r}_1 \cdots d\boldsymbol{r}_4 \times$$
$$\varphi_{0s}(\boldsymbol{r}_1 - \boldsymbol{R}, \nu) a_{\sigma_1 \tau_1}^\dagger(\boldsymbol{r}_1) \cdots \varphi_{0s}(\boldsymbol{r}_4 - \boldsymbol{R}, \nu) a_{\sigma_4 \tau_4}^\dagger(\boldsymbol{r}_4) \quad (8)$$

where $a_{\sigma\tau}^\dagger(\boldsymbol{r})$ is a creation operator of one nucleon with spin $\sigma$, isospin $\tau$ at a spatial point $\boldsymbol{r}$, and $\varphi_{0s}(\boldsymbol{r} - \boldsymbol{R}, \nu)$ is a $0s$ harmonic oscillator (H.O.) wave function centered at $\boldsymbol{R}$ with the form of $\varphi_{0s}(\boldsymbol{r}, \nu) = (2\nu/\pi)^{3/4} \exp[-\nu \boldsymbol{r}^2]$. The coordinate representation of this THSR state $|\Psi_{n\alpha}^{\rm THSR}\rangle$ is expressed as,

$$\langle \boldsymbol{r}_1 \sigma_1 \tau_1, \cdots \boldsymbol{r}_{4n} \sigma_{4n} \tau_{4n} | \Psi_{n\alpha}^{\rm THSR} \rangle = \int d\boldsymbol{R}_1 \cdots d\boldsymbol{R}_n$$
$$\times \exp\left[-\sum_{i=1}^n \sum_k^{x,y,z} \frac{R_{ik}^2}{\beta_k^2}\right] \Psi_{n\alpha}^{\rm B}(\boldsymbol{R}_1, \cdots, \boldsymbol{R}_n), \quad (9)$$

where $\Psi_{n\alpha}^{\rm B}(\boldsymbol{R}_1, \cdots, \boldsymbol{R}_n)$ is the Brink wave function in the $n\alpha$ cluster system [34]. The $n\alpha$ Brink wave function can be written in a form of Slater determinant and another equivalent form as shown below,

$$\Psi_{n\alpha}^{\rm B}(\boldsymbol{R}_1, \cdots, \boldsymbol{R}_n)$$
$$= \det[\varphi_{0s}(\boldsymbol{r}_1 - \boldsymbol{R}_1, \nu)\sigma_1 \tau_1 \cdots \varphi_{0s}(\boldsymbol{r}_{4n} - \boldsymbol{R}_n, \nu)\sigma_{4n}\tau_{4n}] \quad (10)$$
$$= \mathcal{A}[\varphi_{0s}(\boldsymbol{X}_1 - \boldsymbol{R}_1, 4\nu) \cdots \varphi_{0s}(\boldsymbol{X}_n - \boldsymbol{R}_n, 4\nu) \phi(\alpha_1) \cdots \phi(\alpha_n)] \quad (11)$$

where $\boldsymbol{X}_i$ and $\phi(\alpha_i)$ ($i = 1, \cdots, n$) are the center-of-mass (c.o.m.) coordinates and the internal wave function of the $i$-th $\alpha$ cluster, respectively, and $\phi(\alpha_i)$ is described by the $(0s)^4$ H.O. shell-model wave function as follows:

$$\phi(\alpha_i) \propto \exp[-\nu \sum_{j=1}^{4}(\boldsymbol{r}_{4(i-1)+j}-\boldsymbol{X}_i)^2]\sigma_{4i-3}\tau_{4i-3}\cdots\sigma_{4i}\tau_{4i}, \quad (12)$$

and $\mathcal{A}$ is the antisymmetrizer exchanging the nucleons belonging to different $\alpha$ clusters. The integration over the position parameters $\boldsymbol{R}_1,\cdots,\boldsymbol{R}_n$ in Eq. (9) can simply be performed by substituting Eq. (11) into Eq. (9), which leads to

$$\begin{aligned}\Psi_{n\alpha}^{\text{THSR}} &= \langle \boldsymbol{r}_1\sigma_1\tau_1,\cdots\boldsymbol{r}_{4n}\sigma_{4n}\tau_{4n}|\Psi_{n\alpha}^{\text{THSR}}\rangle \\ &= \mathcal{A}\Big\{\prod_{i=1}^{n}\int d\boldsymbol{R}_i\exp\Big[-\sum_{k=}^{x,y,z}\frac{R_{ik}^2}{\beta_k^2}\Big]\varphi_{0s}(\boldsymbol{X}_i-\boldsymbol{R}_i,4\nu)\phi(\alpha_i)\Big\} \\ &\propto \mathcal{A}\Big\{\prod_{i=1}^{n}\exp\Big[-\sum_{k=}^{x,y,z}\frac{2X_{ik}^2}{B_k^2}\Big]\phi(\alpha_i)\Big\} \\ &= g(\boldsymbol{X}_G)\Phi_{n\alpha}^{\text{THSR}},\end{aligned} \quad (13)$$

$$B_k^2 = b^2 + 2\beta_k^2, \quad b = 1/\sqrt{2\nu}$$

with,

$$\begin{aligned}\Phi_{n\alpha}^{\text{THSR}} &\propto \mathcal{A}\Big\{\prod_{i=1}^{n}\exp\Big[-\sum_{k=}^{x,y,z}\frac{2(X_{ik}-X_{Gk})^2}{B_k^2}\Big]\phi(\alpha_i)\Big\} \\ &= \mathcal{A}\Big\{\prod_{i=1}^{n-1}\exp\Big[-\mu_i\sum_{k=}^{x,y,z}\frac{2\xi_{ik}^2}{B_k^2}\Big]\prod_{i=1}^{n}\phi(\alpha_i)\Big\},\end{aligned} \quad (14)$$

and $g(\boldsymbol{X}_G) \propto \exp(-2n\sum_{k=x,y,z}X_{Gk}^2/B_k^2)$, where $\boldsymbol{\xi}_i = \boldsymbol{X}_i-\sum_{j=i+1}^{n}\boldsymbol{X}_j/(n-i)$ and $\mu_i = (n-i)/(n-i+1)$ for $i = 1,\cdots,n-1$, and $\boldsymbol{X}_G = \sum_{j=1}^{n}\boldsymbol{X}_j/n$. $g(\boldsymbol{X}_G)$ and $\Phi_{n\alpha}^{\text{THSR}}$ are the total spurious c.o.m. wave function to be eliminated and the c.o.m. free form of the THSR wave function, respectively. The above coordinate representation of the THSR wave function shows that all the constituent $\alpha$ clusters take an identical motion, occupying the same deformed orbit, $\exp\{-2\sum_{k=x,y,z}(X_k-X_{Gk})^2/B_k^2\}$ with the size parameter $B_k$ ($k=x,y,z$), under the condition that the effect of the antisymmetrizer $\mathcal{A}$ is negligible. This condition is satisfied when the size parameter $B_k$ is much larger than the size parameter of the $\alpha$ cluster, $b$, or equivalently $\beta_k \gg 0$. In this situation, the $\alpha$ clusters can move rather freely inside a mean-field-like potential that is produced self-consistently from a nucleon-nucleon interaction. The parameters $B_k$ or $\beta_k$ then parameterize the width of the mean-field-like potential.

On the contrary, if the width parameters $B_k$ are small enough, the $\alpha$ clusters are resolved by the stronger effect of the antisymmetrizer $\mathcal{A}$, so that there only remains the $\alpha$-cluster-like correlation. Furthermore in the limit of $B_k \to b$ ($k=x,y,z$), or $\beta_k \to 0$ ($k=x,y,z$), the THSR wave function Eq. (13) coincides with the shell-model Slater determinant. This can be easily proved by considering the well known fact that the Brink wave function $\Psi_{n\alpha}^{\text{B}}(\boldsymbol{R}_1,\cdots,\boldsymbol{R}_n)$ coincides with the shell-model Slater determinant in the limit of $\boldsymbol{R}_i \to 0$ ($i=1,\cdots,n$). By using an alternative expression of the delta function, $\lim_{\beta\to 0}(1/\sqrt{\pi\beta^2})\exp(-R^2/\beta^2) = \delta(R)$, this limiting operation $\beta_k \to 0$ ($k=x,y,z$) in Eq. (9) gives the Brink wave function $\Psi_{n\alpha}^{\text{B}}(\boldsymbol{R}_1,\cdots,\boldsymbol{R}_n)$ in the limit of $\boldsymbol{R}_i \to 0$ ($i=1,\cdots,n$), namely the shell-model Slater determinant.

We should note that this wave function has a parity symmetric form, as is clearly understood from this coordinate representation in Eq. (13). We later extend this THSR wave function to parity-violating forms, to discuss $\alpha + {}^{16}\text{O}$ inversion doublet bands of ${}^{20}\text{Ne}$ and Be isotopes in the following sections. While this form of Eq. (13) also allows us to grasp easily its physical picture such as mentioned above, it is very difficult to handle it as it is in practical calculations, since Eq. (13) is written in a general form of $4n$-body microscopic wave function so that the $4n$ nucleons should be explicitly antisymmetrized by the operator $\mathcal{A}$. Instead for practical calculations it is much better to start from the form of the Brink wave function, Eq. (9), since its Slater determinant form, Eq. (10), makes the calculations of matrix elements be much more tractable.

Starting from the form of Eq. (9), we can use the following transformation, to factorize the spurious c.o.m. motion, like Eq. (13),

$$\begin{aligned}\Psi_{n\alpha}^{\text{THSR}} = &\int d\boldsymbol{R}_G \exp\Big[-n\sum_{k}^{x,y,z}\frac{(R_{Gk})^2}{\beta_k^2}\Big] \times \\ &\varphi_{0s}(\boldsymbol{X}_G-\boldsymbol{R}_G,4n\nu)\int d\widetilde{\boldsymbol{R}}_1\cdots d\widetilde{\boldsymbol{R}}_{n-1} \times \\ &\exp\Big[-\sum_{i=1}^{n-1}\mu_i\sum_{k=}^{x,y,z}\frac{\widetilde{R}_{ik}^2}{\beta_k^2}\Big]\Phi_{n\alpha}^{\text{B}}(\widetilde{\boldsymbol{R}}_1,\cdots,\widetilde{\boldsymbol{R}}_{n-1}),\end{aligned} \quad (15)$$

where $\{\widetilde{\boldsymbol{R}}\}$ is the set of Jacobi coordinates transformed from the position-parameter set $\{\boldsymbol{R}\}$, with the relation, $\widetilde{\boldsymbol{R}}_i = \boldsymbol{R}_i - \sum_{j=i+1}^{n}\boldsymbol{R}_j/(n-i)$ for $i = 1,\cdots,n-1$ and $\boldsymbol{R}_G = \sum_{j=1}^{n}\boldsymbol{R}_j/n$, and $\Phi_{n\alpha}^{\text{B}}(\widetilde{\boldsymbol{R}}_1,\cdots,\widetilde{\boldsymbol{R}}_{n-1})$ is the internal part of the $n\alpha$ Brink wave function defined by,

$$\begin{aligned}\Phi_{n\alpha}^{\text{B}}(\widetilde{\boldsymbol{R}}_1,\cdots,\widetilde{\boldsymbol{R}}_{n-1}) = \\ \varphi_{0s}^{-1}(\boldsymbol{X}_G-\widetilde{\boldsymbol{R}}_n,4n\nu)\Psi_{n\alpha}^{\text{B}}(\boldsymbol{R}_1,\cdots,\boldsymbol{R}_n).\end{aligned} \quad (16)$$

While in Eq. (15) the first half which is in the first line is nothing but the spurious total c.o.m. wave function, $g(\boldsymbol{X}_G)$ in Eq. (13), the latter half in the second line is the c.o.m. free THSR wave function defined in Eq. (14), i.e.,

$$\Phi_{n\alpha}^{\text{THSR}} = \int d\widetilde{\boldsymbol{R}}_1\cdots d\widetilde{\boldsymbol{R}}_{n-1}\exp\Big[-\sum_{i=1}^{n-1}\mu_i\sum_{k}^{x,y,z}\frac{\widetilde{R}_{ik}^2}{\beta_k^2}\Big] \times \Phi_{n\alpha}^{\text{B}}(\widetilde{\boldsymbol{R}}_1,\cdots,\widetilde{\boldsymbol{R}}_{n-1}). \quad (17)$$

$$\Phi_{n\alpha}^{JM}(\boldsymbol{\beta}) = \mathcal{N} P_{MK}^J \Phi_{n\alpha}^{\text{THSR}} = \mathcal{N} \int d\Omega D_{MK}^{J*}(\Omega) \mathcal{A} \Big\{ \prod_{i=1}^{n-1} \exp\Big[-\mu_i \sum_{k}^{x,y,z} \frac{2(\widehat{R}(\Omega)\boldsymbol{\xi}_i)_k^2}{B_k^2}\Big] \prod_{i=1}^{n} \phi(\alpha_i) \Big\}$$

$$= \mathcal{N} \int d\Omega D_{MK}^{J*}(\Omega) \int d\widetilde{\boldsymbol{R}}_1 \cdots d\widetilde{\boldsymbol{R}}_{n-1} \exp\Big[-\sum_{i=1}^{n-1}\mu_i \sum_{k}^{x,y,z} \frac{\widetilde{R}_{ik}^2}{\beta_k^2}\Big] \Phi_{n\alpha}^{\text{B}}(\widehat{R}^{-1}(\Omega)\widetilde{\boldsymbol{R}}_1, \cdots, \widehat{R}^{-1}(\Omega)\widetilde{\boldsymbol{R}}_{n-1}),$$

$$= \mathcal{N} \int d\Omega D_{MK}^{J*}(\Omega) \int d\widetilde{\boldsymbol{R}}_1 \cdots d\widetilde{\boldsymbol{R}}_{n-1} \exp\Big[-\sum_{i=1}^{n-1}\mu_i \sum_{k}^{x,y,z} \frac{(\widehat{R}(\Omega)\widetilde{\boldsymbol{R}}_i)_k^2}{\beta_k^2}\Big] \Phi_{n\alpha}^{\text{B}}(\widetilde{\boldsymbol{R}}_1, \cdots, \widetilde{\boldsymbol{R}}_{n-1}), \tag{18}$$

This is written in the form of superposing c.o.m. free Brink wave functions, which allows us to easily perform any calculations. However, one might be afraid that the c.o.m. free Brink wave function $\Phi_{n\alpha}^{\text{B}}(\widetilde{\boldsymbol{R}}_1, \cdots, \widetilde{\boldsymbol{R}}_{n-1})$ in the above equation deviates from the pure Slater determinant, $\Psi_{n\alpha}^{\text{B}}(\boldsymbol{R}_1, \cdots, \boldsymbol{R}_n)$, and hence the above non-Slater-determinant form is difficult to be handled in practical calculations. However, once we know an analytical formula for the matrix element of an operator $\mathcal{O}$, with a pure Slater determinant, like $\langle\Psi^{\text{B}}|\mathcal{O}|\Psi^{\text{B}'}\rangle$, the matrix element to be desired, $\langle\Phi^{\text{B}}|\mathcal{O}|\Phi^{\text{B}'}\rangle$, is also easily derived, since the transformation matrix $U: \{\boldsymbol{R}\} \to \{\widetilde{\boldsymbol{R}}\}$ is also analytically given [7].

Since now the parameter in the THSR wave function $\boldsymbol{\beta}$ can be taken in an anisotropic way, the THSR wave function $\Phi_{n\alpha}^{\text{THSR}}$ in Eq. (17) is not rotationally symmetric. The recovery of the rotational symmetry can be done by employing the following well known angular-momentum projection method: with $\boldsymbol{\beta} \equiv (\beta_x, \beta_y, \beta_z)$, $\Omega$, $\widehat{R}(\Omega)$, $D_{MK}^J(\Omega)$, and $\mathcal{N}$, are Euler angle, a rotational operator, Wigner $D$-function, and a normalization constant, respectively. Then in order to calculate a matrix element of any operator $\mathcal{O}$ for the THSR wave function $\Phi_{n\alpha}^{\text{THSR},(JM)}$, i.e. $\langle\Phi_{n\alpha}^{JM}|\mathcal{O}|\Phi_{n\alpha}^{JM}\rangle$, one should first calculate the matrix element for the intrinsic Brink wave function $\Phi^{\text{B}}$, i.e. $\langle\Phi^{\text{B}}|\mathcal{O}|\Phi^{\text{B}}\rangle$. This is basically represented as a form of Gaussians. Thus integrating over the set of variables $\{\widetilde{\boldsymbol{R}}\}$ is shown to result in multiple Gaussian integral, and therefore essentially it can be performed in an analytical way, while the integration over the Euler angle $\Omega$ can be numerically performed. This aspect will be discussed in more detail in Sec. 3.

All calculations in this section are performed within axially symmetric deformation, where the $z$-axis is taken as the symmetry axis: $\beta_\perp \equiv \beta_x = \beta_y \neq \beta_z$. The above relation of Eq. (18) is simplified and $\Phi_{n\alpha}^{JM}$ can be redefined as,

$$\Phi_{n\alpha}^{JM}(\beta_\perp, \beta_z) = \mathcal{N} \int d\cos\theta d_{M0}^J(\theta) \int d\widetilde{\boldsymbol{R}}_1 \cdots d\widetilde{\boldsymbol{R}}_{n-1} \times$$

$$\exp\Big[-\sum_{i=1}^{n-1}\mu_i \sum_{k}^{x,y,z} \frac{(\widehat{R}_y(\theta)\widetilde{\boldsymbol{R}}_i)_k^2}{\beta_k^2}\Big] \Phi_{n\alpha}^{\text{B}}(\widetilde{\boldsymbol{R}}_1, \cdots, \widetilde{\boldsymbol{R}}_{n-1}), \tag{19}$$

with a redefined normalization constant $\mathcal{N}$.

The energy eigenstates can be represented by superposing the above wave functions, Eq. (19), as follows:

$$\Phi_{n\alpha,\lambda}^{JM} = \sum_{\beta_\perp,\beta_z} f_\lambda^J(\beta_\perp, \beta_z) \Phi_{n\alpha}^{JM}(\beta_\perp, \beta_z). \tag{20}$$

The coefficients $f_\lambda^J$ can then be determined by solving the following Hill-Wheeler equation,

$$\sum_{\beta_\perp',\beta_z'} \langle\Phi_{n\alpha}^{JM}(\beta_\perp, \beta_z)|(H-E)|\Phi_{n\alpha}^{JM}(\beta_\perp', \beta_z')\rangle f_\lambda^J(\beta_\perp', \beta_z') = 0, \tag{21}$$

where the microscopic Hamiltonian $H = T - T_G + \sum_{i<j}^{A} V_{ij}^{(NN)} + \sum_{i<j}^{A} V_{ij}^{(C)}$ is composed of the kinetic energy with the c.o.m. kinetic energy subtracted, $T - T_G$, nucleon-nucleon effective interaction $V_{ij}^{(NN)}$, and Coulomb interaction $V_{ij}^{(C)}$. The important criterion to choose an effective nucleon-nucleon interaction is to reasonably reproduce the $\alpha+\alpha$ scattering phase shift [39].

### 2.1.3 Equivalence of the $3\alpha$ THSR wave function to $3\alpha$ RGM/GCM wave functions

As we discussed in Sec. 2.1.1, the fully microscopic cluster model calculations of $3\alpha$ RGM and $3\alpha$ GCM were performed for $^{12}$C around 40 years ago [2, 13, 14, 35]. The wave function for the Hoyle state reproduces the experimental data available, such as the binding energy, $\alpha$-decay width, monopole transition strength to the ground state, and inelastic electron scattering [2]. We showed in Sec. 2.1.1 that the Brink-GCM wave function for the Hoyle state dominantly has $^8$Be$(0^+) + \alpha$ component, together with a large r.m.s. radius (3.4 fm), compared with the one of the ground state (2.4 fm). Thus the Hoyle state has been considered to have a well-developed $3\alpha$-cluster structure, which are loosely coupled in relative $S$-waves, rather than the $3\alpha$-linear-chain structure. In the $3\alpha$ RGM, or equivalently in the $3\alpha$ Brink-GCM, relative motions between the $\alpha$ clusters are solved without any model assumption, as shown in Eq. (5) in Sec. 2.1.1, in which the $3\alpha$ RGM wave function can be expressed in the following form,

$$\Psi_{3\alpha}^{\rm RGM} = \mathcal{A}[\chi_{3\alpha}(\boldsymbol{\xi}_1, \boldsymbol{\xi}_2)\phi(\alpha_1)\phi(\alpha_2)\phi(\alpha_3)], \quad (22)$$

where the Jacobi coordinates $\boldsymbol{\xi}_i = \boldsymbol{X}_i - \sum_{j=i+1}^2 \boldsymbol{X}_j/(n-i)$ for $i = 1, 2$. On the other hand, as we can easily see from the $r$-space representation of the $3\alpha$ THSR wave function in Eq. (13) for $n = 3$, the relative wave function between the $3\alpha$ clusters in the $3\alpha$ THSR wave function is written like,

$$\chi_{3\alpha}^{\rm THSR}(\boldsymbol{\xi}_1, \boldsymbol{\xi}_2) \propto P_{M=K=0}^{J=0} \times$$
$$\exp\left[-2\sum_{i=1}^3\left\{\frac{(X_{ix}-X_{Gx})^2 + (X_{iy}-X_{Gy})^2}{b^2 + 2\beta_\perp^2} + \frac{(X_{iz}-X_{Gz})^2}{b^2 + 2\beta_z^2}\right\}\right]$$
$$= P_{M=K=0}^{J=0} \exp\left[-2\sum_{i=1}^2 \mu_i\left\{\frac{\xi_{ix}^2 + \xi_{iy}^2}{b^2 + 2\beta_\perp^2} + \frac{\xi_{iz}^2}{b^2 + 2\beta_z^2}\right\}\right]. \quad (23)$$

Table 2 Comparison of the binding energies $E$, r.m.s. radii of matter ($R_{\rm rms}$), and monopole matrix elements ($M(0_2^+ \to 0_1^+)$) between calculated with the THSR wave function given by solving Hill-Wheeler equation, Eq. (20), calculated with the RGM/GCM wave function, and of the corresponding experimental data. $E_{3\alpha}^{\rm th.}$ is the calculated $3\alpha$ threshold energy. Force 1 and 2 denote Volkov No.1 [42] and slightly modified Volkov No.2 forces, respectively [2].

|  |  | Force 1, ($E_{3\alpha}^{\rm th.} = -81.01$ MeV) | | Force 2, ($E_{3\alpha}^{\rm th.} = -82.04$ MeV) | | Exp. |
|---|---|---|---|---|---|---|
|  |  | THSR (Hill-Wheeler) | GCM | THSR (Hill-Wheeler) | RGM |  |
| $E$ (MeV) | $0_1^+$ | $-87.81$ | $-87.9$ | $-89.52$ | $-89.4$ | $-92.2$ |
|  | $0_2^+$ | $-79.97$ | $-79.3$ | $-81.79$ | $-81.7$ | $-84.6$ |
| $R_{\rm rms}$ (fm) | $0_1^+$ | 2.40 | 2.40 | 2.40 | 2.40 | 2.44 |
|  | $0_2^+$ | 4.44 | 3.40 | 3.83 | 3.47 |  |
| $M(0_2^+ \to 0_1^+)$ (fm$^2$) |  | 5.36 | 6.6 | 6.45 | 6.7 | 5.4 |

Obviously the special form is assumed for the relative wave function between the $3\alpha$ clusters in the THSR wave function, compared with the RGM formalism. What we try to discuss here is then how well the THSR wave function describes the $3\alpha$ RGM/GCM wave function that is the full solution of the microscopic $3\alpha$-cluster problem. And as we emphasized repeatedly in the preceding sections, we show that the THSR wave function can perfectly describe the RGM/GCM wave function for the Hoyle state [7, 9, 40, 41].

The energy spectrum and the corresponding wave functions for $^{12}$C in the THSR ansatz explained in the previous section are obtained by solving the Hill-Wheeler equation Eq. (21). Table 2 shows comparison of the calculated energies, r.m.s. radii, and monopole matrix elements with those of the $3\alpha$ RGM/GCM calculation and the experimental data. In the GCM [13] and RGM [14] different effective nucleon-nucleon forces are adopted. In the former Volkov No.1 with Majorana parameter $M = 0.575$ is adopted and in the latter a slightly modified version of Volkov No.2 with $M = 0.59$ [2] is adopted.

Hence we adopt the same effective forces as adopted in the GCM and RGM to give precise comparison of our results with their results. We denote the former as Force 1 and the latter Force 2 in Table 2. One can see that the THSR wave function succeeds in reproducing the properties of the ground state and the Hoyle state. The large value of the r.m.s. radius for the Hoyle state indicates that the state has a volume 3 to 4 times larger than that of the ground state of $^{12}$C. It should be noted that for the Hoyle state the energies calculated with the THSR ansatz are lower than those calculated with the RGM/GCM. In particular, the THSR ansatz gives considerably lower energy, by 0.7 MeV, than the GCM calculation. This is surprising since, according to the framework of the formalism, no model is assumed with respect to the relative motions between the $3\alpha$ clusters for the RGM/GCM, and therefore, due to variational principle, binding energies obtained by RGM/GCM must be lower than those of the THSR ansatz. Also the calculated r.m.s. radius for the Hoyle state is larger than those calculated with the RGM/GCM. In particular in comparison with that of the GCM, THSR ansatz gives much larger r.m.s. radius, by 1 fm. This is because the THSR wave function can much better describe a long tail nature which the Hoyle state located slightly above the $3\alpha$ threshold is expected to have, than the GCM wave function, in which the finite number of spatially localized $3\alpha$-cluster basis functions, (Brink wave functions), are superposed in the practical calculation.

For the ground state the THSR wave function gives slightly lower binding energy than the RGM wave function obtained, though in comparison with the GCM wave function, it gives slightly higher energy. This suggests that even for very compact cluster state like the ground state the THSR ansatz works very well. We will discuss later an extended version of the THSR wave function, in which a $2\alpha + \alpha$-like correlation is taken into account following a "container picture" discussed in Sec. 5, which is shown to give identical value to the one of the GCM [43].

**Fig. 5** Experimental values of inelastic form factor in $^{12}$C to the Hoyle state [44, 45] are compared with our values [46]. In our result, the Hoyle-state wave function is obtained by solving the Hill-Wheeler equation Eq. (21).

The inelastic form factor of $^{12}$C from the ground state to the Hoyle state is another important experimental information available. Rather recently those at forward angle are also measured in Darmstadt [44, 45]. Figure 5 shows the comparison between the data and the result calculated by the THSR wave function. We can see that the curve obtained by the THSR wave function is almost exactly on the experimental data, except for higher momentum region, $q \geq 2.5$ fm$^{-1}$, in which the size of nucleon is probed and hence ordinary microscopic nuclear models cannot be applied.

The THSR wave function obtained by solving the Hill-Wheeler equation Eq. (21) is expressed as a sum, like in Eq. (20), of the original form of the THSR wave function with a single parameter in Eq. (19). The concept of the $\alpha$ condensation originates from the THSR wave function with a single parameter value, which we hereafter call a single THSR wave function. It is important to know how much this $\alpha$-condensation picture kept by the single THSR wave function survives in the superposed wave function in Eq. (20). We then study how the superposed THSR wave function, which is shown to be equivalent to or better than the RGM/GCM wave function, can be approximated by the single THSR wave function.

We first calculate the following energy expectation value by the single THSR wave function in the two-parameter space $\beta_\perp, \beta_z$, with the Force 2,

$$E(\beta_\perp, \beta_z) = \langle \Phi_{3\alpha}^{J=0}(\beta_\perp, \beta_z) | H | \Phi_{3\alpha}^{J=0}(\beta_\perp, \beta_z) \rangle, \quad (24)$$

In Fig. 6(a), we show the energy contour map in the two-parameter space $\beta_\perp$ and $\beta_z$ for the above quantity. A minimum appears at a spherical position, $(\beta_\perp, \beta_z) = (1.5, 1.5$ fm$)$. This corresponds to the ground state which

**Fig. 6** (a): The contour map of the energy surface in two-parameter space $\beta_\perp (= \beta_x = \beta_y)$ and $\beta_z$, which corresponds to the ground state and is defined in Eq. (24). (b): The contour map of the energy surface in two-parameter space $\beta_\perp$ and $\beta_z$, which corresponds to the Hoyle state and is defined in Eq. (25). (c): The contour map of the squared overlap surface with the solution of the Hill-Wheeler equation for the Hoyle state, in two-parameter space $\beta_\perp$ and $\beta_z$, defined in Eq. (27). Figure was reproduced from Ref. [5].

has a compact density structure. The squared overlap between this single wave function and the ground state wave function obtained by solving the Hill-Wheeler equation, $\Phi_{3\alpha,\lambda=1}^{J=0}$ is calculated to be $|\langle\Phi_{3\alpha}^{J=0}(\beta_\perp=1.5,\beta_z=1.5\text{ fm})|\Phi_{3\alpha,\lambda=1}^{J=0}\rangle|^2 = 0.93$. This means that the single THSR wave function with a rather small value of the parameter $(\beta_\perp,\beta_z) = (1.5, 1.5\text{ fm})$ very well approximates the ground state wave function $\Phi_{3\alpha,\lambda=1}^{J=0}$, though the minimum energy $-87.68$ MeV is about 1.8 MeV higher than the energy $-89.52$ MeV obtained by solving the Hill-Wheeler equation.

In this energy surface, however, there only appears one energy minimum, which corresponds to the ground state, and there is no other minimum corresponding to the Hoyle state. This is because the orthogonality to the ground state plays an important role in the formation of the Hoyle state. We then calculate the energy surface given in a configuration space orthogonal to the ground state, which can be defined below,

$$E_\perp(\beta_\perp,\beta_z) = \frac{\langle \widehat{P}_\perp^{\text{g.s.}}\Phi_{3\alpha}^{J=0}(\beta_\perp,\beta_z)|H|\widehat{P}_\perp^{\text{g.s.}}\Phi_{3\alpha}^{J=0}(\beta_\perp,\beta_z)\rangle}{\langle \widehat{P}_\perp^{\text{g.s.}}\Phi_{3\alpha}^{J=0}(\beta_\perp,\beta_z)|\widehat{P}_\perp^{\text{g.s.}}\Phi_{3\alpha}^{J=0}(\beta_\perp,\beta_z)\rangle}, \quad (25)$$

$\widehat{P}_\perp^{\text{g.s.}} = 1 - |\Phi_{3\alpha,\lambda=1}^{J=0}\rangle\langle\Phi_{3\alpha,\lambda=1}^{J=0}|$, is the projection operator onto the functional space which is orthogonal to the ground state. In Fig. 6(b), we show the contour map of this energy surface in the two parameter space $\beta_\perp,\beta_z$. We see an energy minimum at $(\beta_\perp, \beta_z) = (5.2, 1.5\text{ fm})$ in the oblate region of the map and a second energy minimum at $(\beta_\perp,\beta_z) = (2.6, 7.5\text{ fm})$ in the prolate region. The minimum energy value is $-81.75$ MeV. This value is almost the same as the energy of $-81.79$ MeV given by solving the Hill-Wheeler equation. The minimum energy $-81.75$ MeV is close to the second minimum energy of $-81.67$ MeV, and there is a valley with an almost flat bottom connecting these two minima. This means that the energy gain due to the deformation is small. The wave functions at the two minima are almost identical to each other, and indeed, the squared overlap between the wave functions giving the two minima is calculated to be more than 0.9. The fact that the minimum energy is almost the same as the eigenenergy given by solving the Hill-Wheeler equation means that the single THSR wave function at the minimum energy position is also the same as the superposed THSR wave function given by solving the Hill-Wheeler equation.

The single THSR wave function corresponding to this minimum energy, i.e. corresponding to the Hoyle state, can be defined, in the orthogonal configuration space, as,

$$\Phi_{3\alpha,\perp}^{J=0}(\beta_\perp,\beta_z) = \frac{\widehat{P}_\perp^{\text{g.s.}}\Phi_{3\alpha}^{J=0}(\beta_\perp,\beta_z)}{\sqrt{\langle\widehat{P}_\perp^{\text{g.s.}}\Phi_{3\alpha}^{J=0}(\beta_\perp,\beta_z)|\widehat{P}_\perp^{\text{g.s.}}\Phi_{3\alpha}^{J=0}(\beta_\perp,\beta_z)\rangle}}. \quad (26)$$

We then show in Fig. 6(c) the contour map of the squared overlap surface defined below,

$$\mathcal{O}(\beta_\perp,\beta_z) = |\langle\Phi_{3\alpha,\perp}^{J=0}(\beta_\perp,\beta_z)|\Phi_{3\alpha,\lambda=2}^{J=0}\rangle|^2, \quad (27)$$

where $\Phi_{3\alpha,\lambda=2}^{J=0}$ is the Hoyle state wave function obtained by solving the Hill-Wheeler equation. Two maxima appear corresponding to the two minima in the previous figure. The largest value is 0.993 at $(\beta_\perp,\beta_z) = (5.3, 1.5\text{ fm})$. This means that the Hoyle state wave function obtained by solving the Hill-Wheeler equation is almost identical to the single THSR wave function $\Phi_{3\alpha,\perp}^{J=0}(\beta_\perp = 5.3, \beta_z = 1.5\text{ fm})$. Since we keep in mind that the wave function $\Phi_{3\alpha,\lambda=2}^{J=0}$ is practically better than the solutions of the RGM/GCM in Table 2 since the former gives deeper binding energy than those of the RGM/GCM, we can say that the Hoyle state is the best described by the single THSR wave function in the form of Eq. (26). This result gives us a crucial evidence that the Hoyle state is the $3\alpha$ condensate state such as described by the single THSR wave function in Eq. (26).

### 2.1.4 Quantum Monte-Carlo calculation of the Hoyle state

One of the most remarkable recent progresses in nuclear theory is the development of *ab-initio* approach in light nuclei. The so-called lattice Quantum Monte Carlo (QMC) approach [47, 48], which is based on chiral perturbation of an effective field theory, reasonably reproduces the low-lying spectrum of $^{12}$C including the Hoyle state [49, 50], where the authors find an obtuse triangular configuration of the $\alpha$ clusters for the Hoyle state. However this interpretation is quite a lot contradictory to the finding of many other theoretical investigations discussed up to now, that the Hoyle state has a gas-like configuration of loosely coupled $3\alpha$ clusters in relative $S$-wave.

On the other hand, the no-core symplectic model (NC-SpM), which is based on a special symmetry to use only a fraction of the model space extended beyond current no-core shell-model limits, also has a great progress in the description of the cluster states including the Hoyle state [51–53]. By the help of its symmetry guided framework, which allows for selecting physically relevant configurations, the authors in Ref. [52] adjust the energy positions of the Hoyle state and the $2_2^+$ state in $^{12}$C, and discuss the physical observables of the states. The calculated r.m.s. radius of the Hoyle state is 2.93 fm, which is sizably larger than that of the ground state (2.4 fm). The monopole transition strength is also calculated, which is 2.1 $e$fm$^2$, compared with the experimental value, 5.3 $e$fm$^2$ (see below and Table 2).

The pioneering work on the *ab-initio* theory has obviously been done by the Argonne group. In 2000, Wiringa *et al.* published the outstanding result that the $\alpha+\alpha$ cluster structure in $^8$Be is verified [54]. They take a variational approach, consisting of two steps, Variational Monte Carlo (VMC) and Green's Function Monte Carlo (GFMC) with a constrained path algorithm, in which the Argonne $v$18 two-body force (AV18) and Urbana IX (UIX) three-body force, and eventually a new class of three-body potential, Illinois model (IL7), in $p$-shell nuclei, are adopted as realistic nuclear force. Their calculations are now proved to work very well for low-lying states up to $^{12}$C, in spite of the fact that the computational cost increases exponentially with the number of particles [15].

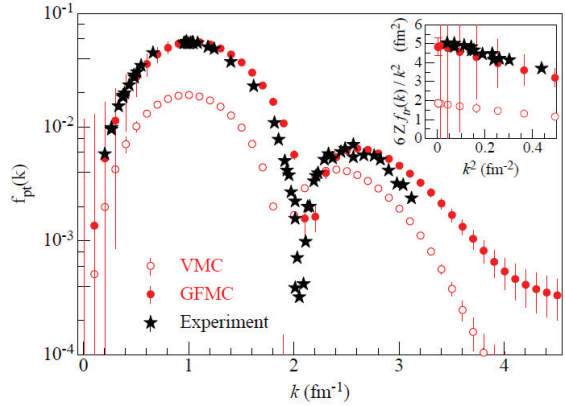

**Fig. 7** Inelastic form factor from ground to Hoyle state from GFMC [15], full circles. The open circles correspond to some approximate calculation, see Ref. [15], and the black stars represent the experimental values [44, 45].

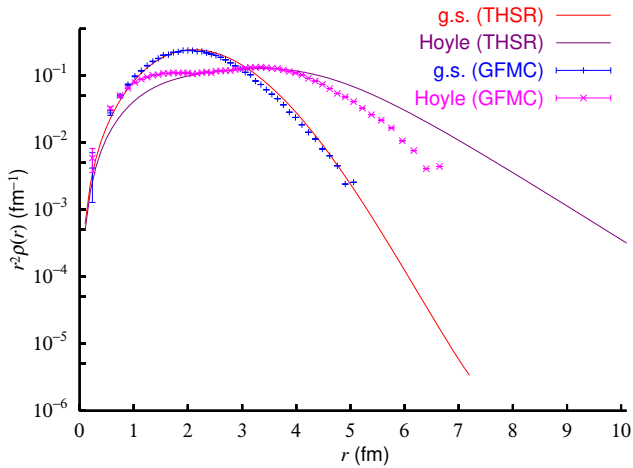

**Fig. 8** Densities of the Hoyle state with GFMC [15] (magenta diamonds), and of ground state (blue crosses), and with THSR, full lines. Figure was reproduced from Ref. [55].

In particular, their result about the Hoyle state is quite amazing. We show in Fig. 7 the inelastic electron scattering form factor, where most of the experimental points are reproduced very accurately [15] as much as the THSR wave function (see Fig. 5). In the inset the calculated $E0$ transition form factor is compared to the experimental data, which leads to $M(E0) = 5.29 \pm 0.14\ e\text{fm}^2$ at $k^2 = 0$, given in [44, 45]. We can see that the GFMC result gives excellent agreement with the data. We should keep in mind that the THSR wave function gives 5.36 $e\text{fm}^2$ and 6.45 $e\text{fm}^2$ in the case of adopting Volkov No.1 and No.2 forces, respectively (see Table 2). The calculated excitation energy of the Hoyle state is still rather high, 10.4 versus 7.65 MeV. In Fig. 8, we compare the density distributions of the Hoyle state and the ground state (weighted with $r^2$) obtained with the THSR wave function and in Ref. [15]. For the ground state both curves are in quite good agreement with each other. For the Hoyle state both curves characteristically give a plateau region between 1 and 4 fm, and both are in good agreement, though in the GFMC calculation it is more pronounced around 1-2 fm than in the THSR ansatz. Beyond 4 fm, the density in the GFMC calculation falls off more rapidly, which is the reason why the r.m.s. radius of the Hoyle state is calculated to be smaller in the GFMC calculation than in the THSR ansatz. The former calculation gives around 3.0-3.5 fm and the latter gives 3.8 fm and 4.4 fm when Volkov No. 1 and No. 2 forces are adopted, respectively (see Table 2).

### 2.1.5 Breathing-like excitation of the Hoyle state

There has been known a broad $0^+$ state above the Hoyle state, at 10.3 MeV with a width of 2.7 MeV. This state has triggered much interest of nuclear physicists, since there has been long controversy between experimentalists and theorists about the spin assignment of this state. Theorists predicted the existence of the second $J^\pi = 2^+$ state around 10 MeV almost 40 years ago by using the microscopic cluster models [13, 14, 35]. Amazingly the $J^\pi = 2^+$ state has been observed recently, as a different state from the broad 10.3 MeV state. The pioneering work on the observation was originally done by Itoh et al. [56], and later it was confirmed in several experiments [57–61]. Furthermore, Itoh et al. pointed out that the broad $0^+$ peak is decomposed into two peaks, giving the $0_3^+$ and $0_4^+$ states at 1.77 MeV and 3.29 MeV above the $3\alpha$ threshold, with the widths of 1.45 MeV and 1.42 MeV, respectively [62]. Theoretically it is very difficult to obtain both $0^+$ states several MeV above the Hoyle state, due to their broad widths. However, with proper treatments of resonances, some theoretical works succeeded in reproducing both states [16–18, 63, 64]. For example, the OCM plus complex scaling method (CSM) [65–69] properly gives two energy poles corresponding to the two $0^+$ states [63, 64]. Though the interpretation of these states is not yet converged, the result of this study suggests that the $0_3^+$ state has a higher nodal structure between $^8$Be + $\alpha$ in relative $S$-wave [63]. Also the AMD and FMD calculations suggest that the $0_4^+$ state has a bent-armed shape of the $3\alpha$ clusters, resembling the $3\alpha$ linear-chain configuration, though in these calculations the $0_3^+$ state is missing [44, 70].

The THSR ansatz could also reproduce the $0_3^+$ and $0_4^+$ states by using a so-called radius constraint method [71] that improves treatment of excited states in resonance region [16–18]. The translationally invariant THSR wave function discussed in this section is what is extended to contain the $2\alpha + \alpha$ correlation [43], as follows:

$$\Phi_{3\alpha}^{J=0}(\boldsymbol{\beta}_1, \boldsymbol{\beta}_2) = \widehat{P}^{J=0}\mathcal{A}\Big\{\prod_{i=1}^{2}\exp\Big[-\mu_i\sum_{k=x,y,z}\frac{2\xi_{ix}^2}{B_{ik}^2}\Big]\prod_{i=1}^{3}\phi(\alpha_i)\Big\}, \quad (28)$$

where $\mu_i = i/(i+1)$ and $B_{ik}^2 = b^2 + 2\beta_{ik}^2$ with $i = 1, 2$ and $k = x, y, z$, and $\widehat{P}^{J=0}$ is the angular-momentum projection operator onto $J = 0$. The degree of freedom of the cluster motions characterized by $\boldsymbol{\beta}$ is divided into $\boldsymbol{\beta}_1$ and $\boldsymbol{\beta}_2$, corresponding to $2\alpha$ ($^8$Be) and $\alpha$, respectively. Thus the case $\boldsymbol{\beta}_1 = \boldsymbol{\beta}_2$ comes back to the original THSR wave function, and this natural extension allows for including the $2\alpha + \alpha$ correlation and its asymptotic configuration. More detail of this extended THSR wave function is discussed in Sec. 5.

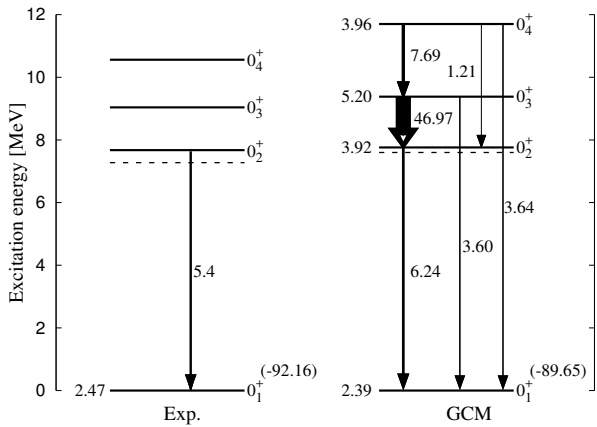

**Fig. 9** The THSR-GCM energy levels, r.m.s. radii for the mass distributions (left side of the energy levels), and the monopole transition strengths (along the transition lines) for the ground state and excited $0^+$ states in $^{12}$C. The dash lines are corresponding to the threshold energies. Figure was reproduced from Ref. [18].

In Fig. 9, the result of solving the Hill-Wheeler equation of the type of Eq. (21), in which the four kinds of parameters, $\boldsymbol{\beta}_1 = (\beta_{1x} = \beta_{1y}, \beta_{1z})$ and $\boldsymbol{\beta}_2 = (\beta_{2x} = \beta_{2y}, \beta_{2z})$ are taken as the generator coordinates, denoted as GCM, is compared with the corresponding experimental data. The so-called radius constraint method [71] mentioned above is also applied to obtain the $0_3^+$ and $0_4^+$ states. We can see that the $0_3^+$ and $0_4^+$ states are reasonably reproduced. Amazing feature of the $0_3^+$ state is the large monopole transition to the Hoyle state, whose strength is around 47 $e$fm$^2$. The r.m.s. radius of this state is calculated to be much larger than the Hoyle state. These results suggest that the $0_3^+$ state is excited from the Hoyle state by monopole-like dilatation.

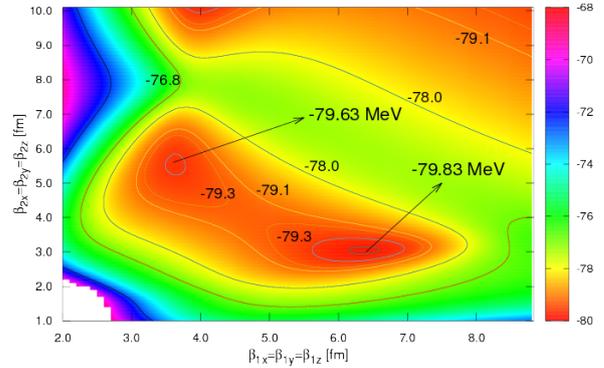

**Fig. 10** The contour plot for the $0_3^+$ state in the spherical $\beta_z$ and $\beta_2$ parameter space, which is obtained from the variation calculations of a constructed single THSR wave function orthogonalized to the ground and Hoyle states of $^{12}$C. Figure was reproduced from Ref. [18].

We should also emphasize that for the formation of the $0_3^+$ state, the orthogonality to the lower-lying states, i.e. the most compact ground state and much more dilute Hoyle state, is crucial. This can be demonstrated as follows: First we introduce the projection operator onto the orthogonal space to the ground state and the Hoyle state, like $\widehat{P}_\perp = 1 - |0_1^+\rangle\langle 0_1^+| - |0_2^+\rangle\langle 0_2^+|$. Second we construct the THSR wave function in the orthogonal space, like $\Psi_\perp(\boldsymbol{\beta}_1, \boldsymbol{\beta}_2) = \widehat{P}_\perp \Phi_{3\alpha}^{J=0}(\boldsymbol{\beta}_1, \boldsymbol{\beta}_2)$. We then calculate the energy of this THSR state, which is defined as,

$$E_\perp = \frac{\langle\Psi_\perp(\boldsymbol{\beta}_1, \boldsymbol{\beta}_2)|H|\Psi_\perp(\boldsymbol{\beta}_1, \boldsymbol{\beta}_2)\rangle}{\langle\Psi_\perp(\boldsymbol{\beta}_1, \boldsymbol{\beta}_2)|\Psi_\perp(\boldsymbol{\beta}_1, \boldsymbol{\beta}_2)\rangle}. \quad (29)$$

In Fig. 10, the contour map of the energy surface in two-parameter space $\beta_{1x} = \beta_{1y} = \beta_{1z}$ and $\beta_{2x} = \beta_{2y} = \beta_{2z}$ is shown. In construction of the projection operator $\widehat{P}_\perp$, optimal single configurations in the THSR states are taken as the $0_1^+$ and $0_2^+$ states (see Ref. [18] for more detail). We can see that two local minima appear, both of which, however, do not indicate two different modes of excitation, but are, in principle, identical state. In fact both minimum states have very large squared overlap of 84%.

Furthermore the calculated squared overlap of these local minimum states with the wave function of the $0_3^+$ state obtained by solving the Hill-Wheeler equation, which is shown in the spectra of Fig. 9, is more than 90%. This means that the $0_3^+$ state appears as the minimum state, obviously indicating that the orthogonalization to the lower states is essential, since otherwise it never appears.

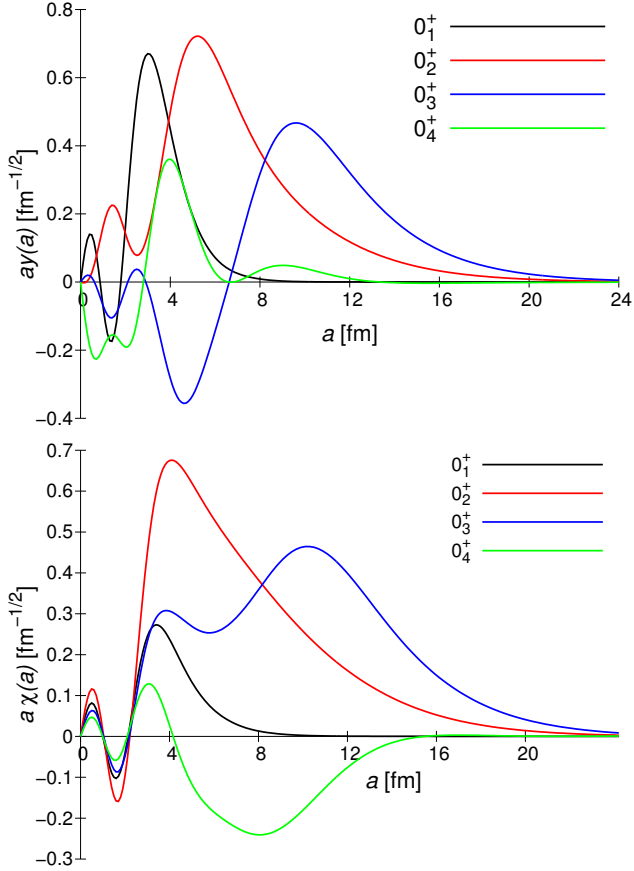

**Fig. 11** (color online). (Upper): The RWAs of the $[L, l]_J = [0, 0]_0$ channel, $\mathcal{Y}_{[0,0]_0}(r)$ in Eq. (30), for the $0_1^+$, $0_2^+$, and $0_3^+$ states. (Lower): The correlation function defined in Eq. (31), for the $0_1^+$, $0_2^+$, and $0_3^+$ states, with $\beta_{2x} = \beta_{2y} = 13.8$ fm and $\beta_{2z} = 13.7$ fm. Figure was reproduced from Ref. [18].

In order to investigate in more detail how the $0_3^+$ state is excited from the Hoyle state with the strong monopole strength, the following quantities are calculated:

$$\mathcal{Y}(a) = \sqrt{\frac{12!}{4!8!}} \left\langle \frac{\delta(\xi_2 - a)}{\xi_2^2} [\Phi_{2\alpha}, Y_0(\hat{\boldsymbol{\xi}}_2)]_0 \phi(\alpha) \Big| \Phi_{\text{GCM}}(0_\lambda^+) \right\rangle, \quad (30)$$

$$\chi(a) = \sqrt{\frac{12!}{4!4!4!}} \left\langle \frac{\delta(\xi_1 - a)}{\xi_1^2} \Phi_{2\alpha-\alpha} Y_0(\hat{\boldsymbol{\xi}}_1) \Big| \Phi_{\text{GCM}}(0_\lambda^+) \right\rangle, \quad (31)$$

where $\Phi_{2\alpha} = \mathcal{N}_0 \widehat{P}^{J=0} \mathcal{A}[\exp\{-\sum_{k=x,y,z} \xi_{1k}^2/B_k^2\} \phi^2(\alpha)]$ and $\Phi_{2\alpha-\alpha} = \widetilde{\mathcal{N}}_0 \widehat{P}^{J=0} \mathcal{A}[\exp\{-\sum_{k=x,y,z} \xi_{2k}^2/B_{2k}^2\} \phi^3(\alpha)]$, with $\mathcal{N}_0$ and $\widetilde{\mathcal{N}}_0$ normalization factors, and $B_k^2 = b^2 + \beta_k^2$ and $B_{2k}^2 = 3/4 b^2 + \beta_{2k}^2$. $\boldsymbol{\xi}_1$ and $\boldsymbol{\xi}_2$ are the Jacobi coordinates. The upper equation is nothing but the RWA of $^8$Be + $\alpha$ channel, where $B_x$, $B_y$, and $B_z$ in $\Phi_{2\alpha}$ is best parameterized to describe $^8$Be nucleus. The lower equation is defined as representing the correlation of the $2\alpha$ clusters associated with the coordinate $\boldsymbol{\xi}_1$.

In Fig. 11(left), the RWAs in the $^8$Be($0^+$)+$\alpha(S)$ channel for the $0_1^+$-$0_4^+$ states are shown. The RWAs for the ground state and the $0_3^+$ state have two and four nodes, respectively. However, for the Hoyle state, the nodal behavior almost disappears and only a remnant of three nodes can be seen as an oscillatory behavior. Since the outermost nodal position corresponds to a radius of repulsive core between the core $^8$Be and the $\alpha$ cluster, due to the effect of the Pauli principle, the disappearance of the nodes for the Hoyle state indicates a dissolution of the $^8$Be core, and hence formation of the loosely coupled $3\alpha$ cluster state.

On the contrary, the $0_3^+$ state recovers the distinct nodal behavior and, with one additional node, forms a higher nodal $^8$Be($0^+$) + $\alpha$ structure. Obviously the relative motion between the $^8$Be and $\alpha$ is excited from the Hoyle state by the monopole transition. This is consistent with the argument in Ref. [63], as mentioned above. On the other hand, in Fig. 11(right), the correlation function of the $2\alpha$ part associated with $\boldsymbol{\xi}_1$ is shown, where the relative motion between the $2\alpha$ and $\alpha$ is chosen to best simulate the $0_3^+$ state. In this calculation, this part is taken as $\Phi_{2\alpha-\alpha}$ with $\beta_{2x} = \beta_{2y} = 13.8$ fm and $\beta_{2z} = 13.7$ fm, since the $0_3^+$ state obtained by solving the Hill-Wheeler equation has the largest squared overlap (around 30%) with the single THSR wave function for this $\boldsymbol{\beta}_2$ value. While for the $0_3^+$ state the two nodes in inner region remain, which come from the repulsive core between the $2\alpha$ clusters, an oscillatory behavior is also clearly seen in outer region more than 4 fm. This indicates that the $2\alpha$ relative motion associated with $\boldsymbol{\xi}_1$ is also excited by the monopole transition from the Hoyle state, although further analyses are necessary to clarify this point of view. This is, however, natural considering the fact that the monopole transition operator is written as $O(E0, ^{12}\text{C}) = \sum_{i=1}^{12} (\boldsymbol{r}_i - \boldsymbol{X}_G)^2 = \sum_{k=1}^{3} \sum_{i \in \alpha_k} (\boldsymbol{r}_i - \boldsymbol{X}_k)^2 + 2\xi_1^2 + 8\xi_2^2/3$ [72]. The strong monopole transition strength from the Hoyle state can be from the two relative motions associated with the coordinates $\boldsymbol{\xi}_1$ and $\boldsymbol{\xi}_2$, giving the picture that the $0_3^+$ state is featured as monopole dilatation from the Hoyle state with a vibrational mode.

## 2.2  $4\alpha$ condensate-like state in $^{16}$O

### 2.2.1 4α OCM calculation

We discuss in this section about the Hoyle-analog state in $^{16}$O calculated by the 4α OCM, in which the 4α problem is fully solved without any model assumption with respect to the α+α relative motions within a semi-microscopic treatment [19]. This approach allows us to obtain a bosonic wave function of the α clusters in an easier way than the fully microscopic approach like the RGM [73]. The RGM equation of motion and the RGM wave function in the 4α-cluster system are written in the following way, respectively,

$$\langle \phi^4(\alpha)|(H-E)|\mathcal{A}[\chi_{4\alpha}(\boldsymbol{\xi}_1,\boldsymbol{\xi}_2,\boldsymbol{\xi}_3)\phi^4(\alpha)]\rangle = 0. \quad (32)$$
$$\phi^4(\alpha) \equiv \phi(\alpha_1)\phi(\alpha_2)\phi(\alpha_3)\phi(\alpha_4).$$

and

$$\Psi_{4\alpha}^{\mathrm{RGM}}(\boldsymbol{r}_1,\cdots,\boldsymbol{r}_{16}) = \mathcal{A}[\chi_{4\alpha}(\boldsymbol{\xi}_1,\boldsymbol{\xi}_2,\boldsymbol{\xi}_3)\phi^4(\alpha)], \quad (33)$$

where $\boldsymbol{\xi}_1$, $\boldsymbol{\xi}_2$, and $\boldsymbol{\xi}_3$ are the Jacobi coordinates between the α clusters and $H$ is the microscopic Hamiltonian composed of 16 nucleons. Eq. (32) is rewritten in the following form,

$$\int d\boldsymbol{\xi}'[\mathcal{H}(\boldsymbol{\xi},\boldsymbol{\xi}') - E\mathcal{N}(\boldsymbol{\xi},\boldsymbol{\xi}')]\chi_{4\alpha}(\boldsymbol{\xi}') = 0, \quad (34)$$

with

$$\left\{\begin{array}{c}\mathcal{H}(\boldsymbol{\xi},\boldsymbol{\xi}')\\ \mathcal{N}(\boldsymbol{\xi},\boldsymbol{\xi}')\end{array}\right\} = \langle \delta(\boldsymbol{\xi}-\boldsymbol{\eta})\phi^4(\alpha)|\left\{\begin{array}{c}H\\1\end{array}\right\}|\mathcal{A}[\delta(\boldsymbol{\xi}'-\boldsymbol{\eta})\phi^4(\alpha)]\rangle \quad (35)$$

where $\boldsymbol{\xi}$, $\boldsymbol{\xi}'$, and $\boldsymbol{\eta}$ denote the sets $\boldsymbol{\xi}_i$, $\boldsymbol{\xi}'_i$, and $\boldsymbol{\eta}_i$, $(i=1,2,3)$, respectively, which are all the Jacobi coordinates between the 4α clusters. In the above equation, the dynamical coordinates of 16 nucleons except for the total c.o.m. coordinate, i.e. internal coordinates of the 4α clusters and Jacobi coordinates $\boldsymbol{\eta}_i$ $(i=1,2,3)$ between the 4α clusters, are integrated out. Therefore, due to the delta functions in the bra and ket, the Jacobi coordinates between the α clusters, $\boldsymbol{\xi}$ and $\boldsymbol{\xi}'$ in the bra and ket, respectively, only remain in $\mathcal{H}$ and $\mathcal{N}$. This equation (35) can further be transformed into the following form,

$$(\mathcal{N}^{-1/2}\mathcal{H}\mathcal{N}^{-1/2} - E)\Phi_J^{(B)} = 0, \quad (36)$$

where $\Phi^{(B)}$ can be defined as a bosonic wave function of the 4α clusters with only the Jacobi coordinates $\boldsymbol{\xi}$ left, as follows:

$$\Phi_J^{(B)}(\boldsymbol{\xi}_1,\boldsymbol{\xi}_2,\boldsymbol{\xi}_3) = \mathcal{N}^{1/2}\chi_{4\alpha} = \int d\boldsymbol{\xi}' \mathcal{N}^{1/2}(\boldsymbol{\xi},\boldsymbol{\xi}')\chi_{4\alpha}(\boldsymbol{\xi}'). \quad (37)$$

In the OCM, the following approximation with respect to the treatment of the antisymmetrization in these kernels is taken,

$$(\mathcal{N}^{-1/2}\mathcal{H}\mathcal{N}^{-1/2} - E)\Phi_J^{(B)} = 0$$
$$\iff \Lambda(\mathcal{N}^{-1/2}\mathcal{H}\mathcal{N}^{-1/2} - E)\Lambda\Phi_J^{(B)} = 0$$
$$\implies \Lambda(\mathcal{H}^{(\mathrm{OCM})} - E)\Lambda\Phi_J^{(\mathrm{OCM})} = 0, \quad (38)$$

where $\Lambda$ is a projection operator onto Pauli allowed space [33, 37] and $\mathcal{H}^{(\mathrm{OCM})}$ is the effective local Hamiltonian of the OCM written below,

$$\Lambda\mathcal{H}^{(\mathrm{OCM})}\Lambda = \sum_i^4 T_i - T_{\mathrm{cm}} + \sum_{i<j}^4 [V_{2\alpha}^{(\mathrm{N})}(i,j) + V_{2\alpha}^{(\mathrm{C})}(i,j) +$$
$$V_{2\alpha}^{(\mathrm{P})}(i,j)] + \sum_{i<j<k}^4 V_{3\alpha}(i,j,k) + V_{4\alpha}(1,2,3,4), \quad (39)$$

where $T_i$, $V_{2\alpha}^{(\mathrm{N})}(i,j)$, $V_{2\alpha}^{(\mathrm{C})}(i,j)$, $V_{3\alpha}(i,j,k)$ and $V_{4\alpha}(1,2,3,4)$ stand for the operators of kinetic energy for the $i$th α cluster, two-body, Coulomb, three-body and four-body forces between the α clusters, respectively. The c.o.m. kinetic energy $T_{\mathrm{cm}}$ is subtracted from the Hamiltonian. $V_{2\alpha}^{(\mathrm{P})}(i,j)$ is the Pauli exclusion operator, by which the Pauli forbidden states between the 2α clusters in $0S$, $0D$, and $1S$ states are eliminated, so that the ground state with the shell-model-like configuration can be described correctly. The two-body α+α force is constructed by folding procedure from effective nucleon-nucleon interaction, for which we adopt Modified-Hasegawa-Nagata [74, 75] (MHN) force, which is known to reproduce the α+α scattering phase shift well. The three-body and four-body forces are phenomenologically introduced to reproduce the ground states of $^{12}$C and $^{16}$O at the same time. Readers can refer to Ref. [19] for more details about the choice of the force parameters and refer to Refs. [31, 33, 37, 73] for details about the OCM formalism.

In Fig. 12, we present the comparison of the experimental $J^\pi = 0^+$ spectrum with the calculated ones by using the 4α OCM [19], together with the results calculated by using the THSR ansatz [20], which will be discussed in the next section. For clarification we denote the six $0^+$ states calculated by the 4α OCM and the four $0^+$ states calculated by the 4α THSR ansatz as $(0_1^+)_{\mathrm{OCM}}$-$(0_6^+)_{\mathrm{OCM}}$ and the $(0_1^+)_{\mathrm{THSR}}$-$(0_4^+)_{\mathrm{THSR}}$, respectively. We can see that the 4α OCM calculation gives a satisfactory one-to-one correspondence with the experimental spectrum. The low-lying states $(0_1^+ - 0_3^+)$ of this spectrum are actually in good agreement with earlier $^{12}$C+α OCM calculations [78, 79] (see also Ref. [80]). We should emphasize that the orthogonality to the low-lying $^{12}$C + α states and the ground state is indispensable for the for-mation of the $4\alpha$-cluster gas-like states. This respect is particularly discussed in Sec. 5.1.

In Table 3 the comparison of the energy $E$, r.m.s. radius $R_{\rm rms}$, monopole matrix element to the ground state $M(E0)$, and $\alpha$-decay width $\Gamma$ between the experimental data and the results of the $4\alpha$ OCM and THSR ansatz is shown. The $\alpha$-decay width is calculated based on the $R$-matrix theory [80, 81]. The $(0_6^+)_{\rm OCM}$ state has the largest r.m.s. radii among the calculated states and this large value 5.4 fm indicates that this state is the $4\alpha$ gas-like state and considered to be the Hoyle-analog state. We can see that the observed widths of the experimental $0_4^+$, $0_5^+$, and $0_6^+$ states are nicely reproduced by the $4\alpha$ OCM calculations. In particular, the observed width of the $0_6^+$ state at 15.1 MeV is very well reproduced by $(0_6^+)_{\rm OCM}$ state, and hence the Hoyle-analog state corresponds to the $0_6^+$ state at 15.1 MeV.

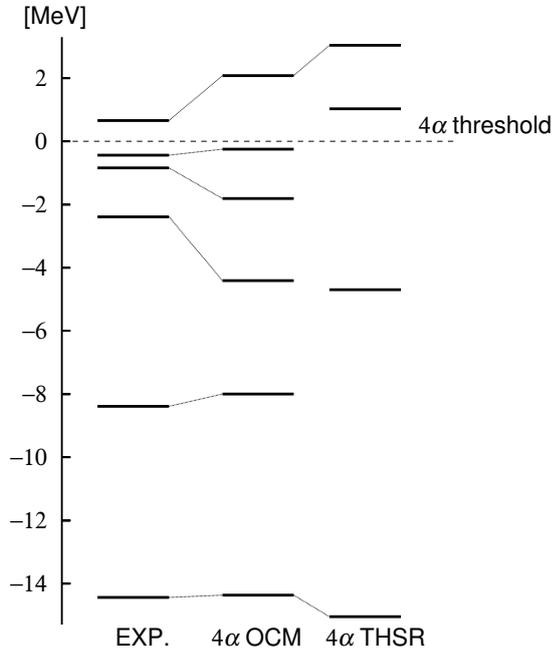

**Fig. 12** Comparison of the $0^+$ energy spectra between experiment and the calculations by using the $4\alpha$ OCM [19] and the $4\alpha$ THSR wave function [20]. Dotted line denotes the $4\alpha$ threshold. Experimental data are taken from Ref. [76] and from Ref. [77] for the $0_4^+$ state. See Table 3 for the values. Figure was reproduced from Ref. [20].

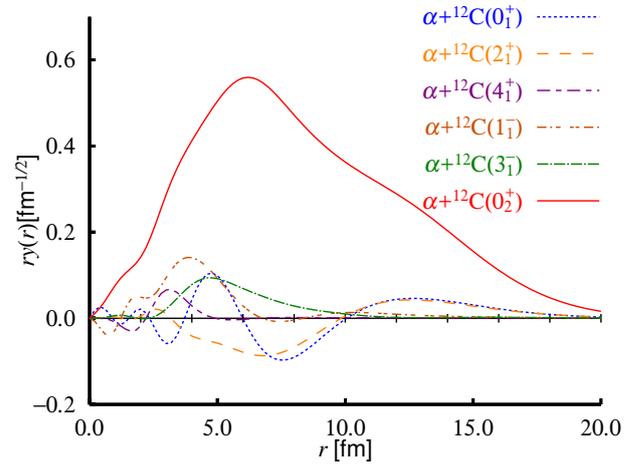

**Fig. 13** RWAs $r\mathcal{Y}_{[L,L]_0}(r)$ defined by Eq. (40) for the $(0_6^+)_{\rm OCM}$ calculated in Ref. [19].

We then investigate the Hoyle-analog state by calculating the RWAs for the $^{12}{\rm C}+\alpha$ channels, which are defined by

$$\mathcal{Y}_{[L,L]_0}(r) = \sqrt{\frac{4!}{3!1!}}$$
$$\times \left\langle \left[\frac{\delta(r'-r)}{r'^2}[\Phi_L^{({\rm OCM})}(^{12}{\rm C}), Y_L(\hat{\boldsymbol{r}}')]_0\phi(\alpha)\right]\bigg|\Phi_{J=0}^{({\rm OCM})}(^{16}{\rm O})\right\rangle, \tag{40}$$

with $\Phi_L^{({\rm OCM})}(^{12}{\rm C})$ and $\Phi_{J=0}^{({\rm OCM})}(^{16}{\rm O})$ being the OCM states of $^{12}{\rm C}$ and $^{16}{\rm O}$, respectively. We show in Fig. 13 the RWAs of the $(0_6^+)_{\rm OCM}$ state. We can see that the Hoyle state component is the most largely included, which indicates that the state is the Hoyle-analog state because the $4\alpha$ condensate state must have a large overlap with the $^{12}{\rm C}(0_2^+)+\alpha$ structure.

**Table 3** Binding energies $E$ measured from the $4\alpha$ threshold energy, r.m.s. radii $R_{\rm rms}$, monopole matrix elements $M(E0)$, and $\alpha$-decay widths $\Gamma$, in units of MeV, fm, $e{\rm fm}^2$, and MeV, respectively.

|  | THSR | | | | $4\alpha$ OCM | | | | Experiment | | | |
| --- | --- | --- | --- | --- | --- | --- | --- | --- | --- | --- | --- | --- |
|  | $E$ | $R_{\rm rms}$ | $M(E0)$ | $\Gamma$ | $E$ | $R_{\rm rms}$ | $M(E0)$ | $\Gamma$ | $E$ | $R_{\rm rms}$ | $M(E0)$ | $\Gamma$ |
| $0_1^+$ | $-15.1$ | 2.5 | | | $-14.4$ | 2.7 | | | $-14.4$ | 2.71 | | |
| $0_2^+$ | $-4.7$ | 3.1 | 9.8 | | $-8.00$ | 3.0 | 3.9 | | $-8.39$ | | 3.55 | |
| $0_3^+$ | | | | | $-4.41$ | 3.1 | 2.4 | | $-2.39$ | | 4.03 | |
| $0_4^+$ | 1.03 | 4.2 | 2.5 | 1.6 | $-1.81$ | 4.0 | 2.4 | $\sim 0.6$ | $-0.84$ | | | 0.6 |
| $0_5^+$ | | | | | $-0.25$ | 3.1 | 2.6 | $\sim 0.2$ | $-0.43$ | 3.3 | | 0.185 |
| $0_6^+$ | 3.04 | 6.1 | 1.2 | 0.14 | 2.08 | 5.6 | 1.0 | $\sim 0.14$ | 0.66 | | | 0.166 |

The $\alpha$-condensate character of the $(0_6^+)_{\rm OCM}$ is also clearly seen in the momentum distribution of the c.o.m. motion of $\alpha$-clusters, which is defined by

$$\rho(k) = \frac{1}{(2\pi)^3} \int d\boldsymbol{r} d\boldsymbol{r}'\, \widetilde{\rho}^{(\rm B)}(\boldsymbol{r}, \boldsymbol{r}')\, e^{-i\boldsymbol{k}\cdot(\boldsymbol{r}-\boldsymbol{r}')}, \quad (41)$$

where $\widetilde{\rho}^{(\rm B)}(\boldsymbol{r}, \boldsymbol{r}')$ is the one-body density matrix constructed by the bosonic OCM wave functions. How to calculate the matrix element and how to choose the internal coordinates as integral variables are in detail discussed in Refs. [82, 83]. We show in Fig. 14(a) the momentum distributions of the $(0_1^+)_{\rm OCM}$-$(0_6^+)_{\rm OCM}$ states. Only the one of the $(0_6^+)_{\rm OCM}$ state has a markedly sharp peak near zero momentum compared with the other $0^+$ states. In Fig. 14(b), we show for comparison the momentum distribution of the Hoyle state by $3\alpha$ OCM (dotted curve), which is also sharply peaked near zero momentum.

In addition to the $\alpha$ condensate state, we also discussed the other $0^+$ states. The structures of the $0_2^+$ and $0_3^+$ states are well established as having the $^{12}{\rm C}(0_1^+)+{\rm alpha}(S)$ and $^{12}{\rm C}(2_1^+)+{\rm alpha}(D)$ cluster structures, respectively. These structures of the $0_2^+$ and the $0_3^+$ states are confirmed in the present $4\alpha$ OCM calculation. The present calculations also show that the $0_4^+$ and $0_5^+$ states mainly have $^{12}{\rm C}(0_1^+)+\alpha(S)$ structure with higher nodal behavior and $^{12}{\rm C}(1^-)+\alpha(P)$ structure, respectively [80].

*2.2.2  $4\alpha$ THSR calculation*

The Hoyle-analog state in $^{16}{\rm O}$ is also investigated with the $4\alpha$ THSR wave function. While in the $4\alpha$ OCM discussed in the previous subsection the $4\alpha$ problem is fully solved without any model assumption with respect to the $\alpha+\alpha$ relative motions within the semi-microscopic treatment, in the $4\alpha$ THSR the THSR-type constraint is imposed on the $\alpha+\alpha$ relative motions. In this calculation, we only consider the spherical case, i.e. $\beta_0 \equiv \beta_\perp = \beta_z$, within a fully microscopic treatment, and as a microscopic Hamiltonian, we adopt the nucleon-nucleon effec-tive interaction named F1 force [84], which includes a three-body term representing a density dependent force.

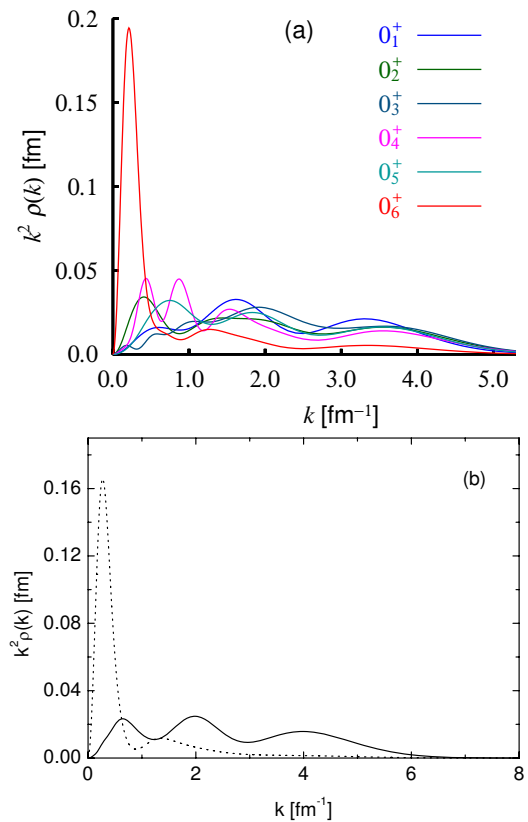

**Fig. 14** Momentum distributions of the c.o.m. motion of $\alpha$-clusters. (a): $0_1^+ \sim 0_6^+$ states of the $4\alpha$ OCM. Figure was reproduced from Ref. [5]. (b): $0_1^+$ and $0_2^+$ states of the $3\alpha$ OCM. Figure was reproduced from Ref. [73].

In Fig. 12, we presented the spectrum calculated with the THSR ansatz, denoted as the $(0_1^+)_{\text{THSR}}$-$(0_4^+)_{\text{THSR}}$ [20]. We can see that in the THSR ansatz the two excited states are missing. This can be understood by the fact that in the OCM the relative motions of $\alpha$ particles are solved in a sufficiently large model space but the THSR ansatz does not possess nonzero spin waves with respect to the $\alpha+\alpha$ relative motions. In particular, due to the OCM calculation, the $(0_3^+)_{\text{OCM}}$ and $(0_5^+)_{\text{OCM}}$ states are shown to have a large fraction of $S$-factor from the $^{12}\text{C}(2_1^+) + \alpha(D)$ and $^{12}\text{C}(1_1^-) + \alpha(P)$ channels, respectively (see Ref. [80]). These states can thus hardly be represented by the present spherical-type THSR wave function. In order to properly describe the $^{12}\text{C} + \alpha$ cluster states, it is necessary to further extend the THSR ansatz in Eq. (14) so as to separate the degree of freedom between the $^{12}\text{C}$ core and the remaining $\alpha$ cluster, by assigning different parameter sets of $\boldsymbol{\beta}$ to the first $3\alpha$'s and the fourth $\alpha$. This corresponds to the extended THSR wave function, which is in detail discussed in Sec. 5.

Thus in the THSR ansatz we consider the $0_3^+$ and $0_5^+$ states to be completely missing and the $(0_2^+)_{\text{THSR}}$ and $(0_3^+)_{\text{THSR}}$ states to be averaged representations of the $0_2^+$ and $0_4^+$ states, respectively. We can therefore say that the $(0_4^+)_{\text{THSR}}$ state corresponds to the $0_6^+$ state with the $4\alpha$ condensate character. In fact, in Table 3 we can see that the $(0_4^+)_{\text{THSR}}$ state has the largest r.m.s. radius of the four states, indicating that it has a dilute gas-like structure. The monopole matrix element and decay width well correspond to those calculated by the $4\alpha$ OCM.

We also calculate the RWAs for the $^{12}\text{C} + \alpha$ channels in the THSR ansatz, which are defined by

$$\mathcal{Y}_{[0,0]_0}(r) = \sqrt{\frac{16!}{12!4!}} \left\langle \left[ \frac{\delta(r'-r)}{r'^2} [\Phi_{3\alpha,\mu}^{L=0}, Y_0(\hat{\boldsymbol{r}}')]_{J=0} \phi(\alpha) \right] \middle| \Phi_{4\alpha,\lambda}^{J=0} \right\rangle, \tag{42}$$

with $\Phi_{3\alpha,\mu}^{J=0}$ and $\Phi_{4\alpha,\lambda}^{J=0}$ being the wave functions of the $^{12}\text{C}(0_\mu^+)$ and $^{16}\text{O}(0_\lambda^+)$ states, respectively, calculated with the (spherical) THSR ansatz in Eqs. (20) and (21). We show in Fig. 15(a)-(d) the amplitudes for the $(0_1^+)_{\text{THSR}}$-$(0_4^+)_{\text{THSR}}$, respectively, in the $^{12}\text{C}(0_1^+) + \alpha$ and $^{12}\text{C}(0_2^+) + \alpha$ channels. The ones of the $(0_4^+)_{\text{THSR}}$ well correspond to the amplitudes for the $(0_6^+)_{\text{OCM}}$ shown in Fig. 13. This again indicates that the $(0_4^+)_{\text{THSR}}$ state is the counterpart of the $(0_6^+)_{\text{OCM}}$ state and hence the Hoyle-analog state. We can also see that the $(0_2^+)_{\text{THSR}}$ and $(0_3^+)_{\text{THSR}}$ states have large amplitudes from the $^{12}\text{C}(0_1^+) + \alpha$ channel and their peak positions are in outer region, beyond 5 fm, which are also consistent with the results of the $4\alpha$ OCM calculation (see Ref. [80]).

In the same way as the $4\alpha$ OCM, we calculate the momentum distribution of the $\alpha$ clusters in the THSR

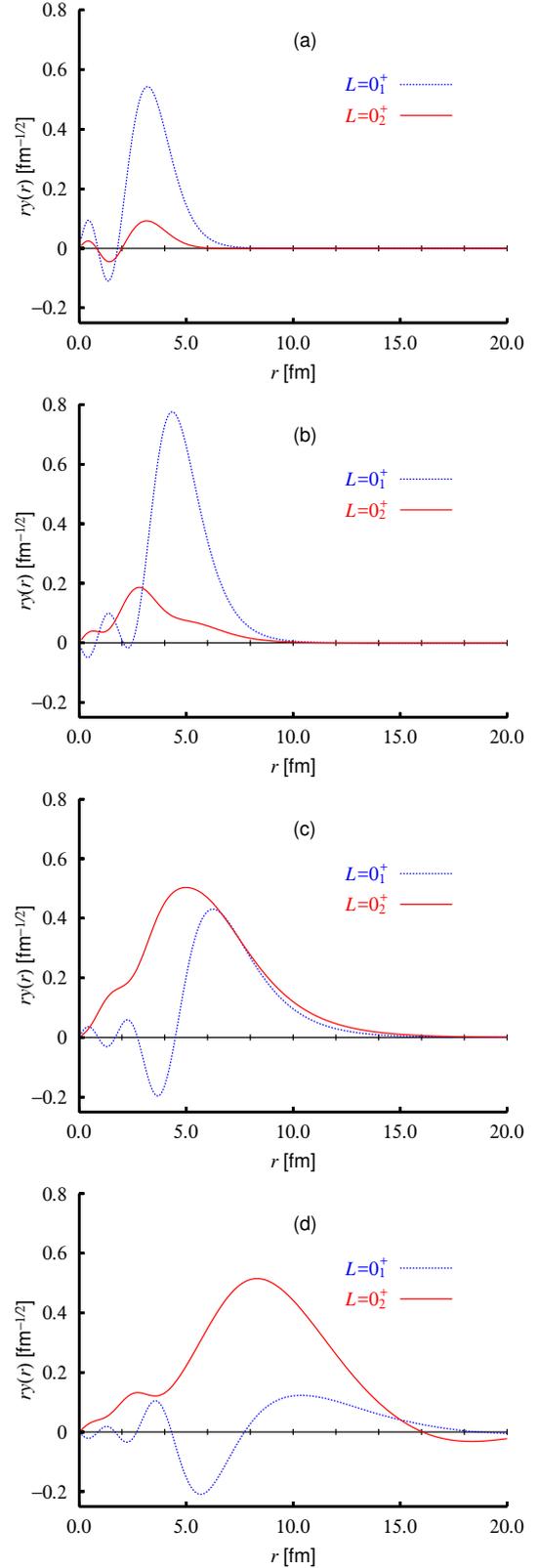

**Fig. 15** RWAs $r\mathcal{Y}_{[0,0]_0}(r)$ defined by Eq. (42) for the $(0_4^+)_{\text{THSR}}$ states in two channels $\alpha + {}^{12}\text{C}(0_1^+)$ (dotted blue curve) and $\alpha + {}^{12}\text{C}(0_2^+)$ (solid red curve) calculated in Ref. [20].

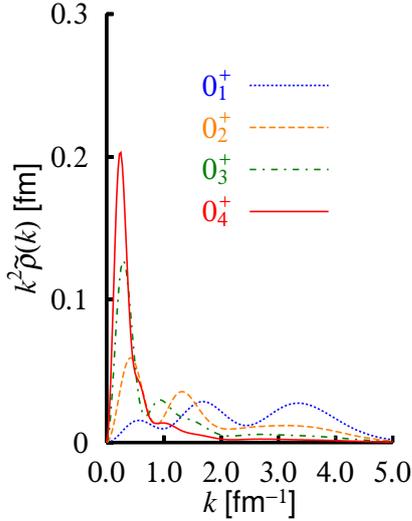

**Fig. 16** Momentum distributions of the c.o.m. motion of $\alpha$-clusters for the $0_1^+ \sim 0_4^+$ states of the $4\alpha$ THSR model. Figure was reproduced from Ref. [20].

## 3 Characters of the THSR wave function

### 3.1 Shell-model limit of the THSR wave function

The original THSR wave function [8] is designed for the study of the $\alpha$ condensation state. With in-depth study for this wave function, it is found that the THSR wave function can describe not only $n\alpha$ gas-like cluster states but also the normal and compact cluster states [11, 22]. One important reason why the THSR wave function can also describes the compact cluster states in nuclei is the THSR wave function has a shell-model limit when its size parameter $B \to b$ (or $\boldsymbol{\beta} \to 0$). This is quite similar with the well-known the shell-model limit of the Brink wave function when the inter-cluster distance approaches 0 as discussed in Sec.2.

To illustrate the shell-model limit of the THSR wave function, let's take the $3\alpha$ THSR wave function as an example,

$$\lim_{B \to b} \Psi_{3\alpha,\text{int}}^{\text{THSR}} = \lim_{B \to b} C_0(B) \exp(-12\nu X_G^2)$$
$$\times \mathcal{A}\left\{\exp\left[-\left(\frac{1}{B^2}\xi_1^2 + \frac{4}{3B^2}\xi_2^2\right)\right]\phi^3(\alpha)\right\} \quad (45)$$
$$= c_0 \mathcal{A}\left\{R_{2,0}(\xi_1, \frac{1}{b^2}) R_{2,0}(\xi_2, \frac{4}{3b^2}) \phi^3(\alpha)\right\} \quad (46)$$
$$= \Phi_{\text{shell}}\left((0s)^4(0p)^8; [444] \,(\lambda, \mu) = (0, 4), J = 0\right), \quad (47)$$

where $\boldsymbol{\xi}_1 = \boldsymbol{X}_2 - \boldsymbol{X}_1$ and $\boldsymbol{\xi}_2 = \boldsymbol{X}_3 - (\boldsymbol{X}_1 + \boldsymbol{X}_2)/2$. $C_0(B)$ and $c_0$ are normalization constants. $R_{N,L}(r, \gamma_0)$ is the radial harmonic oscillator function with size parameter $\gamma_0$ with and $N$ stands for the number of harmonic oscillator quanta, $N = 2n + L$. The obtained Eq. (47) shows that the $3\alpha$ THSR wave function at the limit of $B \to b$ becomes a shell-model wave function. This is why the THSR wave function not only describes well the gas-like Hoyle state but also gives a good description of the shell-model-like ground state of $^{12}$C [9, 11].

Furthermore, if the composed clusters of a nuclear system are closed-shell nuclei like $\alpha$ and $^{16}$O, the THSR wave function for this cluster system also has the simple shell-model limit. In the $^{16}$O+$\alpha$ system, after angular-momentum projection with even $L$, the THSR wave function can be written, $\exp(-20\nu X_G^2)\mathcal{A}\{r^L \exp(-\gamma r^2) Y_{LM}(\hat{r}) \phi(^{16}\text{O}) \phi(\alpha)\}$. The shell-model limit of this THSR wave function can be ex-ansatz. The $4\alpha$ THSR wave function is, however, the fermionic wave function of the 16 nucleons. We therefore have to extract a bosonic degree of freedom from $\Phi_{4\alpha}^{\text{THSR}}$ in Eq. (14), like in Eq. (37) for the RGM wave function. Since in this extraction harmonic oscillator expansion is necessary, which is, however, very difficult for dilute-density states in especially heavier system, the following approximation is taken in Ref. [20],

$$\Phi^{(B)} = \mathcal{N}^{1/2} \chi_{4\alpha}^{\text{THSR}} \approx \frac{\mathcal{N} \chi_{4\alpha}^{\text{THSR}}}{\sqrt{\langle \mathcal{N} \chi_{4\alpha}^{\text{THSR}} | \mathcal{N} \chi_{4\alpha}^{\text{THSR}} \rangle}}, \quad (43)$$

with

$$\chi_{4\alpha}^{\text{THSR}}(\xi_1, \xi_2, \xi_3) = \sum_\beta f_\lambda^{J=0}(\beta) \exp\left[-2 \sum_{i=1}^{3} \mu_i \left\{\frac{\xi_i^2}{b^2 + 2\beta^2}\right\}\right] \quad (44)$$

where $f_\lambda^{J=0}(\beta)$ can be determined by solving the Hill-Wheeler equation Eq. (21) for the spherical case. This approximation was first examined in Ref. [85]. Thus we can calculate the momentum distributions of the c.o.m. motion of $\alpha$-clusters for the THSR ansatz, which is a microscopic treatment. In Fig. 16 those for the $0_\lambda^+$ states with $\lambda = 1, \cdots, 4$ are shown. We can see that the $0_4^+$ state has a sharp peak near zero momentum and its shape is in fairly good agreement with the one of the $(0_6^+)_{\text{OCM}}$ state in Fig. 14, where both curves have a peak height around 2.0 fm and width around 0.5 fm$^{-1}$.

pressed as follows [11],

$$\lim_{B \to b} \Psi_{^{20}\text{Ne},L}^{\text{THSR}} = \lim_{B \to b} C_L(B) \exp(-20\nu X_G^2)$$
$$\times \mathcal{A}\{r^L \exp(-\frac{8}{5B^2}r^2)Y_{LM}(\widehat{r})\phi(^{16}\text{O})\phi(\alpha)\} \quad (48)$$
$$= c_L \mathcal{A}\{R_{8,L}(r, \frac{8}{5b^2})Y_{LM}(\widehat{r})\phi(^{16}\text{O})\phi(\alpha)\} \quad (49)$$
$$= \Phi_{\text{shell}}\left((0s)^4(0p)^{12}(0d1s)^4; [4](\lambda,\mu)=(8,0), LM\right), \quad (50)$$

where $C_L(B)$ and $c_L$ are normalization constants. The derived Eq. (50) shows that at the limit of $B \to b$, the THSR wave function becomes the most important $sd$ shell-model wave function with spatial symmetry [4] and SU(3) symmetry $(\lambda,\mu)=(8,0)$ [86]. This point is very important for the description of shell-model-like states in $^{20}$Ne with the THSR wave function.

What is interesting is the proposed THSR wave function for the $\alpha$ condensation or gas-like states has a shell-model limit when the size parameter tends to zero and this character is more than expected. It should be noted that, on the other limit, that is $B \to +\infty$, the correlations of clusters disappear and the THSR wave function becomes the simple product of the cluster wave functions.

### 3.2 Equivalence of projected prolate and oblate THSR wave functions

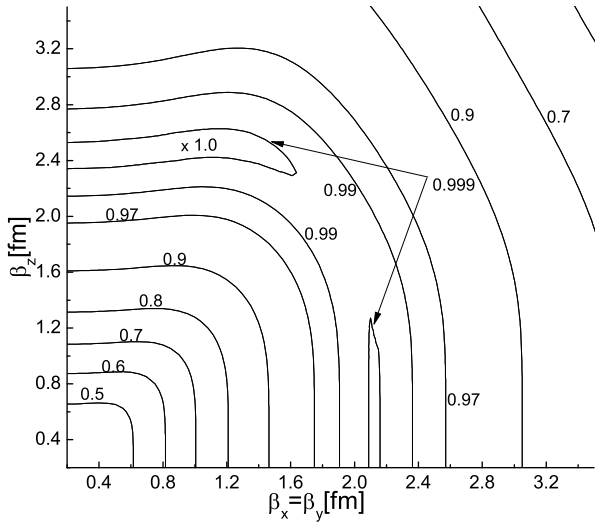

**Fig. 17** Contour map of the squared overlap between the $0^+$ wave function with $\beta_x = \beta_y = 0.9$ fm, $\beta_z = 2.5$ fm and the $0^+$ wave function with variable $\beta_x = \beta_y$ and $\beta_z$ [11]. Numbers attached to the contour lines are squared overlap values.

The THSR is a novel microscopic cluster model and it has some special characteristics, in particular, the equivalence of prolate and oblate THSR wave functions after angular-momentum projections. Figure 17 shows the squared overlaps between a prolate $0^+$ THSR wave function with $0^+$ THSR wave functions with various deformations in $^{20}$Ne [11]. The deformed THSR wave function $\Phi_{\text{Ne}}$ of $^{20}$Ne has the form $\mathcal{A}[\chi(\boldsymbol{r})\phi(\alpha)\phi(^{16}\text{O})]$, where $\chi(\boldsymbol{r})$ is $\exp[-\sum_{k=x,y,z}(8/5B_k^2)r_k^2]$ and $B_k^2 = b^2 + 2\beta_k^2$. We see in this figure that the prolate THSR wave function with $\beta_x = \beta_y = 0.9$ fm, $\beta_z = 2.5$ fm is almost 100% equivalent to oblate THSR wave functions with $\beta_x = \beta_y \approx 2.1$ fm, $\beta_z \approx 0 \sim 1.2$ fm after angular-momentum projections onto the $0^+$ state. The equivalence of prolate and oblate THSR wave functions after angular-momentum projection is also true for other the spin-parity states of $^{20}$Ne.

In spite of the equivalence of prolate and oblate THSR wave functions after angular-momentum projection, we can say that the $^{20}$Ne states expressed by THSR wave functions have all prolate deformation as the actual deformation. This conclusion is obtained from the fact that the expectation values of the quadrupole moments of all the $^{20}$Ne states expressed by THSR wave functions have negative signs. From the well-known formula $Q(J) = -(J/(2J+3))Q(\text{intrinsic})$, we know that when the expectation value $Q(J)$ of the quadrupole moment by the wave function with good spin $J$ is negative, the quadrupole moment of the intrinsic state $Q$ (intrinsic) is positive which means that the deformation of the state is prolate. The THSR wave function after angular-momentum projection has the form $\Phi_{\text{Ne}}^J = \mathcal{A}[\chi J(\boldsymbol{r})\phi(\alpha)\phi(^{16}\text{O})]$ and we can prove that this type of wave function $\Phi_{\text{Ne}}^J$ gives us the following formula for $Q(J)$,

$$Q(J) = -\frac{J}{2J+3}\frac{16}{5}\langle r^2 \rangle, \quad (51)$$

$$\frac{16}{5}\langle r^2 \rangle = \langle \Phi_{\text{Ne}}^J | \sum_{j=1}^{20}(\boldsymbol{r}_j - \boldsymbol{X}_G)^2 | \Phi_{\text{Ne}}^J \rangle - R^2(^{16}\text{O}) - R^2(\alpha), \quad (52)$$

$$R^2(C_k) = \langle \phi(C_k) | \sum_{j \in C_k}(\boldsymbol{r}_j - \boldsymbol{X}_G(C_k))^2 | \phi(C_k) \rangle. \quad (53)$$

This formula shows that $Q(J)$ has negative value and explains why the calculated values of $Q(J)$ by THSR wave functions have all negative signs. Of course the negative sign of $Q(J)$ by THSR wave functions is in accordance with the prolate distribution of nucleon density shown in Fig. 26.

The reason why projected prolate and oblate THSR wave functions are almost equivalent is because the rotation-average of a prolate THSR wave function is almost equivalent to an oblate THSR wave function. The rotation-average wave function $\Phi^{\text{ave}}(\beta_x = \beta_y; \beta_z)$ generated from a prolate THSR wave function $\Phi^{\text{pro}}(\beta_x = \beta_y; \beta_z)$ is defined as follows,

$$\Phi^{\text{ave}}(\beta_x = \beta_y; \beta_z) = \left(\frac{1}{2\pi}\int_0^{2\pi} d\theta e^{i\theta J_x}\right)\Phi^{\text{pro}}(\beta_x = \beta_y; \beta_z). \tag{54}$$

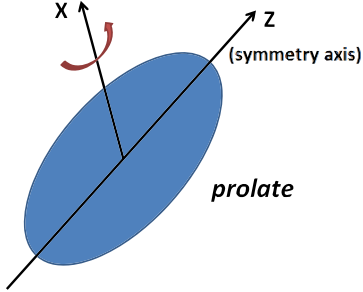

**Fig. 18** Rotation average of a prolate THSR wave function around an axis ($x$-axis) perpendicular to the symmetry axis ($z$-axis) of the prolate THSR wave function. Figure was reproduced from Ref. [11].

If a prolate THSR wave function is rotated around an axis ($x$ axis) perpendicular to the symmetry axis ($z$ axis) of the prolate deformation and we can construct a wave function by taking an average of this rotation, the density distribution of the rotation-average wave function will be oblate (see Fig. 18). In the case of the $0^+$ state, when we construct the rotation-average wave function $\Phi^{\text{ave}}(\beta_x=\beta_y=0.9; \beta_z=2.5$ fm) from the prolate THSR wave function with $(\beta_x, \beta_y, \beta_z)=(0.9, 0.9, 2.5$ fm) which gives the minimum energy for $0^+$, it is almost 100% equivalent to the oblate THSR wave function with $(\beta_x, \beta_y, \beta_z)=(0.9, 2.1, 2.1$ fm),

$$\Phi^{\text{obl}}(\beta_x = 0.9, \beta_y = \beta_z = 2.1\text{fm}) = \Phi^{\text{ave}}(\beta_x = \beta_y = 0.9, \beta_z = 2.5\text{fm}) \tag{55}$$

When we construct the good-spin wave function from $\Phi^{\text{ave}}(\beta_x=\beta_y;\beta_z)$, it is clear that the resulting wave function is the same as the good-spin wave function constructed from $\Phi^{\text{pro}}(\beta_x=\beta_y;\beta_z)$. Therefore in the above example, the oblate THSR wave function $\Phi^{\text{obl}}(\beta_x=0.9; \beta_y=\beta_z=2.1$ fm) is almost 100% equivalent to the prolate THSR wave function $\Phi^{\text{pro}}(\beta_x=\beta_y=0.9; \beta_z=2.5$ fm) after angular momentum projection to the $0^+$ state.

### 3.3 Negative-parity THSR wave function

As we know, the original THSR has a positive-parity nature. In this subsection, we will show that the THSR wave function with a simple extension can be applied to study the negative-parity cluster states without any difficulty.

The first example for studying the negative-parity states in the framework of the THSR is the $^{20}$Ne case. We proposed a new type of microscopic cluster wave function [10], which we call hybrid-Brink-THSR wave function,

$$\Phi_{\text{cluster}}(\boldsymbol{\beta}_i, \boldsymbol{S}_i) = \int d\boldsymbol{R}_1 \ldots d\boldsymbol{R}_n$$
$$\times \exp\Big[-\sum_{i=1}^{n}\sum_{k=}^{x,y,z}\frac{R_{ik}^2}{\beta_{ik}^2}\Big]\Phi_{\text{cluster}}^B(\boldsymbol{R}_1+\boldsymbol{S}_1,\ldots,\boldsymbol{R}_n+\boldsymbol{S}_n) \tag{56}$$

$$\propto \mathcal{A}\Big\{\prod_{i=1}^{n}\exp\Big[-A_i\sum_{k=}^{x,y,z}\frac{(X_{ik}-S_{ik})^2}{2B_{ik}^2}\Big]\phi(C_i)\Big\}, \tag{57}$$

$$\Phi_{\text{cluster}}^B(\boldsymbol{S}_1,\ldots,\boldsymbol{S}_n) \propto \mathcal{A}\Big\{\prod_{i=1}^{n}\exp\Big[-A_i\frac{(\boldsymbol{X}_i-\boldsymbol{S}_i)^2}{2b^2}\Big]\phi(C_i)\Big\}. \tag{58}$$

Here $B_{ik}^2 = \beta_{ik}^2 + 2b^2$ and $\boldsymbol{X}_i$ and $\phi(C_i)$ are the center-of-mass coordinate and the internal wave function of the cluster $C_i$, respectively. Different clusters $C_i$ can have different mass numbers $A_i$ and variational parameters $\boldsymbol{\beta}_i$. The oscillator parameter of the cluster $C_i$ is called $b$, which is taken as the same size value for different clusters. $\Phi_{\text{cluster}}^B$ is the corresponding general Brink model wave function [34].

In Eq. (56), another generator coordinate $\boldsymbol{S}_i$ is introduced to the original THSR wave function. It can be seen from Eq. (57) that this hybrid wave function combines the important characters of the Brink model and the THSR wave function in a very simple way. When $\boldsymbol{S}_i=0$, Eq. (56) corresponds to the THSR wave function and $\boldsymbol{\beta}_i$ or $\boldsymbol{B}_i$ becomes the size parameter. When $\beta_{ik} = 0$, i.e., $B_{ik} = b$ ($k=x,y,z$), this equation is nothing more than the Brink wave function Eq. (58) and $\boldsymbol{S}_i$ is the position parameter of the cluster $C_i$. Thus, due to the violation of the parity symmetry, we can easily projected out the negative-parity components from this hybrid wave function.

As we know, the THSR wave function is a nonlocalized clustering wave function for the cluster structure rather than the localized clustering represented by the Brink model [41]. Since these two different kinds of pictures for clustering are both included in the hybrid-Brink-THSR wave function as two limits, this hybrid wave function provides a very nice way for verifying which picture is more adequate for understanding the relative motions of the cluster structures in nuclei. The details will be discussed in Sec. 4.

To deal with the negative-parity states of $^9$Be, another parity-violating THSR wave function was developed [87], which is much more suitable for the calculations of Monte Carlo techniques. Details will be discussed in Sec.5.5.1. In the constructed THSR wave function of $^9$Be, the spatial wave function of the extra neutron $\Phi_n(\mathbf{r}_n)$ can be written as [87],

$$\Phi_n(\mathbf{r}_n) = \int d\mathbf{R}_n \exp\left(-\frac{R_{n,x}^2}{\beta_{n,xy}^2} - \frac{R_{n,y}^2}{\beta_{n,xy}^2} - \frac{R_{n,z}^2}{\beta_{n,z}^2}\right) e^{im\phi_{\mathbf{R}_n}}$$
$$\times (\pi b^2)^{-3/4} e^{-\frac{(\mathbf{r}_n - \mathbf{R}_n)^2}{2b^2}}. \quad (59)$$

When we change the integration variables $(R_{n,x}, R_{n,y}, R_{n,z})$ to $(-R_{n,x}, -R_{n,y}, -R_{n,z})$, in the integral representation of $\Phi_n(-r_n)$ by Eq. (59), we obtain $\Phi_n(-r_n) = -\Phi_n(r_n)$. It is because the azimuthal angle of $(-R_{n,x}, -R_{n,y}, -R_{n,z})$ is $(\pi + \phi_{\mathbf{R}_n})$ and $\exp[i(\pi + \phi_{\mathbf{R}_n})] = -\exp(i\phi_{\mathbf{R}_n})$. When $m = 0$, Eq. (59) is a standard THSR wave function and has a positive parity as already known. So the total parity of $^9$Be is now determined by $m$,

$$\pi = \pi_\alpha^{(1)} \times \pi_\alpha^{(2)} \times \pi_n = \begin{cases} + & (m = 0) \\ - & (m = \pm 1). \end{cases} \quad (60)$$

From Eq. (59) we can also see that the $z$ component of orbital angular momentum $l_{z,n} = m$ is a good quantum number for the extra neutron. Because of the rotational symmetry of the two-$\alpha$-cluster subsystem about $z$-axis, we have $l_{z,\alpha} = 0$ for $\alpha$ clusters. Thus the $z$-component of the orbital angular momentum $l_z$ of total system is

$$l_z = l_{z,\alpha}^{(1)} + l_{z,\alpha}^{(2)} + l_{z,n} = m. \quad (61)$$

3.4 Single-particle property of the THSR wave function of $^{13}$C

The THSR wave function for the $^{13}$C system employed in Ref. [88] is described as

$$\Phi_l^{(\lambda)}(R) = \sum_B C_B^{(\lambda)}(R)\Phi_l(B, R) \quad (62)$$

with

$$\Phi_l(B, R) = \mathcal{A}\left[\Phi_{3\alpha}^{\text{THSR}}(B)\varphi_l(r_n - X_{3\alpha} - R)\right], \quad (63)$$

where $\Phi_{3\alpha}^{\text{THSR}}(B)$ is the spherical THSR wave function for $^{12}$C with $B = B_x = B_y = B_z$ in Eq. (14) and $\varphi_l(r) = (2\nu/\pi)^{3/4}\exp(-\nu r^2)Y_{lm}(\hat{r})$, with $r_n$ and $X_{3\alpha}$ being the coordinates of the valence neutron and the c.o.m. coordinate of the $3\alpha$ system, respectively. The coefficient of the superposition in Eq. (62), $C_B^{(\lambda)}(R)$, and the corresponding energy eigenvalue $E_l^{(\lambda)}(R)$ can be obtained as a function of the $3\alpha$-$n$ relative distance parameter $R$ by solving the following Hill-Wheeler-type equation,

$$\sum_{B'} \langle \Phi_l(B, R)|H - E_l^{(\lambda)}(R)|\Phi_l(B', R)\rangle C_{B'}^{(\lambda)}(R) = 0. \quad (64)$$

The adiabatic energy curves $E_l^{(\lambda)}(R)$ with $l = 0, 1$ for the lowest ($\lambda = 1$) and the second lowest ($\lambda = 2$) eigenstates are shown in Fig. 19. It is noted that the cases of $\lambda = 1$ and $\lambda = 2$ correspond to the ground state and the Hoyle state for the core $3\alpha$ system, respectively, and hence demonstrate, respectively, that the valence neutron couples with the ground state and the Hoyle state in $l = 0$ or $l = 1$. When the distance parameter $R$ is large enough for the neutron to be far away from the core, we can see that the neutron with $l = 0$ is energetically favored for both $\lambda = 1, 2$ cases. On the contrary, as the $S$-wave neutron approaches the core, it starts to feel the repulsive force from the core, due to the Pauli blocking effect. This effect is considered to be specially important when the core is the ground state. In fact, in Fig. 19, the eigenenergy $E_{l=0}^{(\lambda=1)}(R)$ becomes higher than $E_{l=1}^{(\lambda=1)}(R)$ inside the surface region of the ground state of $^{12}$C, i.e. for $R$ less than around 3 fm. At the center of the core, the neutron sits in the $0p_{1/2}$ orbit rather than in the $1s_{1/2}$ orbit.

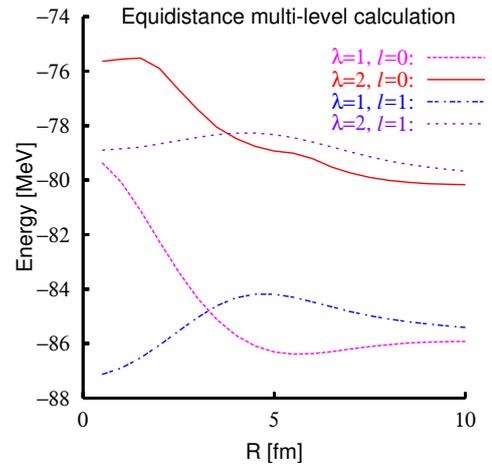

**Fig. 19** Two lowest eigenenergies $E_l^{(\lambda=1)}(R)$ and $E_l^{(\lambda=2)}(R)$ for $l = 0$ and $l = 1$, as a function of the relative distance parameter $R$ between the $3\alpha$ system and valence neutron. Figure was reproduced from Ref. [88].

However, a more interesting result is for the $\lambda = 2$ case, where the core is in the Hoyle state. When the $S$-wave neutron comes closer to the core, it is also seen to feel repulsive core and its energy $E_{l=0}^{(\lambda=2)}$ goes higher than that of the $P$-wave neutron, $E_{l=1}^{(\lambda=2)}$, at a larger value of $R$, according to the larger size for the Hoyle state. Again at the center of the core, the $1s_{1/2}$ orbit for the valence neutron becomes higher than the $0p_{1/2}$ orbit, though the energy spacing between them is much smaller than in the $\lambda = 1$ case. This result means that even for the Hoyle state, which is a dilute object with about 1/3 of the normal density, the effect of the antisymmetrization is not still non-negligible, though the effect is much more smoothed than when the core is in the ground state.

## 4 Localized vs Nonlocalized clustering

### 4.1 Inversion-doublet bands in $^{20}$Ne and their description by the THSR wave function

#### 4.1.1 Energy curve by the hybrid-Brink-THSR wave function

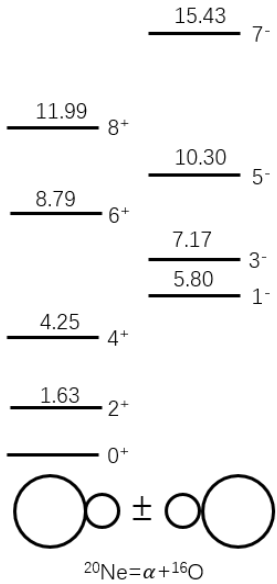

**Fig. 20** The rotational spectra with $K^\pi = 0_1^+$ and $K^\pi = 0_1^-$ bands of $^{20}$Ne [21]. The marked excited energies are in MeV.

As one of the most typical cluster nuclei, the $\alpha+^{16}$O cluster system of $^{20}$Ne has been extensively studied by many nuclear models [89–93]. The observed ground-state band and $K^\pi=0_1^-$ band in $^{20}$Ne were regarded as being an inversion doublet band arising from the hetero-polar di-nucleus configuration [21] of $\alpha+^{16}$O cluster structure. See the schematic diagram in Fig. 20. This kind of inversion doublet band in $^{20}$Ne can be considered as the direct manifestation of cluster structures in $^{20}$Ne. Furthermore, based on the traditional understanding of the cluster structures, in asymmetry two-cluster systems, the inversion doublet band can also be regarded as a clear indication of the existence of the localized cluster structure together with the observation of large cluster decay widths. Actually, in heavier nuclei, many core+$\alpha$ cluster structures [94–96] were identified based on the observed inversion doublet bands. In this situation, it seems that we have to regard states of the inversion doublet band of $^{20}$Ne as having a $\alpha+^{16}$O localized clustering. However, the study from the THSR wave function updates this kind of traditional understanding.

To study the inversion doublet bands in $^{20}$Ne, a hybrid-Brink-THSR wave function wave function was proposed as follows [10],

$$\widehat{\Phi}_{\rm Ne}(\boldsymbol{\beta}, \boldsymbol{S}) = \mathcal{A}\{\exp[-\sum_k^{x,y,z}\frac{8(\boldsymbol{\xi}-\boldsymbol{S})_k^2}{5B_k^2}]\phi(\alpha)\phi(^{16}{\rm O})\} \quad (65)$$

where $B_k^2 = b^2 + 2\beta_k^2$, $\boldsymbol{\xi} = \boldsymbol{X}_2 - \boldsymbol{X}_1$, $\boldsymbol{X}_G = (4\boldsymbol{X}_1 + 16\boldsymbol{X}_2)/20$. $\boldsymbol{X}_1$ and $\boldsymbol{X}_2$ represent the center-of-mass coordinates of the $\alpha$ cluster and the $^{16}$O cluster, respectively. All calculations are performed with restriction to axially symmetric deformation, that is, $\beta_x = \beta_y \neq \beta_z$ and $\boldsymbol{S} \equiv (0, 0, S_z)$. The spin and parity eigenfunctions can be obtained by the angular-momentum projection technique. As the nuclear interaction, we adopt the effective nucleon-nucleon force, Volkov No.1 [42] with the Majorana parameter $M = 0.59$, and the oscillator parameter $b = 1.46$ fm.

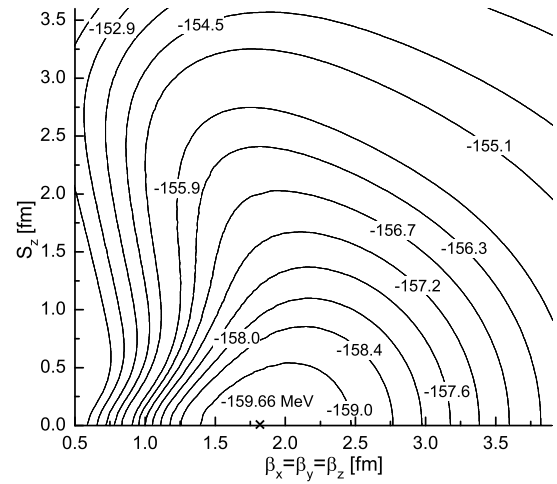

**Fig. 21** Contour map of the energy surface of the intrinsic wave function of $^{20}$Ne in the two-parameter space, $S_z$ and $\beta_x = \beta_y = \beta_z$ [11].

Figure 21 shows the contour map of the energy surface of the intrinsic wave function of $^{20}$Ne in the two-parameter space, $S_z$ and $\beta_x = \beta_y = \beta_z$. We can see the minimum energy $-159.66$ MeV appears at $S_z=0$ and $\beta_x = \beta_y = \beta_z = 1.8$ fm. The value $S_z=0$ means, after variational calculations, the obtained optimum intrinsic hybrid-Brink-THSR wave function becomes the pure THSR wave function in describing the ground state of $^{20}$Ne. This result indicates that the intrinsic THSR wave function based on the concept of nonlocalized clustering is more suitable for describing the ground state of $^{20}$Ne than the Brink wave function.

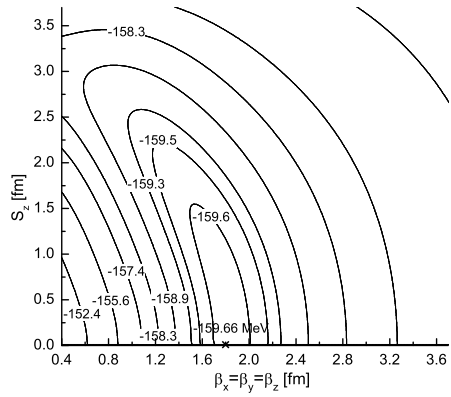
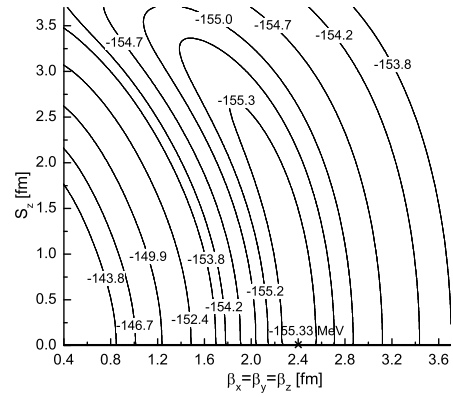
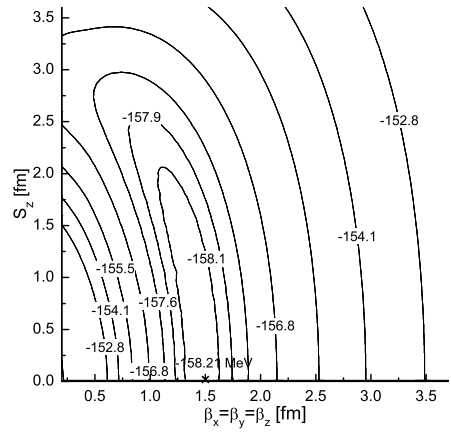
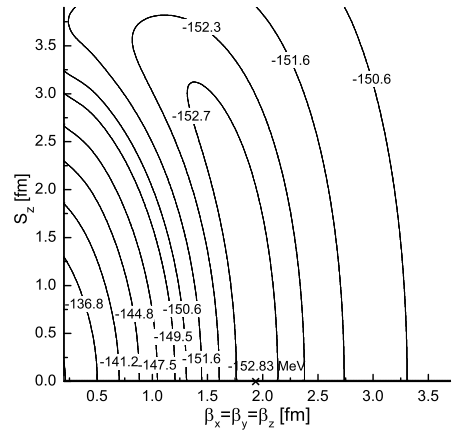

**Fig. 22** Contour maps of energy surfaces of the $J^\pi = 0^+$ state (Upper) and $J^\pi = 2^+$ state (Lower) in the two-parameter space, $S_z$ and $\beta_x = \beta_y = \beta_z$ [11].

**Fig. 23** Contour maps of energy surfaces of the $J^\pi = 1^-$ state (Upper) and $J^\pi = 3^-$ state (Lower) in the two-parameter space, $S_z$ and $\beta_x = \beta_y = \beta_z$ [11].

After angular-momentum projection on the intrinsic hybrid-Brink-THSR wave function with the spherical case $\beta_x = \beta_y = \beta_z$, the optimal projected wave function can be obtained by variational calculations. Figures 22 and 23 show the contour maps for the $0^+$, $2^+$ positive-parity states and $1^-$, $3^-$ negative-parity states of $^{20}$Ne in the two-parameter space, $S_z$ and $\beta_x = \beta_y = \beta_z$, respectively. It is surprising to find that the obtained minimum energies with respect to the projected states all appear at $S_z=0$. For instance, the obtained minimum energy $-159.66$ MeV for $J^\pi = 0^+$ state appears at $S_z = 0$ and $\beta_x = \beta_y = \beta_z = 1.8$ fm. For the $J^\pi = 1^-$ state, the minimum energy $-155.33$ MeV appears at $S_z \approx 0$ and $\beta_x = \beta_y = \beta_z = 2.4$ fm in the contour map. For the other states of inversion doublet bands of $^{20}$Ne, they have the same character after variational calculations. The inter-cluster distance parameter $S_z \to 0$ means that this hybrid-Brink-THSR wave function tends to become a pure THSR wave function in describing the cluster states of the inversion doublet bands in $^{20}$Ne. One should note that although $S_z$ does not give any contribution to the energy gain, it still plays an important role in providing negative-parity states. This is because the parameter $S_z$ is the only component which breaks the parity symmetry, as is clearly seen in the form of the intrinsic wave function $\widehat{\Phi}_{\text{Ne}}(\boldsymbol{\beta}, \boldsymbol{S})$ in Eq. (65).

The above variational calculations with the hybrid-Brink-THSR wave function lead the parameter $S_z$ to zero for the inversion doublet bands. This means that, in spite of the fact that a pure Brink wave function gives a distinct energy minimum point with non-zero $S_z$, the localized clustering picture cannot be supported. We can realize this situation by looking at Fig. 24, which shows the energy curves of the lower excited states of the inversion doublet bands with different widths of the Gaussian relative wave functions in the hybrid model. If $\boldsymbol{\beta}$ is fixed at 0, the hybrid-Brink-THSR wave function becomes the Brink wave function. In this case, $S_z$ is the inter-cluster distance parameter and it is usually regarded as a dynamics parameter for describing the cluster system. For instance, the minimum energy of the ground state of $^{20}$Ne appears at $S_z = 3.0$ fm. For the $J^\pi = 1^-$ state, the optimum position appears at $S_z = 3.9$ fm. The non-zero values of $S_z$ seem to indicate that the $\alpha + ^{16}$O structure of $^{20}$Ne favours localized clustering. This is just the traditional concept of localized clustering. Now, we believe that this argument is misleading. The non-zero minimum point $S_z$ simply occurs since the width of the Gaussian wave function of the relative motion in the Brink model is fixed to a narrow wave packet, characterized by the parameter $b$. If we take non-zero values for $\boldsymbol{\beta}$, namely, $\beta_x = \beta_y = \beta_z = 1.8$ fm, 1.5 fm, 2.4 fm, and 1.9 fm for $J^\pi = 0^+, 2^+, 1^-$, and $3^-$ states, respectively, according to their minimum positions in the contour maps, then we find that the minima appear at $S_z = 0$ in Fig. 24. This indicates that the separation distance parameter $S_z$ does not play any physical role in describing the $\alpha + ^{16}$O cluster structure, even for the negative-parity states. Instead of that, the new parameterization by $\boldsymbol{\beta}$, which characterizes nonlocalized clustering, is more appropriate for describing the cluster structure in $^{20}$Ne.

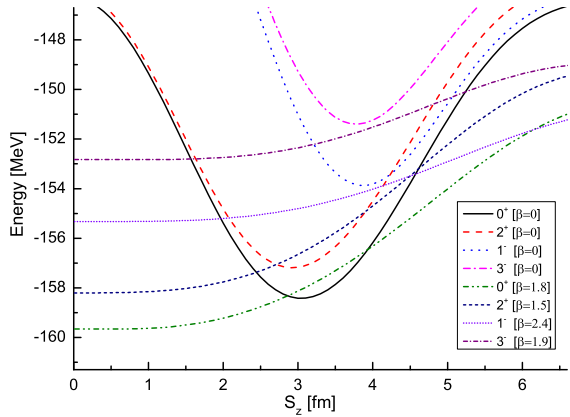

**Fig. 24** Energy curves of $J^\pi = 0^+$, $2^+$, $1^-$, and $3^-$ states with different widths of Gaussian relative wave functions in the hybrid model [11].

### 4.1.2 Equivalence between the $^{16}O+\alpha$ THSR wave function and $^{16}O+\alpha$ RGM/GCM wave functions

Now that the parameter $S_z$ tends to be zero in the obtained hybrid wave functions for the inversion doublet bands in $^{20}$Ne, we can make further variational calculations in the two-parameter space, $\beta_x = \beta_y$ and $\beta_z$ using the projected hybrid-Brink-THSR wave function with the parameter $S_z = 0$ (In practical calculations $S_z$ is fixed at a very small value close to zero for negative-parity states). The obtained minimum energies and the corresponding values of $\beta_x = \beta_y$ and $\beta_z$ using the THSR-type wave functions are listed in Table 4.

**Table 4** $E_{\min}^{J^\pi}(\beta_x = \beta_y, \beta_z)$ are the minimum energies at the corresponding values of $\beta_x = \beta_y$ and $\beta_z$ in the hybrid model. The squared overlaps between the single normalized projected THSR-type wave functions $\hat\Phi_{\min}^{\text{THSR}}$ corresponding to the minimum energies and the normalized Brink GCM wave functions are also listed. Units of energies are MeV.

| State | $E_{\min}^{J^\pi}(\beta_x = \beta_y, \beta_z)$ | $|\langle\hat\Phi_{\min}^{\text{THSR}}|\hat\Phi_{\text{GCM}}^{\text{Brink}}\rangle|^2$ |
|---|---|---|
| $0^+$ | -159.85 (0.9, 2.5) | 0.9929 |
| $2^+$ | -158.53 (0.0, 2.2) | 0.9879 |
| $4^+$ | -155.50 (0.0, 1.8) | 0.9775 |
| $1^-$ | -155.38 (3.7, 1.4) | 0.9998 |
| $3^-$ | -153.07 (3.7, 0.0) | 0.9987 |

On the other hand, the exact solution of the $\alpha + ^{16}$O cluster system can be obtained by superposing the Brink wave functions, that is the Brink-GCM wave function,

$$\sum_j \langle \Phi_{\text{Brink}}^{J^\pi}(R_i)|\hat H - E|\Phi_{\text{Brink}}^{J^\pi}(R_j)\rangle f(R_j) = 0. \quad (66)$$

Here, $\Phi_{\text{Brink}}^{J^\pi}(R_i)$ can be obtained directly from the projected Brink wave function $\Phi_{\text{Brink}}^{J^\pi}(\frac{4}{5}\boldsymbol{R}, -\frac{1}{5}\boldsymbol{R})$ with $\boldsymbol{R} = (0, 0, R_i)$. By solving the Hill-Wheeler equation Eq. (66), we can obtain the following Brink-GCM wave function,

$$\Phi_{\text{GCM}}^{J^\pi} = \sum_i f(R_i)\Phi_{\text{Brink}}^{J^\pi}(R_i). \quad (67)$$

Thus, we can compare the single THSR-type wave function with the exact Brink-GCM wave function by calculating their squared overlap $|\langle\hat\Phi_{\min}^{\text{THSR}}|\hat\Phi_{\text{GCM}}^{\text{Brink}}\rangle|^2$. In Table 4, we can find that the obtained single THSR-type wave functions have 99.29%, 98.79%, 97.75%, 99.98%, and 99.87% squared overlaps for $J^\pi = 0^+, 2^+, 4^+, 1^-$, and $3^-$ states of $^{20}$Ne, respectively, with the corresponding Brink-GCM solutions. These very high squared overlaps mean that the single THSR-type wave functions are almost 100% equivalent to the corresponding RGM/GCM wave functions, thus, these obtained single angular-momentum projected THSR-type wave functions can accurately describe the states of the inversion doublet bands in $^{20}$Ne. Moreover, the concept of nonlocalized clustering indicated by the THSR-type wave function obtained from the hybrid-Brink-THSR wave function is essential to correctly understand the $\alpha + ^{16}$O cluster structure in $^{20}$Ne.

Since we have shown that the single THSR wave functions are almost equivalent to the GCM wave function, it

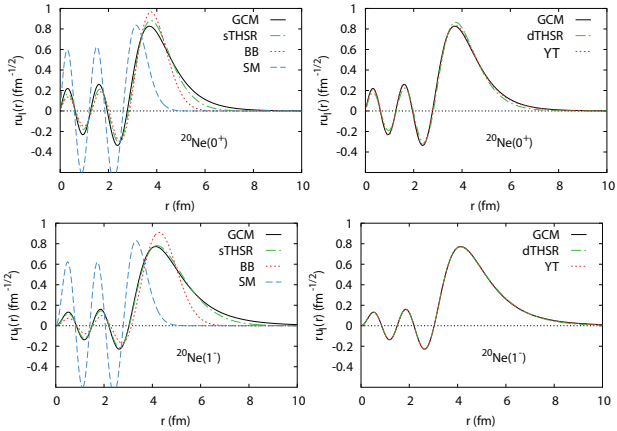

**Fig. 25** Relative wave functions $ru_l(r)$ of the optimized trial wave functions for the $J^\pi = 0^+$ and $1^-$ states of $^{20}$Ne compared with that of the GCM wave function. $ru_l(r) = R_{nl}(b_r; r)$ for the SU(3) shell-model limit is also shown. This figure is from Ref. [97] and see details from it.

## 4.2 Localized clustering from the inter-cluster Pauli principle

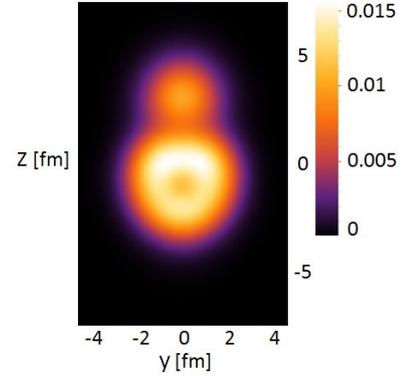

**Fig. 26** Density distribution of the $^{16}$O+$\alpha$ hybrid-Brink-THSR wave function with $S_z = 0.6$ fm and $(\beta_x, \beta_y, \beta_z) = (0.9$ fm, 0.9 fm, 2.5 fm) [11].

is natural that their relative wave functions, or any physical quantities calculated by the wave functions should be also equivalent. In Ref. [97], the author analyzed the $\alpha$-cluster wave functions in cluster states of $^8$Be and $^{20}$Ne by comparing the exact relative wave function obtained by GCM wave function with various types of trial functions. The adopted trial functions are the Brink wave function (BB), the spherical Gaussian THSR wave function (sTHSR), the deformed Gaussian THSR wave function (dTHSR), and a little more mathematical Yukawa tail (YT) function. By comparing with these trial functions and the GCM wave function, it was confirmed again that the THSR wave function can give a very good description of the positive-parity states and negative-parity states in $^{20}$Ne. To be more clear for the behaviors of the localized wave function, nonlocalized wave function, and the exact GCM wave function, we can focus on their relative wave functions, which are essential parts for the description of the motion of clusters. Figure 25 shows the relative wave functions of the optimized trial wave functions and the corresponding GCM wave function, it can be seen clearly whether it is the compact shell-model-like ground state of $^{20}$Ne or the developed negative-parity state, the THSR wave function characterized the nonlocalized clustering is much better than the Brink wave function. Furthermore, in the tail parts of the relative wave functions, the single deformed THSR wave functions can almost give exact description while this is only can be done by superposed many localized Brink wave functions in GCM calculations. The evidence presented thus far supports the idea that the nonlocalized clustering should be one essential feature of cluster motion.

Now we explain how the idea of the parity-violating deformation of localized $^{16}$O+$\alpha$ clustering for the inversion-doublet bands of $^{20}$Ne can be justified within the THSR wave function which assumes nonlocalized clusters. The parity-violating deformation is a property of the intrinsic state which is the instantaneous (or adiabatic) quantum state of the rotation of the nucleus. Since the instantaneous configuration of two clusters is of prolate shape, the prolate THSR wave function is the intrinsic state of the system and the oblate THSR wave function is not the intrinsic state but rather a mathematical object which expresses the rotation-average of the intrinsic state as discussed in Sec. 3.2. The spherical THSR wave function expresses the time average of the fully three-dimensional rotational motion, namely the angular-momentum projected state of the intrinsic state (the prolate THSR wave function). We, however, also need to notice the fact that two clusters can not come close to each other because, as just mentioned, of the inter-cluster Pauli repulsion, which implies that two clusters in the intrinsic state (the prolate THSR wave function) are effectively localized in space. Thus, the prolate THSR wave function has the parity-violating deformation of localized $^{16}$O+$\alpha$ clustering. We can say that dynamics prefers nonlocalized clustering but kinematics makes the system look like localized clustering. Figure 26 shows density distribution of the $^{16}$O+$\alpha$ hybrid-Brink-THSR intrinsic wave function and this kind of localized clustering appears. Of course, this localization is most pronounced in the necessarily strongly prolate two cluster systems. In systems with low density $\alpha$ clusters in number more than two have more space to move independently and are, therefore, less localized in spherical containers.

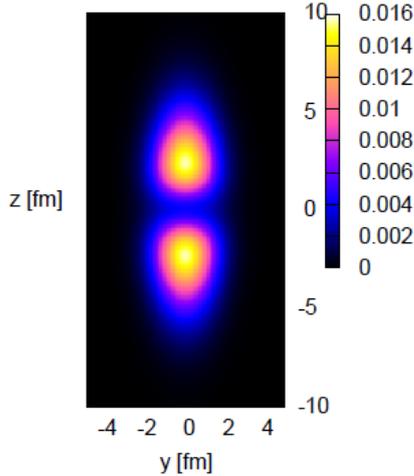

**Fig. 27** Density distribution of a $2\alpha$ prolate THSR wave function with $(\beta_x, \beta_y, \beta_z) = (1.78\text{ fm}, 1.78\text{ fm}, 7.85\text{ fm})$ [11].

The effective localization of clusters in the prolate THSR wave function of the two-cluster system is clearly seen in the density distribution of the prolate THSR wave function. In Fig. 27, it is shown the density distribution of a $2\alpha$ prolate THSR wave function with $(\beta_x, \beta_y, \beta_z)$=(1.78 fm, 1.78 fm, 7.85 fm). Since the THSR wave function before the antisymmetrization operation is obviously composed of nonlocalized clusters, it is evident that the clear spatial localization of clusters shown in this figure is attributed to the inter-cluster Pauli principle.

Recently it has been reported [98] that the density distribution of a $3\alpha$ THSR wave function with strong prolate deformation with $(\beta_x, \beta_y, \beta_z) = (0.01\text{ fm}, 0.01\text{ fm}, 5.1\text{ fm})$ shows clear spatial localization of three $\alpha$ clusters aligned linearly, which is displayed in Fig. 29 in Sec. 5.2. It is to be noted that because of the almost zero values of $\beta_x = \beta_y$, three $\alpha$ clusters are not allowed to expand into the $x$ and $y$ directions, which means that three $\alpha$ clusters are only allowed to make one-dimensional motion along the $z$ direction. Therefore the inter-cluster Pauli principle acts only along the $z$ direction, which is the reason of the spatial localization of the three $\alpha$ clusters. In Ref. [98] it is reported that the $\alpha$ linear-chain Brink-GCM wave function is almost 100% equivalent to a single $3\alpha$ THSR wave function with strong prolate deformation which is just the $3\alpha$ THSR wave function in Fig. 29 with $(\beta_x, \beta_y, \beta_z) = (0.01\text{ fm}, 0.01\text{ fm}, 5.1\text{ fm})$.

All these results provide important insights into the concept of nonlocalized clustering. In two-cluster systems, cluster states generally have effective localization of clusters because of the inter-cluster Pauli repulsion. However, in three or more cluster systems, the spatial arrangement of clusters are not necessarily geometrical, namely clusters can be nonlocalized, although the inter-cluster separations are non-zero simultaneously because of the inter-cluster Pauli repulsion. However, as is discussed in Ref. [98], if a cluster state is forced to have strongly-prolate deformation, the state can have effective localization of clusters like in the case of $3\alpha$ linear-chain structure. When the inter-cluster separations are large, the spatial arrangement of clusters can be non-rigid and gas-like. We will discuss this case in the following section.

## 5 Container model of cluster dynamics

In the previous sections, we showed that a single configuration of the THSR wave function does not only describe very accurately gas-like cluster states like the Hoyle state in $^{12}$C but also non-gaslike localized cluster structure states like the inversion doublet band states in $^{20}$Ne. This means that the THSR ansatz, which was originally introduced to represent the $\alpha$ condensates with a dilute density, can also be applied to more general cluster states. As was discussed in Sec. 4, the important ingredient of the THSR wave function is to be characterized by a nuclear size parameter, $\boldsymbol{\beta}$. This is completely different from conventional cluster model descriptions like the Brink wave function, including the AMD and FMD models, in which the relative distance between clusters, or between nucleons, are the key parameters. Thus the THSR model does not provide geometrical configurations of clusters but is closer to a cluster-mean-field approach, where constituent clusters make independent motions in the cluster-mean-field self-consistent potential, under the constraint of the antisymmetrization of nucleons. Furthermore in this model, a shape and size of the nucleus (characterized by $\boldsymbol{\beta}$) can be varied flexibly, according to dynamical excitation of the nucleus. This model therefore gives the picture that the individual clusters move rather freely in a shape- and size-changeable container. In this section, we investigate the container dynamics of clusters with respect to its size (Sec. 5.1) and shape (Sec. 5.2), rich spectra of $3\alpha$ (Sec. 5.3) and $4\alpha$ (Sec. 5.4), and discuss more extension and application of the THSR ansatz following this picture (Sec. 5.5).

### 5.1 Nuclear size as the generator coordinate for the evolution of cluster structure and extension of the THSR wave function

In Sec. 2.2, we argued that the $(0_2^+)_{\text{THSR}}$, $(0_3^+)_{\text{THSR}}$, and $(0_4^+)_{\text{THSR}}$ states in the THSR ansatz correspond to the $(0_2^+)_{\text{OCM}}$, $(0_4^+)_{\text{OCM}}$, and $(0_6^+)_{\text{OCM}}$ states in the OCM calculation, which have the structures of $^{12}\text{C}(0^+)$ and $\alpha$ in an $S$-wave, of $^{12}\text{C}(0^+)$ and $\alpha$ in a higher nodal $S$-wave, and of the $4\alpha$ condensate, respectively. The THSR wave functions of these states, $\Phi_{4\alpha,\lambda}^{J=0}$ with $\lambda = 1, \cdots, 4$ of Eq. (20), are obtained by solving the Hill-Wheeler equation Eq. (21), with a constraint of $\beta_0 \equiv \beta_\perp = \beta_z$, which are represented as follows: $\Phi^{J=0}_{4\alpha,\lambda} = \sum_{\beta_0} f^{J=0}_\lambda(\beta_0) \Phi^{J=0}_{4\alpha}(\beta_0)$. In this subsection we clarify that the cluster excitation given by solving the Hill-Wheeler equation are associated with an evolution of the size of container, which is characterized by $\beta_0$.

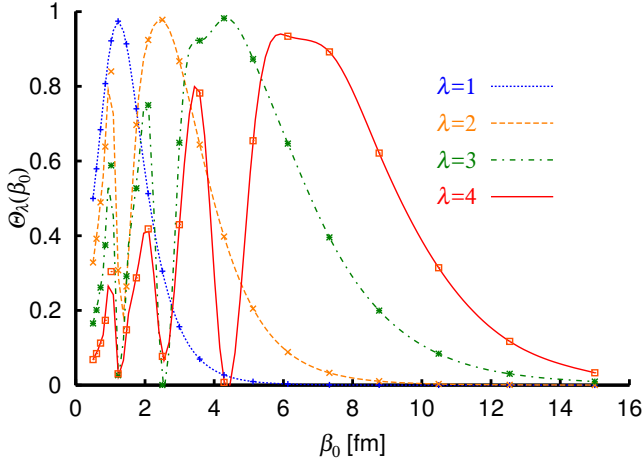

**Fig. 28** Overlap amplitudes $\Theta_\lambda(\beta_0)$ of Eq. (70) between the wave functions $\Phi^{J=0}_{4\alpha,\lambda}$ and $\widetilde{\Phi}_\lambda(\beta_0)$ with $\lambda = 1, \cdots, 4$, as a function of $\beta_0$. Figure was reproduced from Ref. [20].

We first construct orthonormal wave functions from the THSR wave functions as follows:

$$\widetilde{\Phi}_\lambda(\beta_0) = \mathcal{N}_\lambda \widehat{P}_{\lambda-1} \Phi^{J=0}_{4\alpha}(\beta_0), \quad (\lambda = 1, \cdots, 4), \quad (68)$$

where $\mathcal{N}_\lambda$ are normalization constants and $\widehat{P}_{\lambda-1}$ are projection operators defined by

$$\widehat{P}_{\lambda-1} = 1 - \sum_{k=1}^{\lambda-1} |\Phi^{J=0}_{4\alpha,k}\rangle\langle \Phi^{J=0}_{4\alpha,k}|, \quad (\lambda = 1, \cdots, 4). \quad (69)$$

Here the case $\lambda = 1$ gives $\widehat{P}_0 = 1$. The wave function $\widetilde{\Phi}_\lambda(\beta_0)$ depends on the parameter $\beta_0$ and is orthogonal to the wave functions $\Phi^{J=0}_{4\alpha,\lambda'}$ of the $(0^+_{\lambda'})_{\text{THSR}}$ states satisfying $\lambda' \leq \lambda - 1$. We then calculate the following squared overlap amplitudes between the wave functions $\widetilde{\Phi}_\lambda(\beta_0)$ and $\Phi^{J=0}_{4\alpha,\lambda}$,

$$\Theta_\lambda(\beta_0) = \left| \left\langle \Phi^{J=0}_{4\alpha,\lambda} \middle| \widetilde{\Phi}_\lambda(\beta_0) \right\rangle \right|^2, \quad (\lambda = 1, \cdots, 4). \quad (70)$$

In Fig. 28 we show $\Theta_\lambda(\beta_0)$ of Eq. (70) as a function of $\beta_0$. For any wave functions $\Phi^{J=0}_{4\alpha,\lambda}$ ($\lambda = 1, \cdots, 4$), $\Theta_\lambda(\beta_0)$ ($\lambda = 1, \cdots, 4$) take values close to 100% when $\beta_0$ is optimally chosen. This result means that all the $(0^+_\lambda)_{\text{THSR}}$ states are almost represented by single THSR wave functions parametrized by single $\beta_0$ values in the spaces orthogonal to the lower-lying states. We discussed this point of view in Sec. 2.1.3 (see Fig. 6), to show that the Hoyle state can be represented by a single THSR configuration with an optimal $\beta$ parameter value, and hence can be regarded as the $3\alpha$ condensate state. In particular, in this case for the $(0^+_4)_{\text{THSR}}$ state the maximum of $\Theta_4(\beta_0)$ amounts to 96% at $\beta_0 = 6.5$ fm, for which this large value of $\beta_0$ again supports the picture that this state is considered to be the $4\alpha$ condensate state, in which the $4\alpha$ clusters are trapped into a very large (spherical) container characterized by the nuclear size parameter $\beta_0$.

**Table 5** The largest values of the squared overlaps in Fig. 28 are shown for the four THSR states, $(0^+_\lambda)_{\text{THSR}}$ ($\lambda = 1, \cdots, 4$), together with the corresponding $\beta_0$ values.

| | $0^+_\lambda$ | | | |
|---|---|---|---|---|
| | $\lambda = 1$ | $\lambda = 2$ | $\lambda = 3$ | $\lambda = 4$ |
| $\Theta_\lambda(\beta_0)$ | 0.98 | 0.98 | 0.98 | 0.96 |
| $\beta_0$ (fm) | 1.2 | 2.5 | 4.0 | 6.5 |

However, this figure shows that not only the $4\alpha$ dilute gas-like state but also the lower states $(0^+_\lambda)_{\text{THSR}}$ with $\lambda = 1, 2, 3$, in which $^{12}\text{C} + \alpha$ cluster dynamics prevails, are expressed as single THSR wave functions in the proper orthogonal spaces. This corresponds to the case of $^{20}\text{Ne}$ discussed in the previous section, Sec. 4, in which the $^{16}\text{O} + \alpha$ inversion doublet band states are all expressed with almost 100% precision by single THSR configurations. We further show in Table 5 the largest squared overlaps and the corresponding $\beta_0$ values for the four states. We can see that the optimal $\beta_0$ values increase from the $(0^+_1)_{\text{THSR}}$ to the $(0^+_4)_{\text{THSR}}$ states. The r.m.s. radii are also proportional to the $\beta_0$ parameter values. Therefore we can say that the clustering excitations, which are usually described by solving the Hill-Wheeler equation, are very accurately characterized by the variation of the $\beta_0$ value, i.e. dilatation of the container, from the most compact $(0^+_1)_{\text{THSR}}$ state to the most dilute gas-like $(0^+_4)_{\text{THSR}}$ state with the $4\alpha$ condensate structure, through the intermediate density $(0^+_2)_{\text{THSR}}$ and more dilute $(0^+_3)_{\text{THSR}}$ states with the $^{12}\text{C} + \alpha$ cluster structures. The evolution of the cluster excitations can be described by an evolution of the container, from a smaller value of $\beta_0$ to a larger one. This also supports the correctness of the container picture.

We should also note that the projection operator $\widehat{P}_\lambda$, which removes the lower-lying-state components with more compact structure, plays a role as a repulsive force to prevent the $\alpha$ clusters from being resolved or from coming closer to form the more compact structure, due to its orthogonality condition. As a result, higher-lying states are constructed one by one. This is essentially the same as the situation of $^8\text{Be}$, where the antisymmetrization operator $\mathcal{A}$ removes the Pauli forbidden states, to construct a structural repulsive core between the two $\alpha$ clusters.

The more sophisticated treatment of $^{12}\text{C} + \alpha$ configuration in the container picture is to use an extended version of the THSR wave function, which we will discuss in Sec. 5.4.

## 5.2 One-dimensional gas of $\alpha$ clusters — $\alpha$-linear-chain structure

In this subsection, we consider the so-called $\alpha$-linear-chain state, which has been studied for many years [28, 99] and it is one of the most typical representatives of the localized cluster-structure states. A special interest is whether the $\alpha$-linear-chain states, which are ordinary considered as the $\alpha$ clusters geometrically arranged in a line, can be reinterpreted by the container picture provided by the THSR wave function, as was clarified in the case of the $^{20}\text{Ne}$ in Sec. 4 and even of lower-lying states of $^{16}\text{O}$ in Sec. 5.1. The ideal situation of the $n\alpha$ linear-chain structure states are considered to be obtained by the Brink-GCM wave function, which was discussed repeatedly in the precedent sections but is in this case considered in only one dimension. If that is $z$-axis, the (ideal) linear-chain states (LCS) can be obtained below,

$$\Phi_J^{(\text{B-GCM})} = \sum_{\widetilde{R}_{1z},\cdots,\widetilde{R}_{nz}} f_J(\widetilde{R}_{1z},\cdots,\widetilde{R}_{nz})\widehat{P}_{MK}^J \Phi_{n\alpha}^B(\widetilde{R}_{1z},\cdots,\widetilde{R}_{nz}), \tag{71}$$

where $\Phi_{n\alpha}^B(\widetilde{R}_{1z},\cdots,\widetilde{R}_{nz})$ is the Brink wave function in Eq. (16) with $\widetilde{\boldsymbol{R}}_i = (0, 0, \widetilde{R}_{iz})$, and the coefficients $f_J(\widetilde{R}_{1z},\cdots,\widetilde{R}_{nz})$ are determined by solving the Hill-Wheeler equation like Eq. (21) taking the variables, $\widetilde{R}_{1z},\cdots,\widetilde{R}_{nz}$, as the generator coordinates instead of $\boldsymbol{\beta}$ parameter in this equation. In Ref. [98], we calculated the $3\alpha$ and $4\alpha$ LCS in this way and efficiently obtained convergent results by superposing the basis functions in Eq. (71) in a stochastic way, whose numbers amount to 100 and 300 for the $3\alpha$ and $4\alpha$ cases, respectively. Then we calculated the squared overlap of these LCS wave functions and the single configuration of the THSR wave function, i.e. the quantity defined as: $|\langle \Phi_{n\alpha}^J(\boldsymbol{\beta})|\Psi_J^{(\text{B-GCM})}\rangle|^2$. The detailed parameter set chosen in this calculation can be found in Ref. [98]. Most amazingly the squared overlap values are found to be very large, more than 98%, and 93%, for all $J^\pi = 0^+$, $2^+$, and $4^+$ states in $^{12}\text{C}$ and $^{16}\text{O}$ nuclei, respectively. The optimal $\boldsymbol{\beta}$ parameter values in this situation all indicate extremely-prolate deformation, namely, $\beta_\perp \sim 0$ and $\beta_z \gg 0$.

In Table 6, we show the maximum squared overlap values and the corresponding $\boldsymbol{\beta}$ values (denoted as $3\alpha$ and $4\alpha$ LCS). These large overlap values are to be compared with the maximum square overlap values with $3\alpha$ and $4\alpha$ single Brink wave functions, instead of the single THSR wave functions, which are only 78% and 48%, respectively. Therefore for the LCS, the localized $\alpha$-cluster picture aligned in a line, which is provided by the Brink wave function, is less appropriate while the one-dimensional container picture, newly provided by the THSR ansatz, which gives much larger values close to almost 100%, is much more appropriate to understand the LCS. More precisely, the LCS can be represented by a container with extremely prolate shape, in which it seems the $\alpha$ clusters again move rather freely without mutual dynamical correlation.

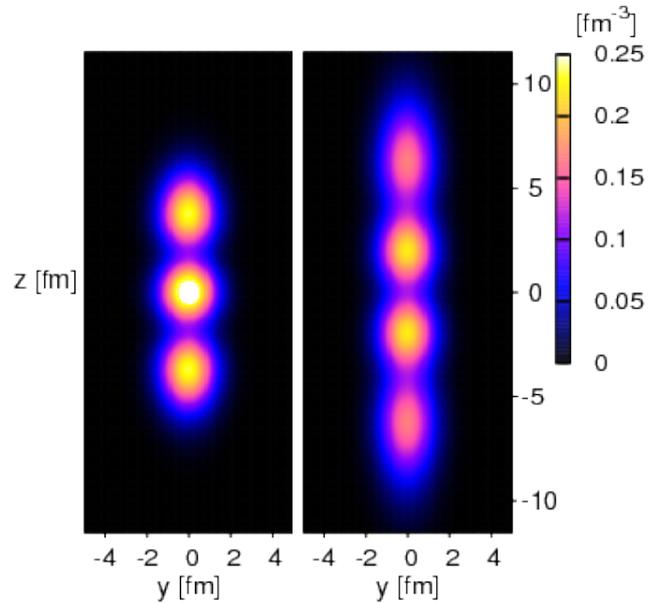

**Fig. 29** Intrinsic density profiles of the $3\alpha$-(Upper) and $4\alpha$-(Lower) linear-chain states from the THSR wave functions before angular-momentum projection at ($\beta_x = \beta_y$=0.1 fm,$\beta_z$=5.1 fm) and at ($\beta_x = \beta_y$=0.1 fm,$\beta_z$=8.2 fm), respectively. Figure was reproduced from Ref. [98].

Nevertheless, just like the case of $^{16}\text{O} + \alpha$ inversion doublet, the nucleon-density distributions of the intrinsic THSR wave functions of the LCS show the localized $\alpha$ clusters, as shown in Fig. 29 for the $3\alpha$ and $4\alpha$ cases. This is because of inter-$\alpha$ Pauli repulsion as discussed in the previous section, Sec. 4. Therefore, this again indicates that the localization of nuclear clustering comes from the kinematic effect, namely the effect of the antisymmetrization, since if we believe the present results, there is no way to make $\alpha$ clusters localized except for the operator $\mathcal{A}$, considering the form of the single THSR wave function of Eq. (18), which essentially represents a free motion of the $\alpha$ clusters in a container with a certain shape and size characterized by $\boldsymbol{\beta}$ parameter.

**Table 6** The maximum squared overlaps between the single THSR wave functions and RGM/GCM wave functions. The corresponding $\boldsymbol{\beta}$ values, where $(\beta_x = \beta_y, \beta_z) = (\beta_\perp, \beta_z)$, are shown in parentheses in a unit of fm.

|     | $^8$Be | $^{12}$C | Max.$(\beta_\perp, \beta_z)$ $^{20}$Ne | $3\alpha$ LCS | $4\alpha$ LCS | $^9_\Lambda$Be |
| --- | --- | --- | --- | --- | --- | --- |
| $0^+$ | 1.000(1.8,7.8) | $0^+_1$: 0.93(1.5,1.5) <br> $(0^+_1$:0.978)* <br> $0^+_2$:0.993(5.3,1.5) | 0.993(0.9,2.5) | 0.987(0.1,5.1) | 0.944(0.1,8.2) | 0.995(1.6,3.0) |
| $2^+$ |  |  | 0.988(0.0,2.2) | 0.989(0.1,5.4) | 0.942(0.1,8.4) | 0.994(0.1,3.0) |
| $4^+$ |  |  | 0.978(0.0,1.8) | 0.981(0.1,6.6) | 0.931(0.1,9.0) | 0.977(0.1,2.1) |
| $1^-$ |  |  | 1.000(3.7,1.4) |  |  |  |
| $3^-$ |  |  | 0.999(3.7,0.0) |  |  |  |

* The value by the use of the extended version of the THSR wave function, with the parameter values $(\beta_{1\perp}, \beta_{1z}, \beta_{2\perp}, \beta_{2z}) = (0.1, 2.3, 2.8, 0.1)$, which we will discuss in Sec. 5.3.1.

We also summarize in Table 6 the squared overlap values between the single THSR configurations and the corresponding Brink-GCM wave functions in $^8$Be [38, 41], $^{12}$C [9, 41, 43], $^{20}$Ne [10, 11, 22], and $^9_\Lambda$Be [100], some of which are already discussed up to now. All these values are very large and close to 100%. These results suggest that the single THSR wave functions can describe gas-like and non-gaslike cluster states in a unified way, allowing us to consider that cluster states in general retain a container structure, where the container corresponds to the cluster-mean-field potential characterized by the size parameter $\boldsymbol{\beta}$ in the THSR wave function. The constituent clusters are confined into the container in a nonlocalized way and occupy an identical orbit, under the effect of the antisymmetrization. This is very different from the conventional cluster model wave function, like the Brink model wave function [34], in which relative motions of clusters are characterized by inter-cluster separation distance parameters, in a localized way.

## 5.3 Rich spectra of $3\alpha$ states above the Hoyle state in $^{12}$C

### 5.3.1 Positive-parity states in $^{12}$C

In Sec. 2.1.5, the nature of the $0^+_3$ state in $^{12}$C was discussed, which was identified as the monopole excitation from the Hoyle state. In this subsection, we argue that there exist besides it a number of other $\alpha$ cluster states above the Hoyle state, which we can qualify as the excited states of the Hoyle state. It is then indispensable to generalize the THSR ansatz, particularly to take into account $^8$Be + $\alpha$-type cluster structure. On the other hand, in Sec. 4, we showed that the single THSR wave functions excellently describe the $^{16}$O + $\alpha$ cluster states in $^{20}$Ne, and they are much better than the single Brink wave functions, whose variational parameter is the inter-cluster separation distance. This result suggests that using the $\boldsymbol{\beta}$ parameter of the THSR ansatz is much better than the Brink-type parameter choice to describe the $^8$Be + $\alpha$ structure in $^{12}$C, or even $^{12}$C + $\alpha$ structure in $^{16}$O.

We thus introduced in Sec. 2.1.5 the extended version of the THSR wave function Eq. (28), in which the two different width parameters $\boldsymbol{\beta}_1$ and $\boldsymbol{\beta}_2$ are associated to the two Jacobi coordinates $\boldsymbol{\xi}_1$ and $\boldsymbol{\xi}_2$, corresponding to $\alpha + \alpha$ and $2\alpha$-$\alpha$ motions, respectively. We show it again below,

$$\Phi^{JM}_{3\alpha}(\boldsymbol{\beta}_1, \boldsymbol{\beta}_2) = \widehat{P}^{JM} \mathcal{A} \Big\{ \prod_{i=1}^{2} \exp\Big[ -\mu_i \sum_{k=x,y,z} \frac{2\xi^2_{ix}}{b^2 + 2\beta^2_{ik}} \Big] \prod_{i=1}^{3} \phi(\alpha_i) \Big\}. \tag{72}$$

This generalization is a natural extension of the original THSR wave function, since taking $\boldsymbol{\beta}_1 = \boldsymbol{\beta}_2$ results in it. This also follows the container concept that is discussed in the previous sections, since the case of $|\boldsymbol{\beta}_1| \leq |\boldsymbol{\beta}_2|$ corresponds to the $^8$Be + $\alpha$ cluster structure, in which the $^8$Be core and the remaining $\alpha$ cluster are confined in a small-size container, while in the core $^8$Be the $2\alpha$ clusters are confined in a large-size container. The shapes and sizes of the two containers are then characterized by the corresponding two parameters, $\boldsymbol{\beta}_1$ and $\boldsymbol{\beta}_2$. We should note that the fact that the THSR wave function includes the $^8$Be + $\alpha$ asymptotic feature for $|\boldsymbol{\beta}_1| \ll |\boldsymbol{\beta}_2|$ allows for the sophisticated approaches to resonances like the CSM [65–69] and the analytic continuation of the coupling constant (ACCC) method [101].

For the excited states above the $3\alpha$ threshold, it is well known that the application of the bound state approximation gives accidental mixing between spurious continuum states and resonances. By using the fact that the root mean square (r.m.s.) radii of spurious continuum states are calculated to be extremely large within the bound state approximation, we developed a new method to remove the spurious continuum components [71]. First we diagonalize the operator of mean square radius as follows:

$$\sum_{\boldsymbol{\beta}'_1,\boldsymbol{\beta}'_2} \left[ \langle \Phi^{JM}_{3\alpha}(\boldsymbol{\beta}_1,\boldsymbol{\beta}_2) | \frac{1}{12} \sum_{i=1}^{12} (\boldsymbol{r}_i - \boldsymbol{X}_G)^2 | \Phi^{JM}_{3\alpha}(\boldsymbol{\beta}'_1,\boldsymbol{\beta}'_2) \rangle \right.$$
$$\left. - \{R^{(\gamma)}\}^2 \langle \Phi^{JM}_{3\alpha}(\boldsymbol{\beta}_1,\boldsymbol{\beta}_2) | \Phi^{JM}_{3\alpha}(\boldsymbol{\beta}'_1,\boldsymbol{\beta}'_2) \rangle \right] g^{(\gamma)}(\boldsymbol{\beta}'_1,\boldsymbol{\beta}'_2) = 0.$$
(73)

We then remove out of the present model space the eigenstates belonging to unphysically large eigenvalues. By taking the following bases,

$$\Phi^{(\gamma)}_{JM} = \sum_{\boldsymbol{\beta}_1,\boldsymbol{\beta}_2} g^{(\gamma)}(\boldsymbol{\beta}_1,\boldsymbol{\beta}_2) \Phi^{JM}_{3\alpha}(\boldsymbol{\beta}_1,\boldsymbol{\beta}_2), \qquad (74)$$

with $\gamma$ satisfying $R^{(\gamma)} \leq R_{\rm cut}$, we diagonalize Hamiltonian as follows:

$$\sum_{\gamma'} \langle \Phi^{(\gamma)}_{JM} | H | \Phi^{(\gamma')}_{JM} \rangle f^{(\gamma')}_\lambda = E_\lambda f^{(\gamma)}_\lambda. \qquad (75)$$

Since the present extended THSR wave function can include $^8$Be+$\alpha$ asymptotic form by taking the large values of the two width parameters $\boldsymbol{\beta}_1$ and $\boldsymbol{\beta}_2$, the $^8$Be+$\alpha$ continuum components, as well as the 3$\alpha$ continuum components, can be successfully removed by imposing the cutoff for the mean square radius $R^{(\gamma)} \leq R_{\rm cut}$. Although we could not obtain the excited states except for the $0^+_2$ and $2^+_2$ states by using the original THSR wave function [40], we can now obtain the other observed $0^+_3$, $0^+_4$, and $4^+_2$ states by using the present extended THSR wave function with a treatment of resonances. This method is also used in Sec. 2.1.5 to obtain the $0^+_3$ state and is there referred to as the radius constraint method.

With this type of generalized THSR wave function, one can get a much richer spectrum of $^{12}$C. In solving the Hill-Wheeler equation with the radius constraint method Eqs. (73)-(75), axial symmetry is assumed and the four $\beta$ parameters are taken as generator coordinates. In Fig. 30(Upper), the calculated energy spectrum is shown. One can see that besides the ground state band, there are many excited states obtained above the Hoyle state. All these states turn out to have large rms radii (3.7 ~ 4.7 fm), and therefore can be considered as excitation states of the Hoyle state. The Hoyle state can, thus be considered as the "ground state" of a new class of excited states in $^{12}$C. In particular, the nature of the series of states ($0^+_2$, $2^+_2$, $4^+_2$) and the $0^+_3$ and $0^+_4$ states have recently been much discussed from the experimental side. As was mentioned in Sec. 2.1.5, the $2^+_2$ state that theoretically has been predicted at a few MeV above the Hoyle state already in the early works of 3$\alpha$ Brink-GCM [13, 35] and 3$\alpha$ RGM [14] was recently confirmed by several experiments [57–61]. A strong candidate for a member of the Hoyle family of states with $J^\pi = 4^+$ was also reported by Freer *et al.* [102]. Itoh *et al.* recently pointed out that the broad $0^+$ resonance at 10.3 MeV should be decomposed into two states: $0^+_3$ and $0^+_4$ [62]. This finding is consistent with theoretical predictions where the $0^+_3$ state is considered as a breathing excitation of the Hoyle state [16–18, 63] and the $0^+_4$ state as the bent arm or linear chain configuration [17, 44, 70].

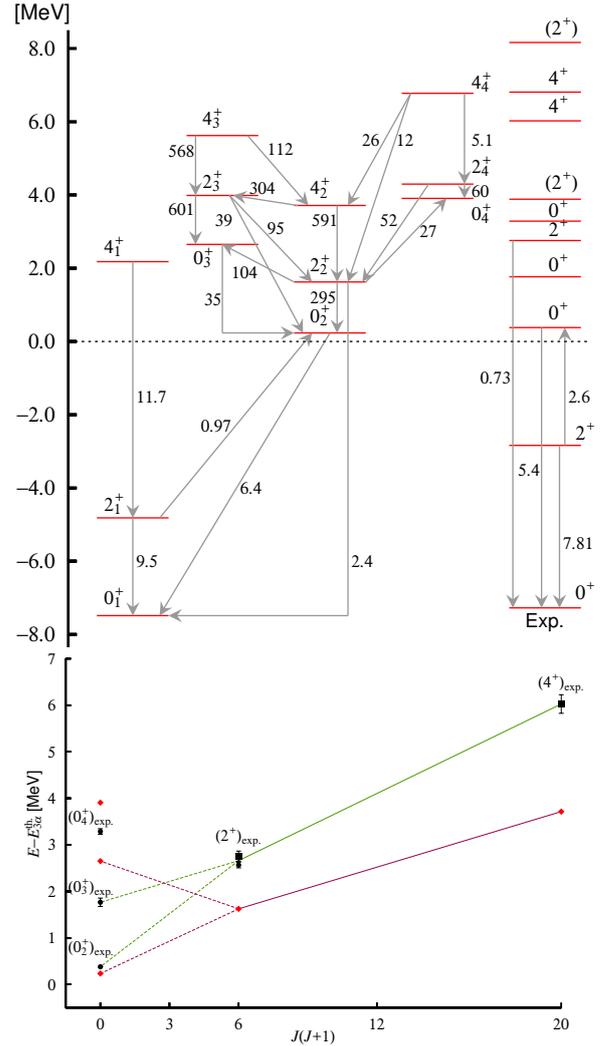

**Fig. 30** (color online). (Upper): Calculated energy levels and electric transition strengths are shown and compared with experiments. (Lower): The observed energy levels for the $0^+_3$, $0^+_4$, and $2^+_2$ states in Ref. [56], and the $2^+_2$ [61] and $4^+_2$ [102] states are denoted by black circles and black squares, respectively. The calculated energy levels for the five states are denoted by red diamonds. Figure (Lower) taken from [16].

In Fig. 30(Upper), the $E2$ transition strengths between $J$ and $J \pm 2$ states and monopole transitions between $0^+$ states are also shown with corresponding arrows. We can note the very strong $E2$ transitions inside the Hoyle band, $B(E2; 4_2^+ \to 2_2^+) = 591\ e^2\text{fm}^4$ and $B(E2; 2_2^+ \to 0_2^+) = 295\ e^2\text{fm}^4$. The transition between the $2_2^+$ and $0_3^+$ states is also very large, $B(E2; 2_2^+ \to 0_3^+) = 104\ e^2\text{fm}^4$. In Fig. 30(Lower), the calculated energy levels are plotted as a function of $J(J+1)$, together with the experimental data. There have been attempts to interpret this as a rotational band of a spinning triangle as this was successfully done for the ground state band [103, 104]. However, the situation may not be as straightforward as it seems. This is because the two transitions $2_2^+ \to 0_2^+$ and $2_2^+ \to 0_3^+$ are of similar magnitude, and hence no clear band head can be identified. It was also pointed out in Refs. [44, 70] that the states $2_2^+, 4_2^+$ form a rotational band not with the $0_2^+$ but with the $0_3^+$ state. The line which connects the two other hypothetical members of the rotational band, in Fig. 30(Lower), has a slope, which points to somewhere in between of the $0_2^+$ and $0_3^+$ states. The similar effect is also argued in the study of the $4\alpha$ condensate and $^{12}\text{C}(0_2^+) + \alpha$ rotational band in $^{16}\text{O}$ [105, 106]. Furthermore we show in Fig. 31 the $S^2$ factors of the $^8\text{Be} + \alpha$ components, which are defined below,

$$S^2_{[I,l]}(J_\lambda^+) = \int dr [r\mathcal{Y}_{[I,l]J}(r)]^2, \qquad (76)$$

where $\mathcal{Y}_{[I,l]J}$ is the RWAs of the $^{12}\text{C}(J_\lambda^+)$ states in the $^8\text{Be}(I) + \alpha(l)$ channel. Except for the $0_4^+$ state, all the states have dominant components from the channels $[0, J]_J$, which is consistent with $^8\text{Be}(0^+)$ and $\alpha$ rotation. However, In the $2_2^+$ and $4_2^+$ states, the components from the channels $[0, 2]_2$ and $[0, 4]_4$, respectively, are dominant. However, in the $2_2^+$ and $4_2^+$ state, the mixtures of the other components are also found to be large. These suggest that to conclude from there this gives rise to a simple rotational band, is premature [16].

Finally we briefly discuss the structure of the $0_4^+$ state in $^{12}\text{C}$. In the AMD and FMD calculations [44, 70], the dominant intrinsic configuration of the $0_4^+$ state is a bent-armed structure of the $3\alpha$ clusters, resembling the linear-chain structure. However, any firm conclusions cannot still be drawn, since in the AMD, FMD, or even the $3\alpha$ Brink-GCM calculation, the four $0^+$ states are not consistently reproduced. As we mentioned above, in these calculations, the $0_3^+$ state is missing and only the $0_4^+$ state was obtained. Indeed, it is reported that the orthogonality to the lower states plays an essential role in the emergence of the linear-chain structure in the $0_4^+$ state [107]. On the other hand, in Fig. 31 the state is shown to have the largest component from $[2, 2]_0$ channel, i.e. the $\alpha$ cluster orbiting in a $D$-wave in the $^8\text{Be}(2^+)$ core. This is consistent with the former calculations in Refs. [16, 63]. We then examine whether the interpretation of the linear-chain-like structure is feasible for the $0_4^+$ state.

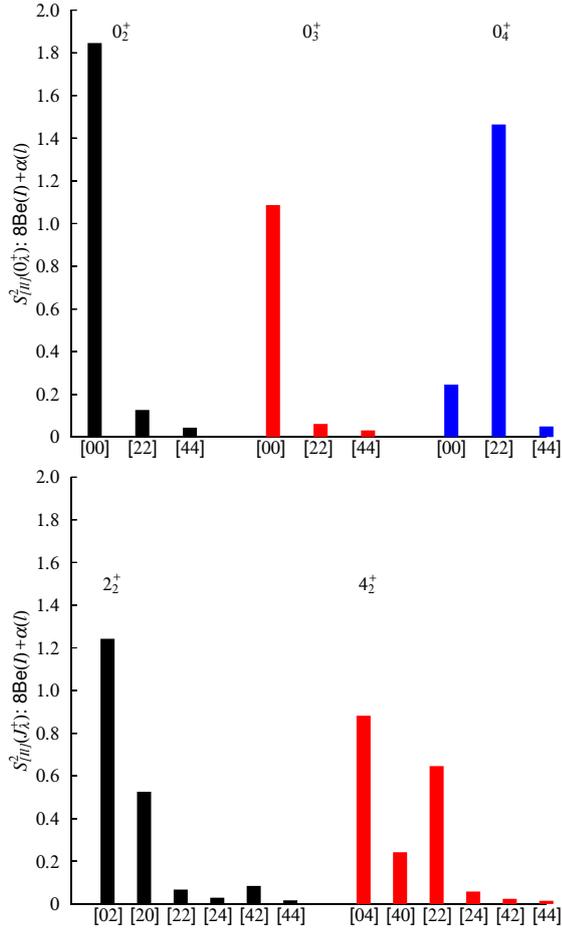

**Fig. 31** Probability distributions for various components in the Hoyle and excitations of the Hoyle state (from [16]).

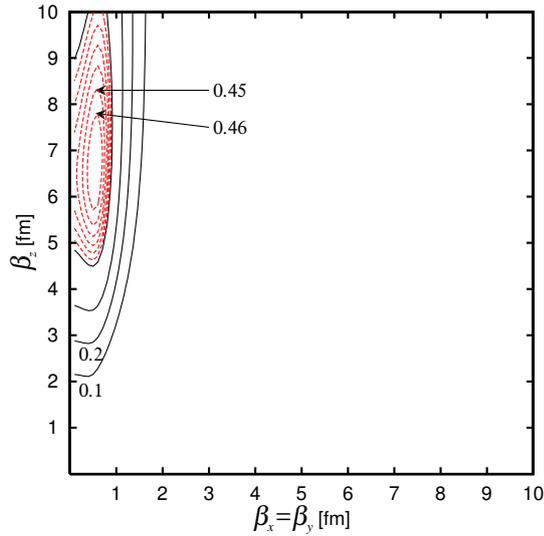

**Fig. 32** (color online) Contour map of the squared overlap between the wave function of the $0_4^+$ state and the single THSR wave function, in two-parameter space, $\boldsymbol{\beta}_1 = \boldsymbol{\beta}_2 = (\beta_x = \beta_y, \beta_z)$. Black solid curves are drawn in a step of 0.1 and red dotted curves, which cover the region of more than 0.41, are in a step of 0.01.

In Fig. 32, we show the contour map of the squared overlap between the $0_4^+$ state and the single THSR configuration in two-parameter space, $\boldsymbol{\beta}_1 = \boldsymbol{\beta}_2 = (\beta_x = \beta_y, \beta_z)$. The contour lines giving more than 0.4 are denoted by dot in red, in a step of 0.01. This contour map has a characteristic feature, where the strongly prolate deformation is only allowed to have a non-negligible squared overlap amplitude. Except for this prolate deformed region, the squared overlap is less than 0.1. The largest value is 0.47, which is not so much large but clearly indicates the $3\alpha$ linear-chain structure. The parameter values giving the maximal squared overlap is calculated to be $\boldsymbol{\beta}_1 = \boldsymbol{\beta}_2 = (0.6, 6.7 \text{ fm})$, which is close to $\boldsymbol{\beta}_1 = \boldsymbol{\beta}_2 = (0.1, 5.1 \text{ fm})$ that was obtained in a rather ideal one-dimension situation in Ref. [98]. In Ref. [98], it is discussed that largely prolate-deformed THSR wave function shows one-dimensional $\alpha$ condensate of $3\alpha$ clusters, which is fairly different from the ordinary picture of the linear-chain state with rigid-body $3\alpha$-cluster configuration arranged in a line in a spatially localized way. Thus we can say that the present $0_4^+$ state has the one-dimensional $\alpha$ condensate structure by around 50%, where the $3\alpha$ clusters are loosely trapped into a prolate-deformed potential like a one-dimensional gas.

### 5.3.2 Container picture for the $3^-$ and $4^-$ states in $^{12}$C

As we know, the negative-parity states in $^{12}$C also have many well-developed cluster states [13, 14, 108]. Quite recently, the $1^-$ state was reported to have the enhanced dipole transition strengths from the ground state and it probably has the dilute cluster structure [109]. On the other hand, the negative-parity states of $^{12}$C are closely related with the intrinsic structure of $3\alpha$ clusters. Bijker et al. [103, 110] proposed a description of cluster states in nuclei in terms of representations of unitary algebras. The $^{12}$C was assumed to have $D_{3h}$ $3\alpha$ cluster structure and many cluster states $0^+, 2^+, 3^-, 4^\pm, 5^-$ were predicted. Particularly, the $5^-$ state at 22.4(2) MeV in $^{12}$C was observed from experiments [104]. Within our microscopic container model, the nonlocalized motion is the essential feature of $3\alpha$ and it seems rigid geometrical arrangements of $^{12}$C were not favored, especially the Hoyle state and other developed positive-parity cluster states. The positive-parity states of $^{12}$C have been well described within the THSR wave function as discussed in Sec. 5.3.1. As a first step to study the negative-parity states in $^{12}$C, we focus on the $3^-$ and $4^-$ states in $^{12}$C in this subsection.

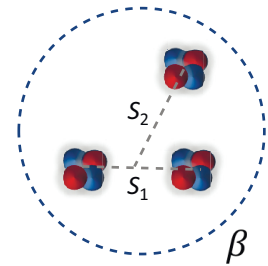

**Fig. 33** Schematic diagram for the container picture for dealing with the negative-parity cluster structure in $^{12}$C. Figure was reproduced from Ref. [111].

To deal with the negative-parity states, we can introduce the shift parameter to violate the parity of the THSR wave function, which has been discussed for dealing with the negative-parity states of two-cluster structure in $^{20}$Ne in Sec. 4. We start from the general two-$\beta$

THSR wave function,

$$\Phi(\boldsymbol{\beta}, \boldsymbol{S}) = \int \boldsymbol{R_1} d\boldsymbol{R_2} \exp[-\frac{(\boldsymbol{R}_1 - \boldsymbol{S}_1)^2}{\beta_1^2} - \frac{(\boldsymbol{R}_2 - \boldsymbol{S}_2)^2}{\beta_2^2}]\Phi^B(\boldsymbol{R}_1, \boldsymbol{R}_2)$$
$$\propto \phi_G \mathcal{A}\{\exp[-\frac{(\boldsymbol{\xi}_1 - \boldsymbol{S}_1)^2}{B_1^2} - \frac{(\boldsymbol{\xi}_2 - \boldsymbol{S}_2)^2}{B_2^2}\phi(\alpha_1)\phi(\alpha_2)\phi(\alpha_3)]\}, \tag{77}$$

$$\Phi^B(\boldsymbol{R}_1, \boldsymbol{R}_2) \propto$$
$$\phi_G \mathcal{A}\{\exp(-\frac{(\boldsymbol{\xi}_1 - \boldsymbol{R}_1)^2}{b^2} - \frac{(\boldsymbol{\xi}_2 - \boldsymbol{R}_2)^2}{\frac{3}{4}b^2})\phi(\alpha_1)\phi(\alpha_2)\phi(\alpha_3)\}, \tag{78}$$

where $B_1^2 = b^2 + \boldsymbol{\beta}_1^2$, $B_2^2 = \frac{3}{4}b^2 + \boldsymbol{\beta}_2^2$. $\boldsymbol{\beta}_1 \equiv (\beta_{1x} = \beta_{1y}, \beta_{1z})$, $\boldsymbol{\beta}_2 = (\beta_{2x} = \beta_{2y}, \beta_{2z})$. $\boldsymbol{\xi}_1 = \boldsymbol{X}_2 - \boldsymbol{X}_1$ and $\boldsymbol{\xi}_2 = \boldsymbol{X}_3 - (\boldsymbol{X}_1 + \boldsymbol{X}_2)/2$. The $\boldsymbol{X}_1, \boldsymbol{X}_2$, and $\boldsymbol{X}_3$ are the center-of-mass coordinates of the three clusters in $^{12}$C. In the above equation, $\Phi^B(\boldsymbol{R}_1, \boldsymbol{R}_2)$ is the $3\alpha$ Brink wave function with the corresponding generator coordinates $\boldsymbol{R}_1$ and $\boldsymbol{R}_2$. The $\boldsymbol{R}_1$ represents the specified distance parameter of two clusters and $\boldsymbol{R}_2$ these two clusters and the third $\alpha$ cluster. The $\boldsymbol{S}_1$ and $\boldsymbol{S}_2$ are introduced shift parameters.

At present, in order to simplify calculations and facilitate the analysis, we adopt the one-deformed THSR wave function by letting $\boldsymbol{\beta}_1 = \sqrt{2}\boldsymbol{\beta}$ and $\boldsymbol{\beta}_2 = \sqrt{3/2}\boldsymbol{\beta}$ in the above equations. And the introduced shift parameter $\boldsymbol{S}_1 \equiv (S_{1x}, S_{1y} = S_{1z} = 0)$ and $\boldsymbol{S}_2 \equiv (S_{2x}, S_{2y}, S_{2z} = 0)$ can arrange arbitrary triangle shapes on $xy$-plane. Also, we have known that these shift parameters play an important role for separating the $2\alpha$ clusters and a third $\alpha$ cluster in the wave function, which can be seen in Fig.33. Thus, by the angular momentum and parity projection technique, we can obtain the negative-parity components from the extended THSR wave function. As for the Hamiltonian, the Volkov No.2 [42] (modified version) with Majorana parameter $M = 0.59$ and harmonic-oscillator size parameter $b = 1.35$ fm, which was used by Kamimura et al. for $3\alpha$ RGM calculation [14].

The first $3^-$ and $4^-$ states in $^{12}$C have been known to have the developed cluster structure for a long time. In the one-$\beta$ container picture, we performed the variational calculations after the angular-momentum projections for the $3^-$ and $4^-$ states in two-parameter $\beta_x = \beta_y$ and $\beta_z$ space. The shift parameter is adopted as $\boldsymbol{S}_1 = (1/2, 0, 0)$, $\boldsymbol{S}_2 = (0, 0, \sqrt{3}/2)$.

Figure 34 shows the variational result for the $3^-$ state. It can be seen that [111] there are two local minimum points appeared in the contour plot. There is a long and narrow valley connecting the two points. The deeper local minimum point appear at $\beta_x = \beta_y$=1.5 fm and $\beta_z$=3 fm. The obtained energy for the $3^-$ state is about -80.9 MeV. Figure 35 shows the contour plot for the $4^-$ state.

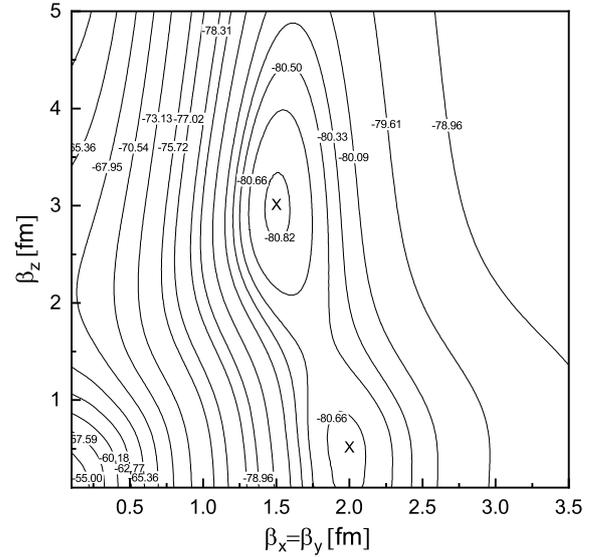

**Fig. 34** Contour plot for the $3^-$ state in the two-parameter space $\beta_x = \beta_y$ and $\beta_z$. Figure was reproduced from Ref. [111].

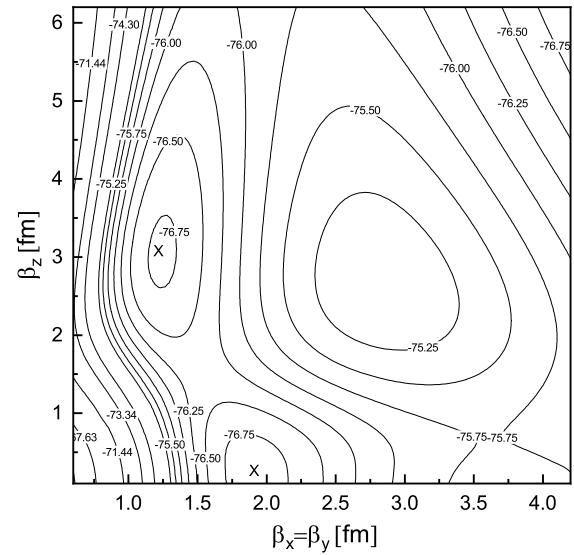

**Fig. 35** Contour plot for the $4^-$ state in the two-parameter space $\beta_x = \beta_y$ and $\beta_z$. Figure was reproduced from Ref. [111].

We found two local minimum points and the deeper one appears at $\beta_x = \beta_y$=1.0 fm and $\beta_z$=3.4 fm. The obtained energy for the $4^-$ state is -76.8 MeV. It should be noted that the THSR wave functions with respect to the two local minima for $3^-$ and $4^-$ states are also quite similar, whose squared overlap values are about 98%. This means the equivalence of the oblate and prolate THSR wave functions after angular-momentum projections, which is similar to the case of $^{20}$Ne in Sec. 4.1.2.

**Table 7** The calculated energies from the single optimum THSR wave functions in Eq. (77), the single optimum Brink wave functions in Eq. (78), and the GCM Brink wave functions for the $3^-$ and $4^-$ states. The values of the squared overlap between the single optimum THSR/Brink wave functions and the GCM Brink wave functions are also shown [111].

| $J^\pi$ | $E_{\min}^{\text{Brink}}(\boldsymbol{R}_1, \boldsymbol{R}_2)$ | $E_{\min}^{\text{THSR}}(\boldsymbol{\beta})$ | $E_{\text{GCM}}^{\text{Brink}}$ | $|\langle \Phi_{\text{GCM}}^{\text{Brink}} | \Phi_{\min}^{\text{Brink}}(\boldsymbol{R}_1, \boldsymbol{R}_2) \rangle|^2$ | $|\langle \Phi_{\text{GCM}}^{\text{Brink}} | \Phi_{\min}^{\text{THSR}}(\boldsymbol{\beta}) \rangle|^2$ |
|---|---|---|---|---|---|
| $3^-$ | $-78.4$ | $-80.9$ | $-81.6$ | 0.78 | 0.96 |
| $4^-$ | $-74.4$ | $-76.9$ | $-77.8$ | 0.72 | 0.92 |

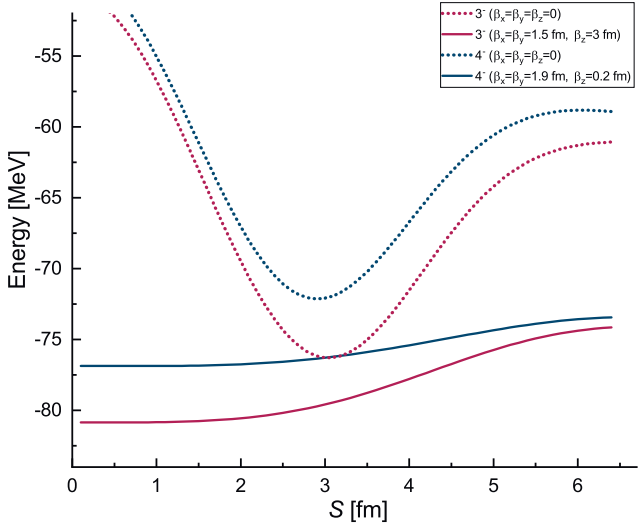

**Fig. 36** Energy curves of the $3^-$ and $4^-$ states in $^{12}$C within the THSR wave function compared with the Brink wave functions. Figure was reproduced from Ref. [111].

Following the study of $^{20}$Ne system in Sec. 4, we further compare the results obtained by the THSR wave functions with those from the Brink cluster model shown in Fig. 36. Firstly, we assume the equilateral triangle shape with the length of the side is $S$ for $3\alpha$ clusters in the Brink model. At the same time, $S$ is also the shift distance on $xy$ plane in our extended THSR wave function. It can be seen that, in the Brink cluster mode, there are two distinct pockets around 3 fm for the $3^-$ and $4^-$ states. It seems that this is a support for the rigid geometrical structure from the inter-cluster distance parameter $S$. However, if we consider the width variables of the relative wave function, that is the spirit of the container picture, the inter-cluster distance parameter is becoming a shift parameter instead of the dynamic variable. It can be seen that the minimum energy points appeared around $S \approx 0$. Furthermore, the obtained energies in the container picture for $3^-$ and $4^-$ states are much more deeper than that obtained from the Brink model. And actually these energies are very closed to the GCM solutions. Moreover, both energy curves are very flat especially in the range $S < 2.$ fm. The obtained energies are almost degenerate without depending the inter-cluster distance parameter $S$.

Fig. 34 and Fig. 35 show that the $3^-$ and $4^-$ states can be considered to have quite a similar intrinsic oblate shape, i.e., $\beta_x = \beta_y = 2.0$ fm and $\beta_z = 0.5$ fm. By fixing the size parameter in the intrinsic wave function, various triangular shapes can be constructed taking the shift parameters $\boldsymbol{S}_1$ and $\boldsymbol{S}_2$ on $xy$ plane. It should be noted that this oblate wave function ($\beta_z \to 0$) actually can be treated as one two-dimensional container. As we have discussed the effective localized clustering in $^{20}$Ne in Sec. 4 and one-dimensional linear chain localized clustering in Sec. 5.2, this kind of localized clustering mainly originates from the Pauli principle. How about the three clusters released from one-dimensional space in the container picture? Figure 37 shows the densities of various intrinsic wave functions with different values of shift parameters. It can be seen that in Fig. 37(a) the length of the shift triangle is too small to violate the positive-parity character of the THSR wave function. To break the symmetry, we take different lengths and orientations of the shift parameters as shown in Fig. 37 (b,c,d). It can be found that, in spite of various triangle geometrical shapes, their corresponding single projected wave functions are quite similar and they are more than 92% equivalent to the corresponding $3^-$ and $4^-$ GCM wave functions. It seems that the $\alpha$ clusters can almost move freely without much influence from Pauli principle in our two-dimensional container and they show the real nonlocalized clustering feature.

### 5.4 Variety of cluster states in $^{16}$O

In this subsection, we discuss various cluster states in $^{16}$O, such as $\alpha+^{12}$C and $4\alpha$ gas-like cluster states, which are obtained via the extended THSR ansatz. The special interest is in how those various cluster states are formed with the increase of excitation energy. As we will discuss in detail in Sec. 6, the $\alpha+^{12}$C cluster structure in $^{16}$O is formed by the activation of cluster degree of freedom in the ground state having a dual property [72, 80], i.e. by the excitation of relative motion between the $\alpha$ and $^{12}$C clusters. The gas-like $4\alpha$ cluster state is then produced as a result of further excitation of the $^{12}$C core, to the $3\alpha$ cluster state, i.e. to the Hoyle state. The path of this cluster evolution is shown in the Ikeda diagram, together with many other paths in many other nuclei (see Sec. 6). We show in Fig. 38 a schematic picture for the path of cluster evolution along the excitation energy in $^{16}$O, which can be regarded as the path of a size and shape evolution of a container.

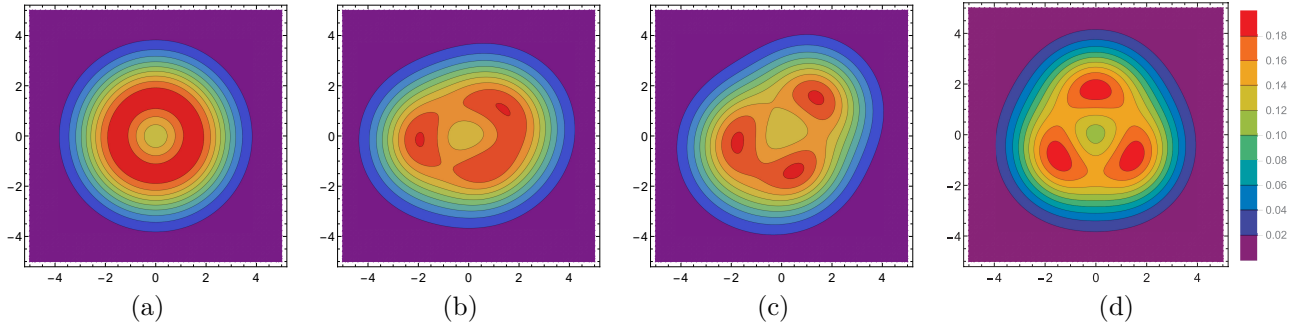

**Fig. 37** Various density profiles from the intrinsic THSR wave functions with different shift parameters. The size parameter is taken as $\beta_x = \beta_y = 2.0, \beta_z = 0.5$. Squared overlaps between the corresponding single projected THSR wave functions $\Phi^{J^\pi}(S_{1x}, S_{2x}, S_{2y})$ and the GCM Brink wave function are given. Figures are taken from Ref. [111]

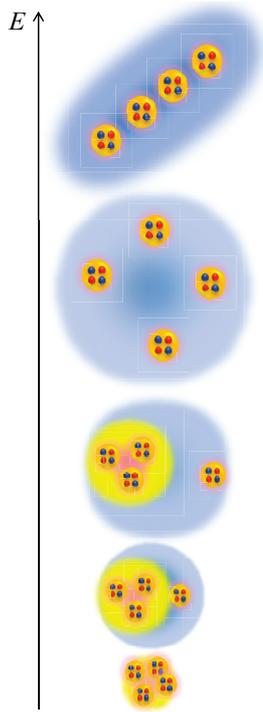

**Fig. 38** (Color online) "Container" evolution picture for cluster structures.

In order to describe the above mentioned cluster evolution in $^{16}$O, the following extended THSR wave function with a double-container (eTHSR) is employed. This is an extension of Eq. (28), which is adopted for the $3\alpha$ system in the previous subsection, to the $4\alpha$ system, and is expressed in the following form:

$$\Phi^{J=0}_{4\alpha}(\boldsymbol{\beta}_1, \boldsymbol{\beta}_2) = \widehat{P}^{J=0} \times$$
$$\mathcal{A}\bigg\{ \exp\bigg[ -2\sum_k^{x,y,z} \bigg( \frac{\mu_1\xi_{1k}^2 + \mu_2\xi_{2k}^2}{b^2 + 2\beta_{1k}^2} - \frac{\mu_3\xi_{3k}^2}{b^2 + 2\beta_{2k}^2} \bigg) \bigg] \prod_{i=1}^4 \phi(\alpha_i) \bigg\},$$
(79)

where $\boldsymbol{\xi}_i$ is again the Jacobi coordinates between the $\alpha$ particles, and $\mu_i = i/(i+1)$, for $i = 1, 2, 3$. While the parameter $b$ characterizes the size of the constituent $\alpha$ particle, the parameters $\boldsymbol{\beta}_1$ and $\boldsymbol{\beta}_2$ characterize the size and shape of a container, in which the $\alpha$ clusters are confined. We also assume the axial symmetry $\beta_{i\perp} \equiv \beta_{ix} = \beta_{iy}$, so as to deal with the four parameters, $\beta_{1\perp}, \beta_{1z}, \beta_{2\perp}, \beta_{2z}$, in the practical calculations.

A schematic picture representing the eTHSR wave function in the present $^{16}$O system is shown in Fig. 39, in which the $3\alpha$ clusters and another $\alpha$ cluster are confined in different containers characterized by the parameters $\boldsymbol{\beta}_1$ and $\boldsymbol{\beta}_2$, respectively. This is contrasted with the Brink wave function, in which the cluster configurations are described by their relative distance parameters.

In Fig. 40, the calculated energy spectrum for $J^\pi = 0^+$ states is shown. The corresponding experimental data and results by the $4\alpha$ OCM calculation [19] discussed in

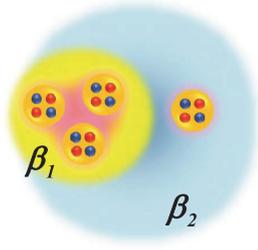

**Fig. 39** (Color online) Schematic representation of the eTHSR wave function, in which the two containers of the $3\alpha$ and $\alpha$ clusters are characterized by the parameters $\beta_1$ and $\beta_2$, respectively.

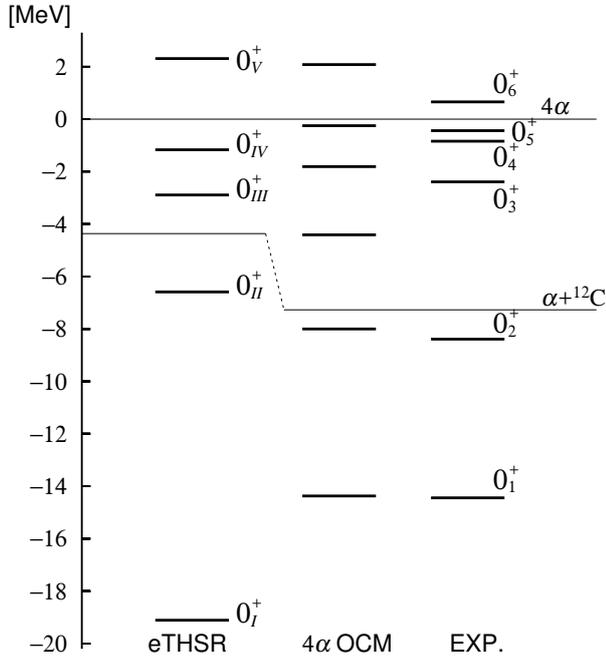

**Fig. 40** Energy spectra of the low-lying $J^\pi = 0^+$ states calculated with the extended THSR ansatzs. The corresponding observed spectrum (Exp.) [76, 77] and result by the $4\alpha$ OCM [19] are also shown. The numbers are r.m.s radii in a unit of fm.

Sec. 2.2.1 are also shown. The solution of Hill-Wheeler equation with the $r^2$ constraint method is shown. The $0_V^+$ state is actually the seventh $0^+$ state obtained by solving the Hill-Wheeler equation, i.e. the fifth and sixth eigenstates are kicked out from the present consideration, since they have larger r.m.s. radii and are regarded as spurious continuum states accidentally mixed with the physical states.

As we discussed in Sec. 2.2.1, the $4\alpha$ OCM calculation gives six $0^+$ states. The $0_6^+$ state has the $4\alpha$ condensate character and the $0_2^+ - 0_5^+$ states all have $\alpha + {}^{12}$C cluster structures. i.e. $\alpha(S) + {}^{12}\text{C}(0_1^+)$, $\alpha(D) + {}^{12}\text{C}(2_1^+)$, $\alpha(S) + {}^{12}\text{C}(0_1^+)$, and $\alpha(P) + {}^{12}\text{C}(1^-)$ cluster structures, respectively. The difference between the $0_2^+$ and $0_4^+$ states are that in the latter the $\alpha$ and $^{12}$C relative motion is further excited and has a higher nodal $S$-wave, to have a larger r.m.s. radius than the former.

Since in the present eTHSR wave function of Eq. (79) the $\alpha$ clusters occupy positive parity orbits, such a state as having the $\alpha(P) + {}^{12}\text{C}(1^-)$ cluster structure, like the $0_5^+$ state in the OCM calculation, is missing. The treatment of negative parity orbit in the THSR ansatz has been discussed in the subsequent Sec 3.3. However, for the other states, a one-to-one correspondence to the experimental data as well as to the $4\alpha$ OCM calculation is consistently obtained. It should be noted that in the OCM calculations [19, 79], the binding energies of the ground states of $^{16}$O and the $^{12}$C are phenomenologically fitted to the corresponding experimental values. In the present calculation, however, there is no adjustable parameter in the microscopic Hamiltonian.

**Table 8** R.m.s. charge radii and monopole matrix elements of the $0_I^+$ - $0_V^+$ states calculated with the eTHSR ansatz, in comparison with the corresponding experimental data. Their units are fm and ($e$fm$^2$), respectively.

|  | eTHSR | | Exp. | |
| --- | --- | --- | --- | --- |
|  | $R_\text{rms}$ | $M(E0)$ | $R_\text{rms}$ | $M(E0)$ |
| $0_I^+$ | 2.7 |  | 2.71(0.02) |  |
| $0_{II}^+$ | 3.2 | 5.9 | 3.55(0.21) |  |
| $0_{III}^+$ | 3.3 | 5.7 | 4.03(0.09) |  |
| $0_{IV}^+$ | 4.9 | 0.8 |  |  |
| $0_V^+$ | 4.9 | 0.7 |  |  |

In Table 8, r.m.s. radii and monopole matrix elements with the ground state are shown. The experimental data available are reasonably reproduced. We can also see that from the $0_I^+$ to the $0_V^+$ states, i.e. as the states are excited, the r.m.s. radius becomes larger and the monopole matrix element becomes smaller. This indicates that the higher the excitation energy is, the more evolved the clustering is. The evolution of the clustering can be described by solving the Hill-Wheeler equation

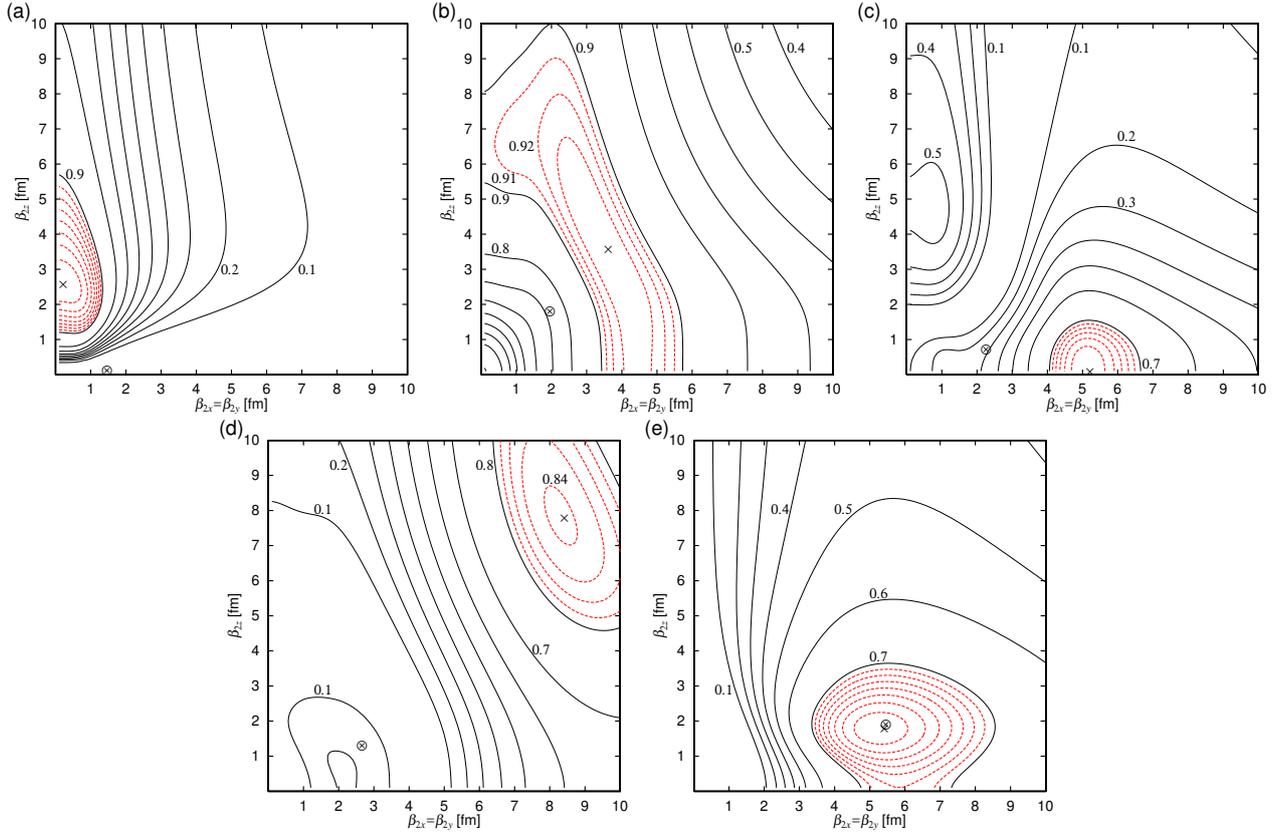

**Fig. 41** (Color online) Contour maps of the squared overlaps between the $0^+_I$ (a), $0^+_{II}$ (b), $0^+_{III}$ (c), $0^+_{IV}$ (d), and $0^+_V$ (e) states, and the single extended deformed THSR wave functions, in two-parameter space $\beta_{2x} = \beta_{2y}$ and $\beta_{2z}$, in which $\boldsymbol{\beta}_1$ parameter values are fixed at optimal ones, denoted by $\otimes$, so that the maxima in four-parameter space $\beta_{1x} = \beta_{1y}$, $\beta_{1z}$, $\beta_{2x} = \beta_{2y}$, $\beta_{2z}$ appear in these figures. The maximum positions are denoted by $\times$. Red dotted contour lines are in a step of 0.01 and black solid ones are in a step of 0.1.

concerning the model parameters $\boldsymbol{\beta}_1$ and $\boldsymbol{\beta}_2$.

This respect is made much clearer by calculating the following squared overlap:

$$\mathcal{O}_\lambda(\boldsymbol{\beta}_1, \boldsymbol{\beta}_2) = |\langle \widetilde{\Phi}^{J=0}_\lambda(\boldsymbol{\beta}_1, \boldsymbol{\beta}_2) | \Psi_\lambda \rangle|^2, \qquad (80)$$

with $\Psi_\lambda$ the $\lambda$th eigenfunction and $\widetilde{\Phi}^{J=0}_\lambda(\boldsymbol{\beta}_1, \boldsymbol{\beta}_2)$ normalized single eTHSR wave function in a space orthogonal to the lower eigenstates, i.e.

$$\widetilde{\Phi}^{J=0}_\lambda(\boldsymbol{\beta}_1, \boldsymbol{\beta}_2) = \mathcal{N}_\lambda \widehat{P}_\lambda \Phi^{J=0}_{4\alpha}(\boldsymbol{\beta}_1, \boldsymbol{\beta}_2), \qquad (81)$$

where $\widehat{P}_\lambda = 1 - \sum_{i=1}^{\lambda-1} |\Psi_i\rangle\langle\Psi_i|$ with $\lambda = I, \cdots, IV$, and $\widehat{P}_V = 1 - \sum_{i=1}^{6} |\Psi_i\rangle\langle\Psi_i|$, and $\mathcal{N}_\lambda$ are the corresponding normalization constants.

This quantity indicates how these five states $\Psi_\lambda$ ($\lambda = I, \cdots, V$) are expressed by single configurations of the eTHSR wave functions, and therefore, gives direct information of whether the container structure is realized or not in these states, and if so, what kind of containers represent the states. In Fig. 41, the contour maps of squared overlap of the states $\Psi_I$-$\Psi_V$ with single configurations in the $\beta_{2\perp}$ and $\beta_{2z}$ parameter space in Eq. (80) are shown. Here the $\boldsymbol{\beta}_1$ parameter, i.e. $(\beta_{1\perp}, \beta_{1z})$ is fixed at the position denoted by $\otimes$ in these figures, so that the maximum value of the squared overlap in the four parameter space $(\beta_{1\perp}, \beta_{1z}, \beta_{2\perp}, \beta_{2z})$ appears at the position denoted by $\times$.

As we have discussed in Sec. 2.1, the single $3\alpha$ THSR wave function can very precisely describe the ground state and excited states of $^{12}$C. It is here useful to show in Table 9 the maximum values of the squared overlap of the $0^+_1$, $2^+_1$ and $0^+_2$ states in $^{12}$C with the single $3\alpha$ THSR configuration.

In Fig. 41, the squared overlaps of the five eigenstates with the single eTHSR configuration defined in Eq. (80) are shown. In the ground state, shown in Fig. 41(a), $3\alpha$ clusters are put into an oblate-deformed and very compact container with $\beta_{1\perp} \gg \beta_{1z}$, while the remaining $\alpha$ cluster is put into a prolate-deformed and very compact container with $\beta_{2\perp} \ll \beta_{2z}$. This means that the first $3\alpha$ clusters move in a $xy$-plane and the last $\alpha$ cluster moves in $z$-direction. This supports the idea that the ground state has a tetrahedral shape of the $4\alpha$ clusters proposed by several authors [112, 113]. Our calculation indicates that this configuration is contained in the $0_I^+$ state by 98%.

**Table 9** Maxima of the squared overlaps for the $0_1^+$, $2_1^+$ and $0_2^+$ states in $^{12}$C in two-parameter space $\beta_\perp$ and $\beta_z$. The corresponding $B_\perp$ and $B_z$ values are also shown.

|  | $\mathcal{O}_{max}$ | $(\beta_\perp, \beta_z)$ | $(B_\perp, B_z)$ |
| --- | --- | --- | --- |
| $^{12}$C($0_1^+$) | 0.93 | (1.9, 1.8 fm) | (3.0, 2.9 fm) |
| $^{12}$C($2_1^+$) | 0.90 | (1.9, 0.5 fm) | (3.0, 1.6 fm) |
| $^{12}$C($0_2^+$) | 0.99 | (5.6, 1.4 fm) | (8.0, 2.4 fm) |

In the $0_{II}^+$ state, shown in Fig. 41(b), the $3\alpha$ clusters are in a spherical container with $\beta_{1\perp} \sim \beta_{1z}$. The fourth $\alpha$ cluster is put into a larger size container with spherical shape, i.e. $\beta_{2\perp} \sim \beta_{2z} > \beta_{1\perp} \sim \beta_{1z}$. In particular, the parameter set $(\beta_{1\perp}, \beta_{1z}) = (1.8, 1.8$ fm$)$ is almost the same as that for $^{12}$C in Table 9, i.e. $(\beta_\perp, \beta_z) = (1, 9, 1.8$ fm$)$. This means that the first $3\alpha$ clusters are confined in a compact container to form the ground state of $^{12}$C, since the $^{12}$C($0_1^+$) state can be very precisely described by the single configuration with these parameter values. The fourth $\alpha$ cluster moves in a larger spherical container, because of $(\beta_{2\perp}, \beta_{2z}) = (3.5, 3.6$ fm$)$, which gives the largest squared overlap 94%. This is the new interpretation of the $\alpha + {}^{12}$C cluster structure, whose traditional understanding is that the $\alpha$ cluster orbits in an $S$-wave around the $^{12}$C($0_1^+$) state.

The $0_{III}^+$ state, shown in Fig. 41(c), is similar to the $0_{II}^+$ state but both containers are not spherical but deformed. The $\boldsymbol{\beta}_1$ parameter takes almost the same value as that of the isolated $^{12}$C($2^+$) state, as shown in Table 9, which means that the first $3\alpha$ clusters form the $^{12}$C($2^+$) state, since the state is described by the single parameter value of $\boldsymbol{\beta}$. The configuration of the remaining $\alpha$ cluster $(\beta_{2\perp}, \beta_{2z}) = (5.1, 0.1$ fm$)$, giving the largest value 76%, means that the $\alpha$ cluster moves in a deformed and larger container. This is present understanding of the $0_3^+$ state, which is conventionally considered to have the $\alpha(D) + {}^{12}$C($2^+$) structure.

In the $0_{IV}^+$ state, shown in Fig. 41(d), one can see that the $3\alpha$ clusters are put in slightly larger container than that for the $^{12}$C($0_1^+$) state, which is slightly deformed in a oblate shape. The fourth $\alpha$ cluster, however, moves in a much larger and almost spherical container, like a satellite. This configuration expresses the $0_{IV}^+$ state dominantly by 84%. This means that the second container characterized by $\boldsymbol{\beta}_2$ is further evolved from that in the $0_{II}^+$ state. We can say that this state corresponds to the $0_4^+$ state in the former $4\alpha$ OCM calculation, which predicts the $\alpha + {}^{12}$C($0_1^+$) higher nodal structure for the state.

The $0_V^+$ state, shown in Fig. 41(e), is the most striking. All the $\alpha$ clusters occupy an identical orbit, with $(\beta_{1\perp}, \beta_{1z}, \beta_{2\perp}, \beta_{2z}) = (5.3, 1.9, 5.3, 1.8$ fm$)$. This is qualified to call the $\alpha$ condensation. This configuration is contained in this state by 78%, which is still very large. Furthermore, this container is very close to the one of the Hoyle state, with $(\beta_{1\perp}, \beta_{1z}) = (5.6, 1.4$ fm$)$ in Table 9. This means that the $0_V^+$ state is regarded as the Hoyle analog state, in which the fourth $\alpha$ cluster is also put into the container occupied with the $3\alpha$ clusters in the Hoyle state. The large size of this container indicates that the $4\alpha$ clusters are loosely coupled with each other and configured like a gas. Note that the $4\alpha$ condensate state is also predicted by the $4\alpha$ OCM calculation slightly above the $4\alpha$ threshold, as the $0_6^+$ state.

These results tell us that the evolution of cluster structures is described by the container evolution with respect to its size and shape. The reason why the container evolution arises is the orthogonality to the lower states, which is explicitly taken into account in the definition of the single configuration $\widetilde{\Phi}_k^{J=0}$ in Eq. (80). The orthogonality condition prevents a higher state configuration from overlapping with the lower-states more compact configurations. It thus plays a role as a repulsive core and is considered to give the container evolution.

### 5.5 Application to other nuclear systems

#### 5.5.1 Neutron-rich isotopes of beryllium

Recent years, the study for the cluster structures in neutron-rich nuclei is becoming a very interesting and important subject [114–116]. Compared with the typical cluster structures in $n\alpha$ nuclei, neutron-rich nuclei could have some novel kinds of cluster structures due to the dynamics of the excess neutrons. This kind of clustering, which is sometimes enhanced by valence neutrons, can have some molecular cluster structures with some kind of valence neutron orbit and this never appears in stable nuclei. For example, $^{14}$C is reported to have a $3\alpha$ equilateral-triangular shape [56] and linear-chain structure [117] with attractive interaction between the excess neutron and $\alpha$ particles.

As typical examples, the Be isotopes have been studied intensively with cluster models, molecular-orbit models, and AMD [118] these years. In AMD calculations, the existence of clusters are not a priori assumption like the traditional cluster models. However, AMD calculations show the appearance of $2\alpha$ cluster structures surrounded by valence neutrons with some kind of orbit in Be isotopes, which clearly supports the existence of clustering structures in neutron-rich nuclei. Figure 42 shows the density distributions of the intrinsic states of the band heads of $^{10}$Be and $^{12}$Be obtained by AMD. It can be seen that various cluster structures appear in different states in the Be isotopes. For example, in $^{10}$Be, there is a larger clustering deformation for the $0_2^+$ state than the ground state. From AMD calculations, we know that in the ground state of $^{10}$Be two valence neutrons occupy $\pi$ orbits while those of the $0_2^+$ state occupy $\sigma$ orbits.

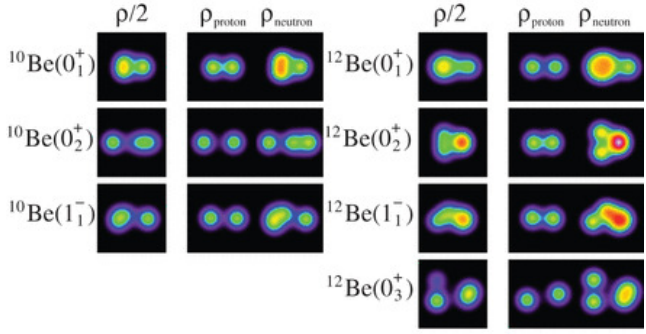

**Fig. 42** Density distributions of the intrinsic states of the band heads of $^{10}$Be and $^{12}$Be obtained by AMD [118].

Now, the THSR wave function has been extended to describe the general cluster states from the $\alpha$ condensation states or gas-like states to the compact shell-model-like states. The concept of nonlocalized clustering has been proved to be very successful in the $n\alpha$ nuclei. Considering the different features of the clustering in stable nuclei and neutron-rich nuclei, a natural question is, is it possible to describe the molecular cluster structures of neutron-rich nuclei using the THSR wave function?

Recently, the Be isotopes have been investigated using the THSR wave functions [87, 119]. The constructed THSR wave function for the $^9$Be can be written [87],

$$\left|\Phi^{^9\text{Be}}\right\rangle = (C_\alpha^\dagger)^2 c_{n,\sigma}^\dagger \left|\text{vac}\right\rangle, \quad (82)$$

$$c_{n,\sigma}^\dagger = \int d\boldsymbol{R}_n \ \exp[im\phi_{\boldsymbol{R}_n}] \exp[-\frac{R_{n,x}^2 + R_{n,y}^2}{\beta_{n,xy}^2} - \frac{R_{n,z}^2}{\beta_{n,z}^2}]$$

$$\times \int d\mathbf{r}_n (\pi b^2)^{-3/4} e^{-\frac{(\mathbf{r}_n - \mathbf{R}_n)^2}{2b^2}} a_{n,\sigma,}^\dagger(\mathbf{r}_n), \quad (83)$$

where $C_\alpha^\dagger$ and $c_{n,\sigma}$ are creation operators of $\alpha$-particle and neutron, respectively. $\mathbf{r}_n$ represents the coordinate of the extra neutron and its corresponding generate coordinate is $\mathbf{R}_n$. $a_{\sigma,\tau}^\dagger(\mathbf{r}_n)$ is the creation operator of the extra neutron with spin $\sigma$. Here, spin is assumed to be up. The $\phi_{\mathbf{R}_n}$ is the azimuthal angle of $\mathbf{R}_n$, which is introduced here for obtaining the negative-parity states of $^9$Be as discussed in Sec. 3.3.

As for the Hamiltonian, the effective nucleon-nucleon Volkov No.2 force [42] is employed and G3RS [120, 121] is used as the spin-orbit interaction. In practical calculations, the Monte Carlo method is used for the multi-integration calculations from the Gaussian integration and angular-momentum projection.

In Table 10, it is shown the calculated results of ground-state rotational band by the single THSR wave functions and GCM Brink wave functions. By superposing different configurations for the $2\alpha+n$ cluster structures, the GCM Brink wave function usually can provide us with a very exact solution for this kind of cluster system. Now, from Table 1, it can be seen that the single THSR wave functions give a very similar energies compared with the results from the GCM Brink wave functions. Especially, there are very high-percentage squared overlap between the single THSR wave functions and GCM Brink wave functions, e.g., as for the ground state $^9$Be, their squared overlap is as high as 0.96. That means the constructed THSR wave functions in Eq. (82) give a very good description of the ground state, $5/2^-$, and $7/2^-$ states and the correlations between $\alpha$ clusters and the extra nucleon are well grasped in the constructed THSR wave function.

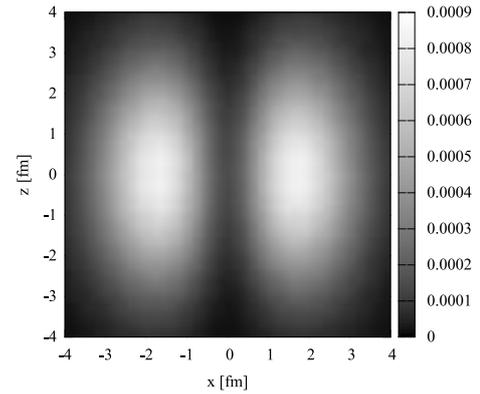

**Fig. 43** Density distribution of the extra neutron of the intrinsic ground state of $^9$Be [87].

Figure 43 shows the density distribution of the extra neutron. It can be seen that the extra neutron in $^9$Be can spread more than 6 fm along the $z$ direction and it has the overlap with the distribution of two $\alpha$ clusters. This is just the well-known $\pi$ orbit, which has also been obtained by other cluster models. Different with the traditional molecular orbit models, in the THSR wave function, the extra nucleon is assumed to make a nonlocalized motion inside the nucleus rather than occupy some kind of molecular orbit. Finally, the $\pi$-orbit emerges from the antisymmetrization which cancels out nonphysical distributions in a very natural way. The reproduction of nuclear molecular orbit structure provides another support for the extension of the nonlocalized clustering concept to $^9$Be.

Next, we show some calculated results about the $^{10}$Be using the THSR wave function [119]. One constructed

**Table 10** The calculated energies for the first $3/2^-$, $5/2^-$, and $7/2^-$ states using the single THSR wave functions and the GCM Brink wave functions. The values in parentheses are the corresponding excitation energies. $|\langle\Psi^{\text{THSR}}|\Psi^{\text{GCM}}\rangle|^2$ represents the squared overlap between the single THSR wave function and the GCM Brink wave function [87]. Units of energies are MeV.

| State | $E^{\text{THSR}}$ | $E^{\text{GCM}}$ | $|\langle\Psi^{\text{THSR}}|\Psi^{\text{GCM}}\rangle|^2$ |
| --- | --- | --- | --- |
| $3/2^-$ | -55.4 (0.0) | -56.4 (0.0) | 0.96 |
| $5/2^-$ | -53.0 (2.4) | -53.8 (2.6) | 0.95 |
| $7/2^-$ | -48.6 (6.8) | -49.4 (7.0) | 0.93 |

THSR wave function for the $^{10}$Be can be written,

$$\left|\Phi^{^{10}\text{Be}}\right\rangle = (C_\alpha^\dagger)^2 c_{n,\uparrow}^\dagger c_{n,\downarrow}^\dagger |\text{vac}\rangle. \tag{84}$$

Here, $C_\alpha^\dagger$ is the creation operator of the $\alpha$ cluster. $c_{n,\uparrow}^\dagger$ and $c_{n,\downarrow}^\dagger$ are the creation operators of the two valence neutrons with spin up and down, respectively. See details in Ref. [119].

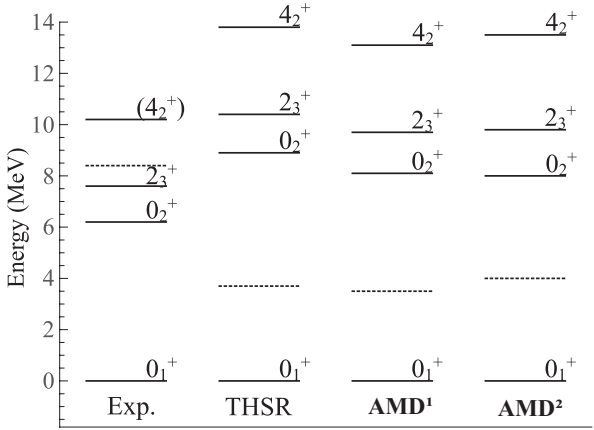

**Fig. 44** Calculated energies of the $0_2^+$, $2_3^+$, and $4_2^+$ states [119] comparison with the AMD and experimental values. AMD[1] represents the AMD+DC method [122]. AMD[2] represents the $\beta-\gamma$ constrained AMD method [123]. The dashed lines are the corresponding $\alpha+\alpha+n+n$ threshold energies.

As we know, $0_2^+$ state in $^{10}$Be was reported to be a developed molecular cluster state from AMD calculations [118]. The calculated energy spectrum of the $0_2^+$ rotational band with the THSR wave function are shown in Fig. 44. It can be seen that the obtained results agree well with those of other theoretical models [122, 123]. It should be noted that again the only one single THSR wave function can describe well the $0_2^+$ state of $^{10}$Be while they need a large number of basis for other cluster models. This shows that the constructed THSR wave function is also very suitable for the description of the molecular cluster structures in the neutron-rich nuclei.

By using the single and also simple THSR wave functions, some low-lying cluster states of the $^9$Be and $^{10}$Be are described very well. This means the THSR wave function can also be extended to the non-$n\alpha$ nuclei and the concept of nonlocalized clustering is also suitable for the neutron-rich nuclei. In the future, to give a unified description of the various molecular cluster structures in neutron-rich nuclei, how to construct some kind of more general microscopic cluster wave function according to the THSR spirit is still a challenging task.

### 5.5.2 $\Lambda$-hypernuclei

Light $\Lambda$ hypernuclei have been studied by using the microscopic and semi-microscopic cluster models for a long time [124]. Since a $\Lambda$ particle is not affected by antisymmetrization on nucleons, it plays a role as an impurity when it is injected into a core nucleus. Many studies in the past made an emphasis on structural change of the core nucleus with the impurity, while the pursuit of knowledge of $\Lambda$-nucleon interaction through the structure study of $\Lambda$ hypernuclei has also been one of the most important issues. A well-known feature in $\Lambda$ hypernuclei is that the additional $\Lambda$ particle contracts the nucleon spatial distribution. In the following subsections, the THSR ansatz is shown to be easily applicable to $\Lambda$ hypernuclei and to better demonstrate the so-called shrinkage effect as contraction of a container of the core.

**(1) $2\alpha+\Lambda$ cluster structure in $^9_\Lambda$Be** We propose here a new-type microscopic cluster model wave function, which is applicable to hypernuclei and we hereafter refer to as the Hyper-THSR (H-THSR) wave function [100, 125]. This is based on the above mentioned deformed $n\alpha$ THSR wave function.

The H-THSR wave function describing the $2\alpha+\Lambda$ hypernucleus with good angular momentum is then introduced as follows:

$$\Phi_J^{\text{H-THSR}}(\boldsymbol{\beta}) = \hat{P}_{MK}^J \Phi_{2\alpha}^{\text{THSR}}(\boldsymbol{\beta}) \sum_\kappa f_\Lambda(\boldsymbol{\beta},\kappa)\varphi_\Lambda(\kappa), \tag{85}$$

where $\hat{P}_{MK}^J$ is the angular-momentum projection operator and the $\Lambda$ particle simply couples to the $2\alpha$ core nucleus in an $S$ wave. Its radial part is expanded in terms of Gaussian basis functions, $\varphi_\Lambda(\kappa) =$ $(\pi/2\kappa)^{-3/4}\exp(-\kappa r_{2\alpha-\Lambda}^2)$ with respect to the width parameter $\kappa$, where $\boldsymbol{r}_{2\alpha-\Lambda} = \boldsymbol{X}_\Lambda - (\boldsymbol{X}_1+\boldsymbol{X}_2)/2$ with $\boldsymbol{X}_\Lambda$ a position vector of the $\Lambda$ particle. In the practical calculations, we assume an axially-symmetric deformation for $\Phi^{\mathrm{THSR}}_{2\alpha}(\boldsymbol{\beta})$, $\beta_x = \beta_y \equiv \beta_\perp \neq \beta_z$.

The coefficients of the expansion $f_\Lambda(\beta_\perp,\beta_z,\kappa)$ are then determined by solving the following Hill-Wheeler-type equation of motion,

$$\sum_{\kappa'} \langle \hat{P}^J_{M0}\Phi^{\mathrm{THSR}}_{2\alpha}(\beta_\perp,\beta_z)\varphi_\Lambda(\kappa)|\hat{H} - E(\beta_\perp,\beta_z)|$$
$$\hat{P}^J_{M0}\Phi^{\mathrm{THSR}}_{2\alpha}(\beta_\perp,\beta_z)\varphi_\Lambda(\kappa')\rangle f_\Lambda(\beta_\perp,\beta_z,\kappa') = 0. \quad (86)$$

This H-THSR wave function is characterized only by the parameters $\beta_\perp$ and $\beta_z$, which correspond to a spatial extension of the whole nucleus. It is well known that the $\Lambda$ particle invokes spatial core shrinkage in many hypernuclei. It should be mentioned that such a core shrinkage effect is expected to be taken into account very naturally by this parameterization of $\boldsymbol{\beta}$, in general hypernuclei, since it specifies the dilatation of the whole nucleus. In this section, only the $S$-wave component of the $\Lambda$ particle is considered, for simplicity. Thus the monopole-like shrinkage is expected to be described nicely by this wave function. The extension to inclusion of the other angular-momentum channels is discussed in the next section for the study of $^{13}_\Lambda$C.

In order to compare the single H-THSR wave function with the Brink-GCM wave function, we calculate the following squared overlap:

$$\mathcal{O}_J(\beta_\perp,\beta_z) =$$
$$\frac{|\langle \Phi^{\mathrm{H-THSR}}_J(\beta_\perp,\beta_z)|\Phi^{\mathrm{B-GCM}}_J\rangle|^2}{\langle \Phi^{\mathrm{H-THSR}}_J(\beta_\perp,\beta_z)|\Phi^{\mathrm{H-THSR}}_J(\beta_\perp,\beta_z)\rangle\langle \Phi^{\mathrm{B-GCM}}_J|\Phi^{\mathrm{B-GCM}}_J\rangle}. \quad (87)$$

The above wave function $\Phi^{\mathrm{B-GCM}}_J$ is the Brink-GCM wave function projected onto the model space with the angular-momentum channel $(L,\lambda) = (J,0)$, in which $L$, $\lambda$ are the angular momentum between the $2\alpha$ clusters and the one between the $\Lambda$ particle and $^8$Be, respectively.

In Fig. 45, we show the contour maps of this quantity in the two-parameter space $\beta_\perp$ and $\beta_z$, calculated for the $J^\pi = 0^+$ (top), $2^+$ (middle), and $4^+$ (bottom) states. For these states, we can see two maxima in prolate and oblate regions (denoted as $\times$ and $+$), as in the case of the energy surfaces. The values of the maxima are extremely large and close to unity. The values for the $J^\pi = 0^+, 2^+$ states are 99.5% and 99.4%, respectively. For the $J^\pi = 4^+$ state, the maximum value is slightly down to 97.7%. This reduction may originate from a mixture of spurious scattering-state components. These practically 100% squared overlap values of course mean that, at least on the subspace with $L = J$ and $\lambda = 0$, the Brink-GCM wave function obtained by solving the $2\alpha + \Lambda$ Hill-Wheeler equation is equivalent to the single configuration of the H-THSR wave function. We should note that the Brink-GCM wave function in this subspace $(J, \lambda) = (L, 0)$ is shown to be almost the same as the one in the full angular-momentum-channel space. This result also means that the size parameter $\boldsymbol{\beta}$, which specifies the monopole-like dilatation of the whole nucleus, quite well takes into account the effect of the spatial core shrinkage by the additional $\Lambda$ particle. We can thus conclude that in $^9_\Lambda$Be the $2\alpha$ clusters are trapped into a container, which is specified by the optimal value of the size parameter $\boldsymbol{\beta}$, under the influence of the antisymmetrizer $\mathcal{A}$ acting on the nucleons, such as realized in the form of the wave function Eq. (85).

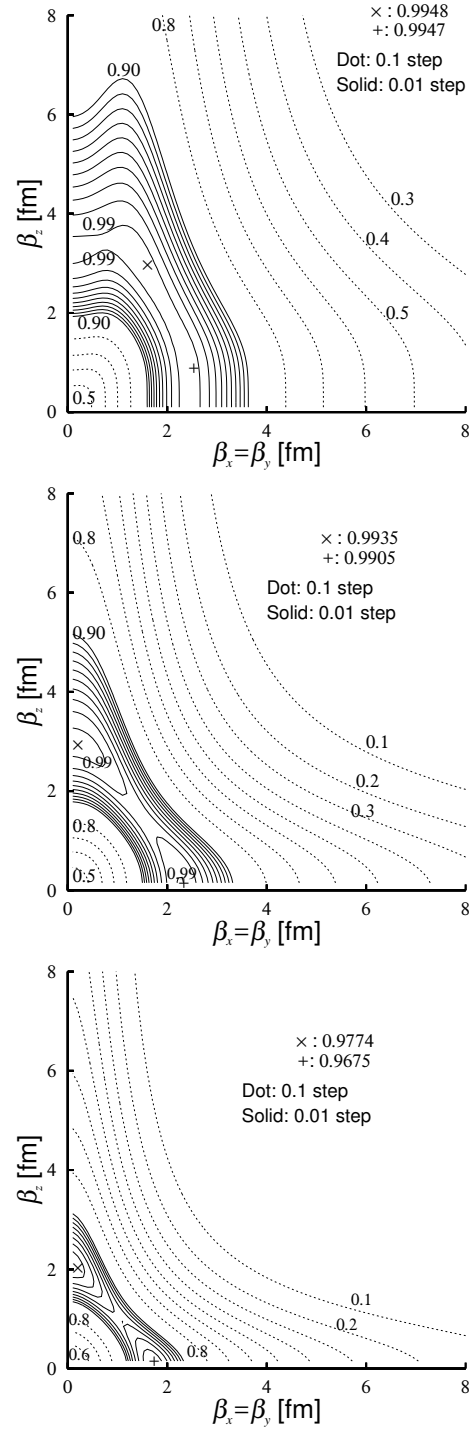

**Fig. 45** Contour maps of the squared overlap surfaces for $J^\pi = 0^+$(top), $2^+$(middle), and $4^+$(bottom) states in two-parameter space $\beta_x = \beta_y (\equiv \beta_\perp), \beta_z$, defined by $\mathcal{O}_J(\beta_\perp,\beta_z)$ in Eq. (87). Two maxima are denoted by $\times$ and $+$. YNG-ND interaction is adopted for the $\Lambda N$ interaction [126].

We here discuss the intrinsic structure of $^9_\Lambda$Be, together with that of $^8$Be. Using the intrinsic wave function defined as the THSR wave function before the angular-momentum projection in Eq. (85), we can calculate the following intrinsic density of nucleons,

$$\rho_N(\boldsymbol{r}) = \frac{\langle \Phi^{\text{H-THSR}}_{2\alpha-\Lambda}(\beta_\perp, \beta_z)| \sum_{i=1}^{8} \delta(\boldsymbol{r}_i - \boldsymbol{X}_G - \boldsymbol{r})|\Phi^{\text{H-THSR}}_{2\alpha-\Lambda}(\beta_\perp, \beta_z)\rangle}{\langle \Phi^{\text{H-THSR}}_{2\alpha-\Lambda}(\beta_\perp, \beta_z)|\Phi^{\text{H-THSR}}_{2\alpha-\Lambda}(\beta_\perp, \beta_z)\rangle}. \tag{88}$$

This density is normalized as usual to the total number of nucleons, $\int d\boldsymbol{r}\rho_N(\boldsymbol{r}) = 8$.

In Fig. 46, we show the intrinsic density profiles of $^9_\Lambda$Be defined by Eq. (88) and of $^8$Be. The single optimal $\boldsymbol{\beta}$ values giving the minimum energies are adopted, i.e. $(\beta_\perp, \beta_z) = (1.5, 2.8\text{fm})$ for $^9_\Lambda$Be and $(1.8, 7.8\text{fm})$ for $^8$Be. While both ones clearly show the $2\alpha$-cluster structure with the prolate-deformed shape, $^8$Be has a gas-like tail of the $2\alpha$ clusters and $^9_\Lambda$Be not. In $^9_\Lambda$Be the $\Lambda$ particle gives rise to strong shrinkage and the gas-like tail in $^8$Be disappears. The rms radius of the $^8$Be core is accordingly changed from $R_{\text{rms}} = 2.87$ fm for $^8$Be to $R^{(c)}_{\text{rms}} = 2.31$ fm for $^9_\Lambda$Be. Nevertheless the $2\alpha$-cluster structure definitely remains in this very compact object, which is produced by the competition between the quite strong effect of inter-$\alpha$ Pauli repulsion originating from the antisymmetrizer $\mathcal{A}$ and the fairly strong attractive effect among the $2\alpha$ and $\Lambda$ particles, and therefore the $2\alpha$ clusters in this intrinsic state are effectively localized in space. This means that even for the states which are described by nonlocalized-type wave function with container structure, localized nature of clustering can appear in density distribution due to the Pauli principle.

We next simulate the shrinkage effect by varying $\Lambda N$ interaction artificially, with overall factor $\delta$ multiplied with $V^{(\Lambda N)}$. The $\boldsymbol{\beta}$ values to give the energy minima after projection onto $J = 0$ space, which are listed in Table 11, are adopted. As the $\Lambda N$ interaction is strength-ened, from $\delta = 0.4$ to $0.8$, the $\alpha+\alpha$ distance is shortened along $z$-direction. The corresponding $\boldsymbol{\beta}$ values, $\Lambda$ binding energies, the maximum squared overlaps in Eq. (87) with the use of the same artificial $\Lambda N$ potential, rms radii of the $^8$Be core, are also shown in Table 11. $\delta = 0$ and 1 correspond to the cases of $^8$Be and $^9_\Lambda$Be, respectively. Starting from the case of $^8$Be with very large value of $\beta_z$ and small value of $\beta_\perp$, i.e. $\beta_\perp = 1.8$ fm and $\beta_z = 7.8$ fm, only the $\beta_z$ value drastically gets smaller as the increase of $\Lambda N$ interaction, and eventually for $^9_\Lambda$Be, $\beta_z$ becomes much smaller while $\beta_\perp$ is almost unchanged, i.e. $\beta_\perp = 1.5$ fm and $\beta_z = 2.8$ fm.

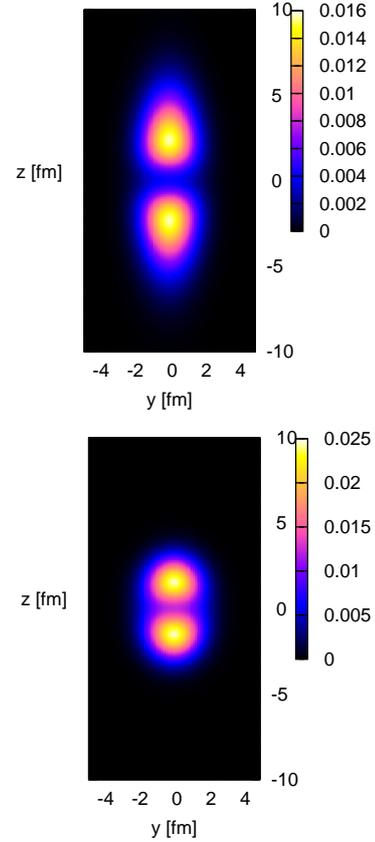

**Fig. 46** Intrinsic density profiles of $^9_\Lambda$Be (upper) defined by Eq. (88), on $yz$ plane with $x = 0$. For comparison, the one of $^8$Be is also shown (lower).

Table 11 The minimum binding energies of $E(\beta_\perp, \beta_z)$, the corresponding $\Lambda$ binding energies $B_\Lambda$, the rms radii of the $^8$Be core $R_{\rm rms}^{(c)}$, the rms distances between the core and $\Lambda$ particle $R_{\rm rms}^{(c-\Lambda)}$, and the rms radii of $^9_\Lambda$Be, $R_{\rm rms}$, at the minimum positions, calculated by using artificial $\Lambda N$ interaction $V^{(\Lambda N)} \to \delta \times V^{(\Lambda N)}$. The maximum squared overlap values of $\mathcal{O}_J(\beta_\perp, \beta_z)$ defined by Eq. (87), are also shown. $\delta = 0$ corresponds to the results of $^8$Be.

| $\delta$ | $E(\beta_\perp, \beta_z)$ | $B_\Lambda$ | $R_{\rm rms}^{(c)}$ | $R_{\rm rms}^{(c-\Lambda)}$ | $R_{\rm rms}$ | $\mathcal{O}_{J=0}(\beta_\perp, \beta_z)$ |
|---|---|---|---|---|---|---|
| 0.0 | $-54.45(1.8, 7.8)$ | | 2.87 | | | 1.00 (1.8, 7.8) |
| 0.4 | $-54.67(1.8, 5.4)$ | 0.22 | 2.65 | 7.13 | 3.36 | 0.994(1.7, 5.9) |
| 0.6 | $-56.09(1.7, 3.9)$ | 1.64 | 2.47 | 3.78 | 2.62 | 0.993(1.7, 4.2) |
| 0.8 | $-58.29(1.6, 3.2)$ | 3.84 | 2.38 | 2.95 | 2.43 | 0.994(1.6, 3.4) |
| 1.0 | $-61.00(1.5, 2.8)$ | 6.55 | 2.31 | 2.57 | 2.33 | 0.995(1.6, 3.0) |

We should note that at each step of the change of $\Lambda N$ interaction, from dilute $2\alpha$ cluster structure to compact localized $2\alpha$ cluster structure, all are described precisely by single H-THSR wave functions, which have optimal sizes of deformed container, i.e. those with large-size container to small-size container. We should emphasize that at every step of $\delta$, the squared overlaps with the full Brink-GCM solutions are more than 99.3%. This may imply that the THSR-type container picture is essentially important in understanding every type of cluster structures, from the compact to dilute ones. We should also emphasize that for $^8$Be a single optimal Brink wave function can only approximate the Brink-GCM wave function by 0.722 [41], while for $^9_\Lambda$Be the former can approximate the latter by 0.940 [100].

**(2) Multi-cluster dynamics in $^{13}_\Lambda$C** We discuss $^{13}_\Lambda$C in this subsection by using the H-THSR ansatz. $^{13}_\Lambda$C is theoretically investigated and the second $1/2^+$ state is found as a bound state, which is considered to correspond to the $^{12}$C$(0_2^+) \otimes \Lambda$ state [127, 128], although experimentally the second $1/2^+$ state in $^{13}_\Lambda$C has not yet been observed [129]. The H-THSR wave function introduced in this study is slightly different from the one used in the previous subsection for the study of $^9_\Lambda$Be, as follows:

$$\Phi_{3\alpha}^{\rm H-THSR}(\boldsymbol{\beta}_1, \boldsymbol{\beta}_2, \boldsymbol{\beta}_\Lambda) = \mathcal{A}\bigg[\exp\bigg\{-\sum_{i=1}^2 \mu_i \sum_{k=x,y,z} \frac{2\xi_{ik}^2}{b^2 + 2\beta_{ik}^2}\bigg\} \phi^3(\alpha)\bigg] \varphi_\Lambda(\boldsymbol{\beta}_\Lambda), \quad (89)$$

with

$$\varphi_\Lambda(\boldsymbol{\beta}_\Lambda) = \exp\bigg(-\mu_\Lambda \sum_{k=x,y,z} \frac{2\xi_{\Lambda k}^2}{b^2 + 2\beta_{\Lambda k}^2}\bigg), \quad (90)$$

where the $\varphi_\Lambda(\boldsymbol{\beta}_\Lambda)$ is the wave function of the $\Lambda$ particle with its width parameter, $\boldsymbol{\beta}_\Lambda$, which can be out of the antisymmetrization, $\mu_\Lambda = 3m_\Lambda/(12m_N + m_\Lambda)$ with $m_N$ and $m_\Lambda$ being nucleon and $\Lambda$-particle masses, respectively, and $\boldsymbol{\xi}_\Lambda = \boldsymbol{X}_\Lambda - (\boldsymbol{X}_1 + \boldsymbol{X}_2 + \boldsymbol{X}_3)/3$. In this wave function, the $\Lambda$ particle is also confined in a deformed container, so that the present H-THSR wave function is more extended than the previous one, which only contains the $S$-wave component.

Also in this calculation, the axially symmetric deformation is taken into account, i.e. $\beta_{ix} = \beta_{iy} \neq \beta_{iz}$ and $\beta_{\Lambda x} = \beta_{\Lambda y} \neq \beta_{\Lambda z}$. After performing the angular-momentum projection for the H-THSR wave function in Eq. (89), the energy eigenstates can be obtained by solving the Hill-Wheeler equation by taking the six parameters as generator coordinates.

In Fig. 47, the calculated $J^\pi = 0^+$ spectra of $^{12}$C and $^{13}_\Lambda$C and the observed $J^\pi = 0^+$ spectrum of $^{12}$C are shown. We discussed the $0^+$ states of $^{12}$C in detail in Secs. 2.1.5 and 5.3 (see Figs. 9 and 30). Let us keep in mind that the $0_3^+$ state is concluded to be the excited state from the Hoyle state by monopole vibration and the $0_4^+$ state dominantly has a $3\alpha$ linear-chain configuration. While the $0_3^+$ state is considered to be a family of the Hoyle state, the $0_4^+$ state has a quite different structure from the $0_2^+$ and $0_3^+$ states. In $^{13}_\Lambda$C, first we should mention that we simply denote the spectrum of $^{13}_\Lambda$C as $J^\pi = 0^+$ states, neglecting the intrinsic spin $1/2$ of the $\Lambda$ particle, for simplicity. From this figure,

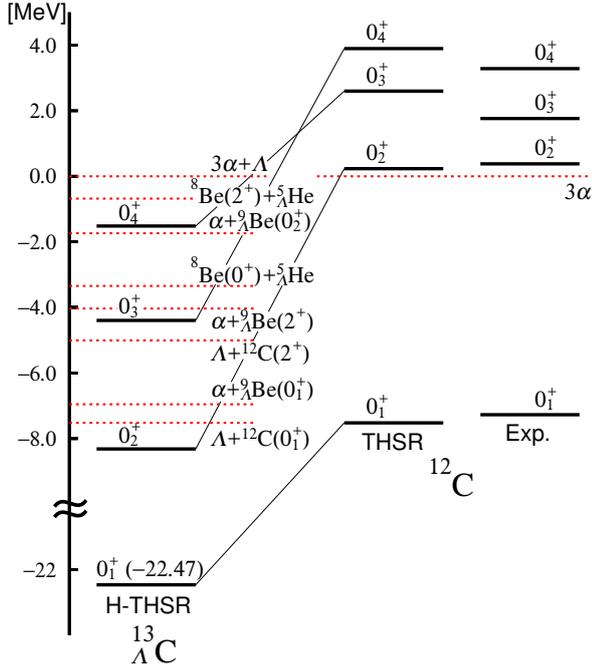

**Table 12** R.m.s. radii of $^{13}_\Lambda$C ($R_{\rm rms}$) and of the $^{12}$C core ($R^{(c)}_{\rm rms}$), r.m.s. distance between the $\Lambda$ particle and $^{12}$C core, for the $0^+_1 - 0^+_4$ states of $^{13}_\Lambda$C, and for comparison, r.m.s. radius of the $0^+_1 - 0^+_4$ states in $^{12}$C, are shown in a unit of fm. $B_\Lambda$ values for the $0^+_1 - 0^+_4$ states in $^{13}_\Lambda$C, together with the experimental data, are also shown.

| | $^{13}_\Lambda$C | | | | $^{12}$C |
|---|---|---|---|---|---|
| | $R^{(c)}_{\rm rms}$ | $R^{(c-\Lambda)}_{\rm rms}$ | $R_{\rm rms}$ | $B_\Lambda$ (MeV) | $R_{\rm rms}$ |
| $0^+_1$ | 2.2 | 2.1 | 2.2 | 15.0 (exp : 11.69) | 2.4 |
| $0^+_2$ | 2.8 | 3.4 | 2.9 | 8.5 | 3.7 |
| $0^+_3$ | 3.1 | 4.8 | 3.2 | 8.3 | 4.2 ($0^+_4$) |
| $0^+_4$ | 4.3 | 4.8 | 4.3 | 4.1 | 4.7 ($0^+_3$) |

**Fig. 47** (color online) Spectra of $^{12}$C and $^{13}_\Lambda$C calculated with THSR and H-THSR wave functions. Experimental spectrum of $^{12}$C is also shown [58]. Spectrum of $^{12}$C calculated with THSR wave function is taken from Refs. [16, 17].

one can see that $^{13}_\Lambda$C gives a much larger number of decay channels than $^{12}$C, i.e. $\Lambda + ^{12}$C$(0^+_1)$, $\alpha + ^{9}_\Lambda$Be$(0^+_1)$, $\Lambda + ^{12}$C$(2^+_1)$, $\alpha + ^{9}_\Lambda$Be$(2^+_1)$, $^{8}$Be$(0^+) + ^{5}_\Lambda$He, $\alpha + ^{9}_\Lambda$Be$(0^+_2)$, and $^{8}$Be$(2^+) + ^{5}_\Lambda$He. Up to the $3\alpha + \Lambda$ threshold, we have the four $J^\pi = 0^+$ states, corresponding to the four $J^\pi = 0^+$ states in $^{12}$C. We can see that the $0^+_2$ in $^{13}_\Lambda$C is obtained as a bound state which is located below the lowest threshold, $\Lambda + ^{12}$C$(0^+_1)$, by 0.8 MeV. The $0^+_3$ and $0^+_4$ states are connected to the $0^+_4$ and $0^+_3$ states of $^{12}$C, respectively, as you can also see from the figure. This assignment is because the $0^+_3$ and $0^+_4$ states in $^{13}_\Lambda$C largely include $^{12}$C$(0^+_4) + \Lambda$ and $^{12}$C$(0^+_3) + \Lambda$ components, respectively [125].

In fact, the $0^+_3$ state in $^{13}_\Lambda$C is found to keep essentially the same features as the $0^+_4$ state in $^{12}$C, which are shown in Sec. 5.3. For example, we obtain very similar result as in Fig. 32, also for the $0^+_4$ state in $^{13}_\Lambda$C, where the squared overlap with the single THSR configuration is calculated. The maximum squared overlap value is calculated to be 47%, which is almost the same as $0^+_4$ state in $^{12}$C, but the $\beta$ parameter values to give the maximum is shifted from $(\beta_x = \beta_y, \beta_z) = (0.6, 6.7 \text{ fm})$ in the $0^+_4$ state of $^{12}$C, to $(\beta_x = \beta_y, \beta_z) = (0.4, 4.6 \text{ fm})$ in the $0^+_3$ state of $^{13}_\Lambda$C. The linear-chain configuration in $^{13}_\Lambda$C then shrinks along $z$-direction, compared with the case of $^{12}$C, due to the additional $\Lambda$ particle.

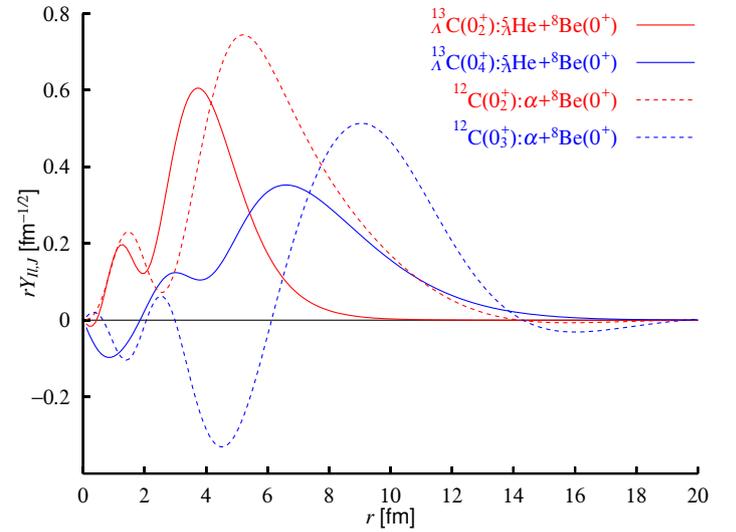

**Fig. 48** (color online) RWAs of the $0^+_2$ (dotted curve in red) and $0^+_3$ (dotted curve in blue) state of $^{12}$C in the channel $\alpha + ^{8}$Be$(0^+)$ and of the $0^+_2$ (solid curve in red) and $0^+_4$ (solid curve in blue) states of $^{13}_\Lambda$C in the channel $^{5}_\Lambda$He$+ ^{8}$Be$(0^+)$ are shown.

In Table 12, we show $\Lambda$ binding energy $B_\Lambda$, the root mean square (r.m.s.) radius of the core $R_{\text{rms}}^{(c)}$, the r.m.s. distance between the core and $\Lambda$ particle $R_{\text{rms}}^{(c-\Lambda)}$, and the r.m.s. radius of $^{13}_\Lambda$C, $R_{\text{rms}}$, together with the r.m.s. radii of isolated $^{12}$C calculated in Ref. [16]. Here the $\Lambda$ binding energy is defined as the binding energy measured from the corresponding $^{12}$C($0^+$) + $\Lambda$ threshold. From the calculated r.m.s. radii for these states, we can expect that the $0_4^+$ state is qualified to be a gas-like cluster state, since only this state has a larger r.m.s. radius 4.3 fm than the one of the Hoyle state 3.7 fm. From the RWA analysis, we find that the $0_4^+$ state has the largest component of $\alpha + ^9_\Lambda$Be($0_2^+$), in which $^9_\Lambda$Be($0_2^+$) has a dilute $\alpha + ^5_\Lambda$He structure [125]. Therefore this state is a loosely coupled $2\alpha + ^5_\Lambda$He state, i.e. a Hoyle-analog state. In this table, we can also see that the $0_4^+$ state suffers from much less shrinkage effect from the additional $\Lambda$ particle than the other $0_2^+$ and $0_3^+$ states, except for the ground state that has saturation density and gives almost no shrinkage effect. The $B_\Lambda$ value for this state is also smaller than the one of the other states. This is because for the $0_4^+$ state the $\Lambda$ particle couples with only one $\alpha$ cluster, to form $^5_\Lambda$He. The little overlap with nucleons reduces the shrinkage as well as the gain of the $\Lambda$ binding energy. In fact, this state appears around the $2\alpha + ^5_\Lambda$He threshold energy, for the threshold rule. The slightly higher energy position, 1.6 MeV above the $2\alpha + ^5_\Lambda$He threshold, allows for the diluteness of this state.

On the other hand, the dominant component included in the $0_2^+$ state is $\alpha + ^9_\Lambda$Be($0_1^+$) configuration. The $\Lambda$ particle then overlaps with the $2\alpha$ clusters and therefore gives much stronger shrinkage effect and larger $\Lambda$ binding energy. The binding energy of the $^9_\Lambda$Be is so large that it is energetically favored that the injected $\Lambda$ particle shrinks the core and forms $\alpha + ^9_\Lambda$Be($0_1^+$) cluster configuration. In fact, this state is located at around the $\alpha + ^9_\Lambda$Be($0_1^+$) threshold as well as around the $\Lambda + ^{12}$C($0_1^+$) threshold (see Fig. 47), which follows the so-called threshold rule.

We further show in Fig. 48 the RWAs in the $^8$Be($0^+$)+ $^5_\Lambda$He channel of the $0_2^+$ (solid curve in red) and the $0_4^+$ (solid curve in blue) states in $^{13}_\Lambda$C. In comparison, the corresponding RWAs of the Hoyle state (dotted curve in red) and the $0_3^+$ state (dotted curve in blue) in $^{12}$C for the $^8$Be($0_1^+$) + $\alpha$ channel are also shown (see also Fig. 11(Upper)). As we discussed in Sec. 2.1.5, in the Hoyle state (dotted curve in red) the nodal behavior appears as oscillation because of dissolution of $^8$Be core, while in the $0_3^+$ state in $^{12}$C (dotted curve in blue) there exist four nodes. On the other hand, in $^{13}_\Lambda$C, the most prominent feature is the disappearance of nodes for the $0_4^+$ state, though the innermost node only remains. This means that the $^8$Be core is dissolved in the $0_4^+$ state, completely unlike the case of the $0_3^+$ state in $^{12}$C. We should also be aware that for the $0_4^+$ state a long tail is still very much developed, which is similar to the Hoyle state. This is also completely different from the $0_2^+$ state, in which the whole amplitude is pushed inside with no long tail any more. All these results give a strong support that the $0_4^+$ state has a gas-like structure of $2\alpha + ^5_\Lambda$He clusters, as a Hoyle analog state. It is important to mention that this Hoyle analog state uniquely appears in $^{13}_\Lambda$C and we cannot expect in $^{13}$C $2\alpha + ^5$He gas, since $^5$He is not bound [130], whereas $^5_\Lambda$He is bound by 3.1 MeV.

### 5.6 More generalization and manipulation of the THSR wave function

#### 5.6.1 Two-center THSR wave function

Recently, to investigate the correlation between the valence proton and $2\alpha$ clusters in $^9$B, a two-center THSR wave function was proposed [131]. Based on the designed THSR wave function for the description of the $^9$Be($2\alpha+n$) cluster system [87], one similar THSR wave function for $^9$B($2\alpha+p$) can be written as,

$$\Phi_0(\boldsymbol{\beta}_\alpha, \boldsymbol{\beta}_p) = \int d\boldsymbol{R}_1 \, d\boldsymbol{R}_2 \exp[-\sum_{i=1}^{2}(\frac{R_{i,x}^2 + R_{i,y}^2}{\beta_{\alpha,xy}^2} + \frac{R_{i,z}^2}{\beta_{\alpha,z}^2})]$$
$$\int d\boldsymbol{R}_p \exp[-\frac{R_{p,x}^2 - R_{p,y}^2}{\beta_{p,xy}^2} - \frac{R_{p,z}^2}{\beta_{p,z}^2}]$$
$$\times e^{im\phi_{\boldsymbol{R}_p}} \Phi_{2\alpha+p}^B(\boldsymbol{R}_1, \boldsymbol{R}_2, \boldsymbol{R}_p), \quad (91)$$

where $\boldsymbol{\beta}_\alpha$ is the size parameter for the nonlocalized motion of $2\alpha$ clusters and $\boldsymbol{\beta}_p$ is the size parameter for the valence proton around the $2\alpha$ clusters. The $e^{im\phi_{\boldsymbol{R}_p}}$ is the phase factor for the treatment of the negative-parity states. $\Phi_{2\alpha+p}^B(\boldsymbol{R}_1, \boldsymbol{R}_2, \boldsymbol{R}_p)$ is the Brink wave function, in which the ($\boldsymbol{R}_1$, $\boldsymbol{R}_2$), and $\boldsymbol{R}_p$ are corresponding generator coordinates for the $\alpha$ clusters and the valence proton, respectively.

In the $2\alpha+n$ system, many configurations should be considered for some developed cluster states due to the nonlocalized motion of the valence proton. To describe these cluster states and also discuss the correlations of valence proton and $\alpha$ clusters in a more accuracy way, the following THSR wave functions are constructed [131],

$$\Phi_j(\boldsymbol{\beta}_\alpha, \boldsymbol{\beta}_p) = \int d\boldsymbol{R}_1 \, d\boldsymbol{R}_2 \exp[-\sum_{i=1}^{2}(\frac{R_{i,x}^2 + R_{i,y}^2}{\beta_{\alpha,xy}^2} + \frac{R_{i,z}^2}{\beta_{\alpha,z}^2})]$$
$$\times \int d\boldsymbol{R}_p \exp[-\frac{R_{p,x}^2 - R_{p,y}^2}{\beta_{p,xy}^2} - \frac{R_{p,z}^2}{\beta_{p,z}^2}]$$
$$\times e^{im\phi_{\boldsymbol{R}_p+\boldsymbol{R}_j}} \Phi_{2\alpha+p}^B(\boldsymbol{R}_1, \boldsymbol{R}_2, \boldsymbol{R}_p+\boldsymbol{R}_j), \quad (j=1,2)$$
$$(92)$$

It should be noted that, compared with the THSR wave function in Eq. (91), the correlation of the valence proton and $\alpha$ clusters are considered in the $\Phi_j(\boldsymbol{\beta}_\alpha, \boldsymbol{\beta}_p)$ by slightly changing the generator coordinate of the valence proton in the Brink wave function in the construction.

Finally, the THSR wave function of $^9$B can be constructed as a superposition of both cluster-correlated configuration $\Phi_j(\boldsymbol{\beta}_\alpha, \boldsymbol{\beta}_p)$ and large spreading configuration $\Phi_0(\boldsymbol{\beta}_\alpha, \boldsymbol{\beta}_p)$

$$\Psi = c(\Phi_1(\boldsymbol{\beta}_\alpha, \boldsymbol{\beta}'_p) + \Phi_2(\boldsymbol{\beta}_\alpha, \boldsymbol{\beta}'_p)) + d\Phi_0(\boldsymbol{\beta}_\alpha, \boldsymbol{\beta}_p). \quad (93)$$

Here $\Phi_1$ and $\Phi_2$ can include cluster-correlated configurations of $^9$B describing the correlated motion of valence proton around each $\alpha$-cluster. The wave function $\Phi_0$ is corresponding to the large spreading configuration of $^9$B in which the valence proton orbits around the $^8$Be core. The $c$ and $d$ are coefficient parameters of the two-center THSR wave function.

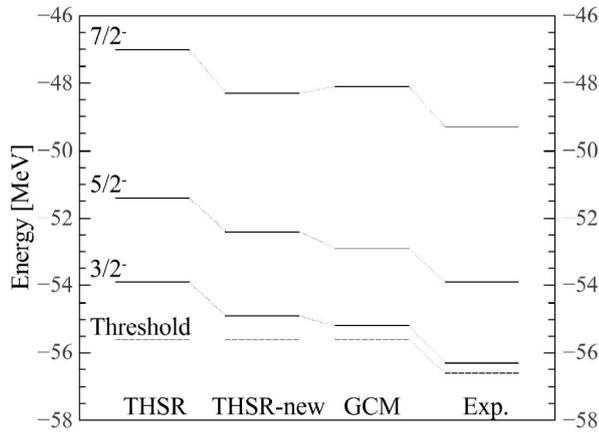

**Fig. 49** Theoretical and experimental results of the energy spectrum of $^9$B [131].

Figure 49 shows the obtained energy spectrum of the $^9$B using the two-center THSR wave function and the comparisons with those obtained by the THSR wave function $\Phi_0$, GCM Brink wave function, and the experimental data. It can be seen that the significant improvements for the binding energy of $^9$B using the two-center THSR wave function which are in much better agreements with the experimental values. Furthermore, by analyzing the weight (c,d) in the two-center THSR wave function, it can be seen that the cluster-correlated and the large spreading configurations both have significant contributions to the dynamics of the ground state of $^9$B. Details discussions can be found in Ref. [131].

### 5.6.2 Cluster systems with different size variables

In nuclear cluster physics, the widths of Gaussian packets describing clusters are very important dynamics variables especially for the asymmetrical cluster systems of nuclei, which have been discussed in many works for many years [132–135]. However, the introduced width variables will lead to the serious center-of-mass problem, i.e., the intrinsic wave function and the center-of-mass wave function cannot be separated exactly, which is a long-standing problem in nuclear cluster physics.

In RGM [135], the trial cluster wave function is constructed in term of the translational invariant coordinates and the width variables can be naturally included. Nevertheless, the practical calculation is becoming very tedious. A more practical method for the description of cluster structure is GCM combined with in the traditional Brink wave function [34] and other microscopic cluster wave functions [136], in which the cluster wave function can be expressed as the superposition of Slater determinants. However, the center-of-mass problem still remains. Quite recently, we found that, in the THSR framework, cluster systems with different size variables can be described in a simple way, in which an exact solution to the center-of-mass problem was given. The key to separate exactly the center-of-mass part and the translation-invariant intrinsic wave function part is the following constructed integral operator [137],

$$\widehat{\boldsymbol{G}}_n(\beta_0) = \int d\boldsymbol{R}_1 \cdots d\boldsymbol{R}_n \exp[-\sum_{i=1}^n \frac{A_i R_i^2}{\beta_0^2 - 2b_i^2}]\widehat{\boldsymbol{D}}(\boldsymbol{R}). \quad (94)$$

Here, the $\widehat{\boldsymbol{D}}$ is a shift operator. With the auxiliary generator coordinate $\boldsymbol{R}$ ($\boldsymbol{R}_1 \dots \boldsymbol{R}_n$), this $\widehat{\boldsymbol{G}}_n(\beta_0)$ operator can perform the integral transformation for exactly separating the center-of-mass part from a many-body wave function. This handled wave function can be antisymmetrizd or not, nuclear cluster or nucleon wave function.

By means of the integral operator, a new THSR-type wave function can be naturally obtained, in which the center-of-mass part can be exactly separately. Take a two-cluster system ($C=C_1+C_2$) as an example. The THSR-type wave function can be constructed as follows,

$$\Psi_c(\beta_0, b_{c_1}, b_{c_2})$$
$$= \int d\boldsymbol{R}_1 d\boldsymbol{R}_2 \exp[-\frac{A_1 R_1^2}{\beta_0^2 - 2b_{c_1}^2} - \frac{A_2 R_2^2}{\beta_0^2 - 2b_{c_2}^2}]\Phi_c^B(\boldsymbol{R}_1, \boldsymbol{R}_2) \quad (95)$$

$$\propto \exp[-\frac{A}{\beta_0^2}X_{cm}^2] \mathcal{A}\{\exp[-\frac{A_1 A_2}{A\beta_0^2}X_{rel}^2] \phi_{c_1}^{int}(b_{c_1})\phi_{c_2}^{int}(b_{c_2})\}. \quad (96)$$

$$\Phi_c^B(\boldsymbol{R}_1, \boldsymbol{R}_2) \propto \mathcal{A}\{\exp[-\frac{A_1}{2b_{c_1}^2}(\boldsymbol{X}_1 - \boldsymbol{R}_1)^2 - \frac{A_2}{2b_{c_2}^2}(\boldsymbol{X}_2 - \boldsymbol{R}_2)^2]$$
$$\times \phi_{c_1}^{int}(b_{c_1})\phi_{c_2}^{int}(b_{c_2})\}. \quad (97)$$

Here the clusters $C_1$ and $C_2$ have the atomic mass number $A_1$ and $A_2$, respectively. $\boldsymbol{X}_1$ and $\boldsymbol{X}_2$ are the corresponding center-of-mass coordinate, respectively. $\boldsymbol{X}_{cm} = (A_1\boldsymbol{X}_1 + A_2\boldsymbol{X}_2)/A$ and $\boldsymbol{X}_{rel} = \boldsymbol{X}_2 - \boldsymbol{X}_1$. $b_{c_1}$

and $b_{c_2}$ are the width parameters or the size parameters of cluster $C_1$ and $C_2$. $\Phi_c^B$ is the general Brink wave function with different widths of clusters. It can be seen that compared with the previous THSR wave function in Sec. 4 for $^{20}$Ne, the main change of Eq. (95) is just one more integral for the generator coordinate $\boldsymbol{R}$. The computation is very straightforward and simple, in particularly, compared with the other RGM-based approaches.

Following the two-cluster wave function in Eq. (95), the $^{20}$Ne wave function can be written directly,

$$\Psi_{\mathrm{Ne}}(\beta_0, b_\alpha, b_{^{16}\mathrm{O}})$$
$$\propto \exp[-\frac{20}{\beta_0^2}X_{\mathrm{cm}}^2]\,\mathcal{A}\{\exp[-\frac{16}{5\beta_0^2}X_{\mathrm{rel}}^2]\,\phi_\alpha^{\mathrm{int}}(b_\alpha)\phi_{^{16}\mathrm{O}}^{\mathrm{int}}(b_{^{16}\mathrm{O}})\}. \tag{98}$$

Here $\boldsymbol{X}_{\mathrm{cm}} = (4\boldsymbol{X}_1 + 16\boldsymbol{X}_2)/20$ and $\boldsymbol{X}_{\mathrm{rel}} = \boldsymbol{X}_2 - \boldsymbol{X}_1$. $\boldsymbol{X}_1$ and $\boldsymbol{X}_2$ are the center-of-mass coordinate of $\alpha$ cluster and $^{16}$O cluster, respectively. $b_\alpha$ and $b_{^{16}\mathrm{O}}$ are the width parameters or the size parameters of $\alpha$ cluster and $^{16}$O cluster, respectively. The $\Phi_{\mathrm{Ne}}^B(\boldsymbol{R}_1, \boldsymbol{R}_2)$ is a generalized Brink wave function [34, 138] for $^{20}$Ne system with different width variables. As we know, in the traditional Brink wave function, the widths of different clusters usually are fixed at the same values, i.e., $b_\alpha = b_{^{16}\mathrm{O}}$ in Eq. (97), to avoid the center-of-mass problem.

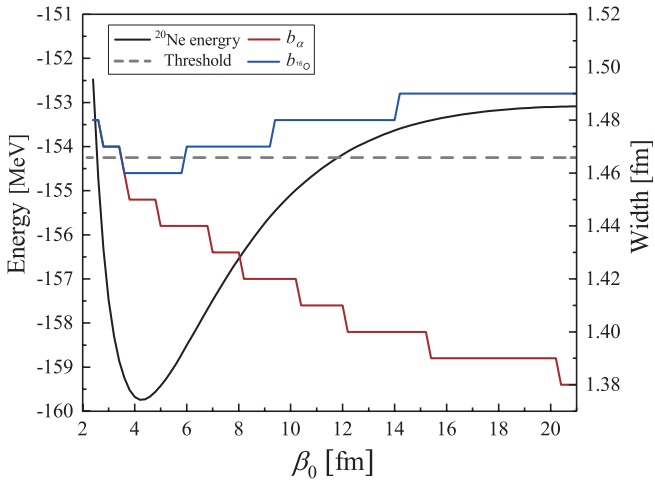

**Fig. 50** Variational energy calculations for $^{20}$Ne in the three-parameter space, the width variable of $\alpha$, the width variable of $^{16}$O, and the size parameter $\beta_0$ [137]. It should be noted that the widths' lines are not smooth due to the choice of the adopted 0.01-fm meshpoints.

By making variational calculations for widths $b_\alpha$, $b_{^{16}\mathrm{O}}$, and $\beta_0$ variables, the energy curve of $^{20}$Ne and the variation of width variables can be obtained. Figure 50 shows the energy curve of $^{20}$Ne along with the obtained optimum widths $b_\alpha$ and $b_{^{16}\mathrm{O}}$ using the intrinsic wave function in Eq. (98). Clearly, it can be seen that with the increase of the size parameter $\beta_0$ indicating the cluster system expands, the gap between $b_\alpha$ and $b_{^{16}\mathrm{O}}$ is becoming larger and larger. Another saying, imagine a situation where the stable double-closed-shell nuclei $\alpha$ and $^{16}$O are approaching each other to form a stable nuclear state (the ground state of $^{20}$Ne), in the process Fig. 50 tells us the $^{16}$O is slightly shrinking while the $\alpha$ nucleus is becoming more loosely bound. It has been known that the ground state of $^{20}$Ne has a very compact cluster state. In this case, the obtained minimum energy for the ground state is $-159.74$ MeV. The corresponding wave function is characterized by a rather small value of the size parameter $\beta_0$=4.2 fm and slightly different values of widths $b_\alpha$=1.45 fm and $b_{^{16}\mathrm{O}}$=1.46 fm. As a comparison, by using the traditional angular-momentum projected Brink wave function in a variational calculation with the common width variable and inter-cluster distance parameter $R$, the minimum energy $E(b_\alpha=b_{^{16}\mathrm{O}} = 1.47$ fm, $R= 3.0$ fm)$= -158.43$ MeV is obtained. This is more than 1 MeV higher than the optimum energy obtained by our new intrinsic wave function.

In the new THSR-type wave function, not only the size parameter for the relative motion of clusters but also the width of the clusters are also included. Actually, without considering the width variables, many physical quantities like radius, differential cross section, phase shift actually cannot be well reproduced and it has been discussed for a long time in the RGM approach [1, 139, 140]. Therefore, the gas-like cluster structure, halo nuclei, the distortion effects of clusters in light nuclei are very promising to be investigated in a more realistic way in this new framework.

### 5.6.3 Monte-Carlo integration in the THSR wave function

As we discuss in Sec. 2, the original THSR wave function has a very simple mathematics form and it is also related with the Slater determinant-based Brink wave function. In the practical calculations, the THSR matrix element of Hamiltonian can be directly expressed as,

$$\langle\Psi_{n\alpha}(\boldsymbol{\beta})|\hat{H}|\Psi_{n\alpha}(\boldsymbol{\beta}')\rangle =$$
$$\int d\boldsymbol{R}_1\ldots d\boldsymbol{R}'_n \exp[-\sum_{i=1}^n(\frac{R_i^2}{\beta^2}+\frac{R_i'^2}{\beta'^2})] \tag{99}$$
$$\langle\Phi_{n\alpha}^B(\boldsymbol{R}_1,\cdots,\boldsymbol{R}_n)|\hat{H}|\Phi_{n\alpha}^B(\boldsymbol{R}'_1,\cdots,\boldsymbol{R}'_n)\rangle. \tag{100}$$

To compute the above THSR matrix element, the first step is to obtain the analytical kernels of the Brink wave function, which has been studied for many years and there are many well-developed techniques and methods for this kind of analytical calculations [37]. The second step is to perform the integrals on the obtained analytical expressions. Because the weight function has a simple Gaussian form, the integrals for the generator coordi-nates can also be done the analytical way, which does not lose any accuracy. For example, the calculations for the $2\alpha$, $3\alpha$, $4\alpha$, and $^{20}$Ne($\alpha+^{16}$O) systems discussed in previous sections are all performed in the exact analytical way.

However, in some heavier multi-cluster systems, e.g., $6\alpha$ cluster system, this kind of analytical calculations for the THSR kernels are becoming very difficult. Firstly, even the calculations for analytical expressions for the Brink kernel is not so easy due to its huge size. Moreover, analytically, the computations for the high-dimensional integral is very expensive. For example, based on the Eq. (99), to study the $6\alpha$ cluster system, we need to perform $6\times3\times2$ Gaussian integrals for the generator coordinates to the extremely large analytical expressions from the Brink kernels, which will lead to time-consuming computations. To overcome the high-dimensional integral problem, a natural method is to use the Monte Carlo technique for the integrals. As we know, Monte Carlo integration does not depend on the number of integral dimensions, which can be considered as a very economic advantage of Monte Carlo integration against the analytical integrals. Besides, once we use the Monte Carlo technique for the integral, the Brink kernel part is not necessary to be obtained analytically. Thus, the complexity of the calculations can be reduced.

To study the possible $\alpha$ condensation states around the heavy core nucleus, Itagaki and his collaborators introduce the Monte Carlo techniques [141, 142] for the treatment of the THSR wave function. The original THSR wave function can be reformulated as,

$$\Psi_{n\alpha}(\boldsymbol{\beta}) =$$
$$\int d\boldsymbol{R}_1\ldots d\boldsymbol{R}_n \exp[-\sum_{i=1}^{n}\frac{R_i^2}{\beta^2}]\, \mathcal{A}[\Phi_{\alpha_1}(\boldsymbol{R}_1)\cdots\Phi_{\alpha_n}(\boldsymbol{R}_n)] \quad (101)$$

$$= \mathcal{A}\{\prod_{i=1}^{n}[\int d\boldsymbol{R}_i \ \exp[-\frac{R_i^2}{\beta^2}]\Phi_{\alpha_i}(\boldsymbol{R}_i)]\}. \quad (102)$$

Here $\Phi_{\alpha_i}$ is the wave function for the $i$th $\alpha$ cluster centered at $\boldsymbol{R}_i$, which has a Gaussian form $\text{Exp}[-\nu(\boldsymbol{r}-\boldsymbol{R}_i)^2]$ with the width $\nu$. To simplify the calculation of this THSR wave function, the integral over the generator coordinate parameters $\boldsymbol{R}_i$ in the original THSR wave function can be replaced by the summation of many Slater determinants as follows,

$$\Psi_{n\alpha}(\beta) \approx \sum_{k=1}^{m} \mathcal{A}[\Phi_{\alpha_1}(\boldsymbol{R}_1)\cdots\Phi_{\alpha_n}(\boldsymbol{R}_n)]. \quad (103)$$

The generator coordinate parameter $\{\boldsymbol{R}_i\}$ can be randomly produced, however, the random numbers are generated by the weight function $W(\boldsymbol{R}_i)$ with the same Gaussian shape:

$$W(\boldsymbol{R}_i) \propto \exp[-\frac{R_i^2}{\beta^2}]. \quad (104)$$

With the increasing ensemble number, the distribution of $\{\boldsymbol{R}_i\}$ will be close to the Gaussian with the width parameter $\boldsymbol{\beta}$. Therefore, this kind of approximation method can be considered that the integration of the THSR wave function is performed by using a Monte Carlo technique. Clearly, when the number of the superposed Slater determinants is becoming a very large number, the obtained wave function will agree with the exact THSR wave function. Most importantly, this numerical treatment for the integral is much easier than the analytical integral, especially the high-dimensional integral problem.

To check the reliability of the Monte Carlo technique for the integral of the THSR wave function, Itagaki et al. studied from the well-studied $^8$Be and $^{12}$C [141]. It was found that as for the $^8$Be($\alpha+\alpha$) cluster system, except for the line with small $\beta$ (=2 fm) value, all other lines have a good converge character to the $2\alpha$ threshold energy even with a small number of superposed wave functions. By using the Monte Carlo, the Hoyle state was also studied and the results are consistent with the original THSR calculations. This means, this kind of Monte Carlo technique is reliable for the calculations of the THSR wave function.

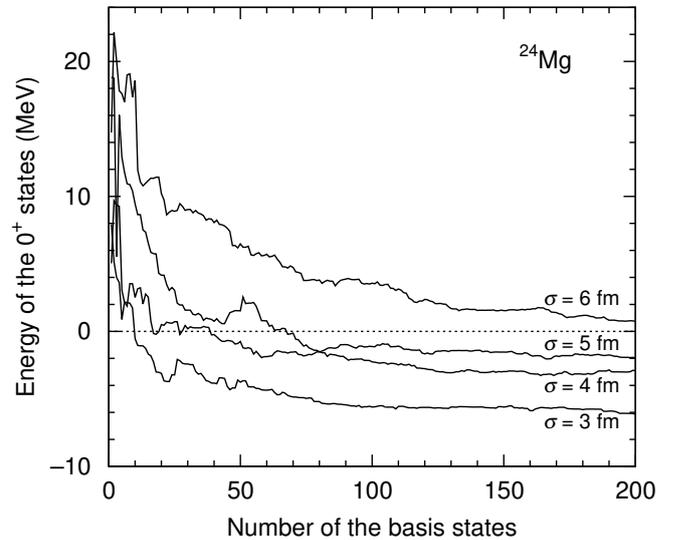

**Fig. 51** The energy convergence of two $\alpha$ clusters around the $^{16}$O core ($^{24}$Mg) for the case of $\beta = 3, 4, 5,$ and 6 fm. The energy is measured from the $^{16}$O+$2\alpha$ threshold [143].

By using the Monte Carlo technique, the THSR wave function can be extended to study the heavier cluster system. Itagaki et al., studied the $\alpha$-condensed state with a heavy core. One simple case is the $\alpha$-condensed state around the $^{16}$O core in $^{24}$Mg. Figure 51 shows the energy convergence of $^{24}$Mg measured from the $^{16}$O+2$\alpha$ threshold, where two $\alpha$-clusters occupy the $\alpha$-condensed state around the $^{16}$O core. The solid and dotted lines ($\beta$ = 2, 3 fm) converge to the energies below the threshold, however, the dashed- and dash-dotted-lines ($\beta$ = 4, 5 fm) converge almost around the threshold energy. Here, the converged energies of the these lines show that they are rather insensitive to the $\beta$ values. Thus, the $\alpha$-condensed state is considered to be realized in $^{24}$Mg around the $^{16}$O+$\alpha$ +$\alpha$ threshold energy.

Recently, the Monte Carlo technique is also applied to study the non-n$\alpha$ nuclei [87]. As for the non-n$\alpha$ nuclei, the constructed THSR wave function has a different weight function compared with the original n$\alpha$ THSR wave function, which is not suitable for the analytical integral. However, using the Monte Carlo technique, the weight function is not necessary to be limited the Gaussian form and the calculation is becoming much easier. The Monte Carlo technique provides us with a flexible way to extend the THSR wave function to heavier multi-cluster systems.

### 5.6.4 New development of analytical treatment of the THSR wave function

We are extending the THSR framework for dealing with various nuclear systems, which is one of the key ansatz for the nuclear cluster aggregate. Therefor, let us consider the practical analytical treatments of the THSR ansatz beyond the original container model which has been restricted to one container by Gaussian confinement including deformation. Here we present three kinds of possible extension as follows: Firstly, the introduction of a distant parameter between two aggregates of nuclear cluster in the integrand of the wave function. This makes it possible to estimate the parity dependence including the odd state in the container. The kind of wave function for $\alpha$+$^{16}$O has been discussed in Sec. 4. Secondly, the introduction of the same distant parameter as the previous one in the integral operator of the THSR ansatz. This can give two containers with different centers each other. It is very interesting to clarify the interaction between two cluster-aggregates through two containers, which has been applied for the calculations for the $n-3\alpha$ system in Sec. 3.4. Thirdly, the partial condensation of the nuclear cluster aggregate. For instance, let us see 3$\alpha$ clusters around $^{16}$O core. The integral operator acts only on outer 3$\alpha$ clusters. Itagaki et al. applied this ansatz to $^{24}$Mg [141] which is represented by 2$\alpha$−$^{16}$O core. They carried out multiple integral by the Monte Carlo method. Here we devote our efforts to analytical derivation of exchange kernel coming from the Pauli principle, for $2\alpha-\alpha$ system as an example for three kinds of extension. We will concentrate on the norm kernel for three kinds of extensions for simplicity.

**(1) Original THSR ansatz for** $3\alpha$ **system**   At first, we should give a brief explanation of treatment of the original 3$\alpha$ THSR ansatz in order to compare possible extensions of the THSR ansatz. The THSR ansatz consists of two parts; that is, the integrand which is fully antisymmetrized wave function on the parameter coordinate space, and the integral operator which acts on each $\alpha$ cluster on the space. The integrand is described as follows:

$$\Phi(\boldsymbol{R}_1, \boldsymbol{R}_2, \boldsymbol{R}_3) = \prod_p \begin{vmatrix} \phi_{p(1)}(\boldsymbol{R}_1) & \phi_{p(1)}(\boldsymbol{R}_2) & \phi_{p(1)}(\boldsymbol{R}_3) \\ \phi_{p(2)}(\boldsymbol{R}_1) & \phi_{p(2)}(\boldsymbol{R}_2) & \phi_{p(2)}(\boldsymbol{R}_3) \\ \phi_{p(3)}(\boldsymbol{R}_1) & \phi_{p(3)}(\boldsymbol{R}_2) & \phi_{p(3)}(\boldsymbol{R}_3) \end{vmatrix}_p, \quad (105)$$

where suffix $p$ means the spin and the isospin part. and $\boldsymbol{R}_i$ represents the coordinate of the $i^{th}$ $\alpha-$ cluster in the parameter space $\boldsymbol{R}$. The spatial part of a single nucleon wave function is written by

$$\phi_{p(i)}(\boldsymbol{R}_j) = \frac{1}{(\pi b^2)^{3/4}} \exp\left\{-\frac{1}{2b^2}\left\{\boldsymbol{r}_{p(i)} - \boldsymbol{R}_j\right\}^2\right\}, \quad (106)$$

where $i, j = 1, 2, 3$. The variable $\boldsymbol{r}_{p(i)}$ stands for the real coordinate for $i^{th}$ nucleon with $p^{th}$ spin-isospin state. We call the integrand the Brink-type wave function.

The integral operator, which is a key stone in the THSR ansatz, and which acts on the parameter coordinates of 3$\alpha$ clusters, is given by

$$\hat{C}(\beta) = \frac{1}{(\pi\beta^2)^{3\times 3/2}} \int \prod_{i=1}^{3} d\boldsymbol{R}_i \exp\left(-\frac{R_i^2}{\beta^2}\right), \quad (107)$$

where the integral is carried out over the infinite parameter space. The THSR wave function is represented by

$$\Psi(\beta) = \hat{C}(\beta) \Phi(\boldsymbol{R}_1, \boldsymbol{R}_2, \boldsymbol{R}_3) \quad (108)$$

The size parameter of the container is denoted as $\beta$.

Let us obtain the norm kernel by the analytical expression:

$$\langle \Psi(\beta) | \Psi(\beta') \rangle = \hat{C}(\beta) \hat{C}(\beta')$$
$$\times \langle \Phi(\boldsymbol{R}_1, \boldsymbol{R}_2, \boldsymbol{R}_3) | \Phi(\boldsymbol{R}'_1, \boldsymbol{R}'_2, \boldsymbol{R}'_3) \rangle \quad (109)$$

The overlap of the wave function (105) is given after carrying the integral with respect to real variable $\boldsymbol{r}_{p(i)}$:

$$\langle \Phi(\boldsymbol{R}_1, \boldsymbol{R}_2, \boldsymbol{R}_3) | \Phi(\boldsymbol{R}'_1, \boldsymbol{R}'_2, \boldsymbol{R}'_3) \rangle = \begin{vmatrix} G_{11} & G_{12} & G_{13} \\ G_{21} & G_{22} & G_{23} \\ G_{31} & G_{32} & G_{33} \end{vmatrix}^4, \quad (110)$$

where the overlap between single nucleon wave functions is written by

$$G_{ij} = \langle \phi_{p(k)}(\boldsymbol{R}_i) | \phi_{p(k)}(\boldsymbol{R}'_j) \rangle$$
$$= \exp\left\{-\frac{1}{4b^2}(\boldsymbol{R}_i - \boldsymbol{R}'_j)^2\right\}, \quad (111)$$

where the suffix $p$ is for all the spin-isospin states, where $k = 1, 2, 3$. Therefor, all the terms, which come from the permutation, in the integrand of Eq.(109) are written by the Gaussian function. For instance, the first term in the integrand corresponding to $(G_{11}G_{22}G_{33})^4$ has the exponential form:

$$\exp\left\{-\left(\tilde{R}\right)^t \begin{pmatrix} (\tilde{A}) & (\tilde{B}) \\ (\tilde{B})^t & (\tilde{A}') \end{pmatrix} \left(\tilde{R}\right)\right\}, \quad (112)$$

where matrices are given by

$$\left(\tilde{R}\right)^t = (\boldsymbol{R}_1, \boldsymbol{R}_2, \boldsymbol{R}_3, \boldsymbol{R}'_1, \boldsymbol{R}'_2, \boldsymbol{R}'_3)$$

$$\left(\tilde{A}\right) = \begin{pmatrix} \frac{1}{\beta^2} + \frac{1}{b^2} & 0 & 0 \\ 0 & \frac{1}{\beta^2} + \frac{1}{b^2} & 0 \\ 0 & 0 & \frac{1}{\beta^2} + \frac{1}{b^2} \end{pmatrix}$$

$$\left(\tilde{A}'\right) = \begin{pmatrix} \frac{1}{\beta'^2} + \frac{1}{b^2} & 0 & 0 \\ 0 & \frac{1}{\beta'^2} + \frac{1}{b^2} & 0 \\ 0 & 0 & \frac{1}{\beta'^2} + \frac{1}{b^2} \end{pmatrix}$$

$$\left(\tilde{B}\right) = -\begin{pmatrix} \frac{1}{b^2} & 0 & 0 \\ 0 & \frac{1}{b^2} & 0 \\ 0 & 0 & \frac{1}{b^2} \end{pmatrix}, \quad (113)$$

where only the matrix $(\tilde{B})$ is varied depending on the exchange property, but the matrices $(\tilde{A}), (\tilde{A}')$ are unchanged during all the permutations.

The multiple integral with respect to the parameters, $\boldsymbol{R}_1, \boldsymbol{R}_2, \boldsymbol{R}_3, \boldsymbol{R}'_1, \boldsymbol{R}'_2, \boldsymbol{R}'_3$ is easily obtained because the integrand is written by the summation of the Gaussian form. The result of the first term including the coefficients is as follows:

$$N_1(\beta, \beta') = \frac{1}{(\beta^2 \beta'^2)^{3 \times 3/2}} \left| \begin{pmatrix} (\tilde{A}) & (\tilde{B}) \\ (\tilde{B})^t & (\tilde{A}') \end{pmatrix} \right|^{-3/2} =$$

$$\left\{ \left(1 + \frac{\beta^2}{b^2}\right)^3 \left(1 + \frac{\beta'^2}{b^2}\right)^3 \left| \begin{pmatrix} (\tilde{\tilde{A}}) & (\tilde{\tilde{B}}') \\ (\tilde{\tilde{B}})^t & (\tilde{\tilde{A}}) \end{pmatrix} \right| \right\}^{-3/2}, \quad (114)$$

where

$$\left(\tilde{\tilde{A}}\right) = \begin{pmatrix} 1 & 0 & 0 \\ 0 & 1 & 0 \\ 0 & 0 & 1 \end{pmatrix}$$

$$\left(\tilde{\tilde{B}}\right) = -\begin{pmatrix} x & 0 & 0 \\ 0 & x & 0 \\ 0 & 0 & x \end{pmatrix} \quad \text{with} \quad x = \frac{\frac{\beta^2}{b^2}}{1 + \frac{\beta^2}{b^2}}$$

$$\left(\tilde{\tilde{B}}'\right) = -\begin{pmatrix} x' & 0 & 0 \\ 0 & x' & 0 \\ 0 & 0 & x' \end{pmatrix} \quad \text{with} \quad x' = \frac{\frac{\beta'^2}{b^2}}{1 + \frac{\beta'^2}{b^2}}.$$

We concentrate our thought on a computing procedure of the analytical derivation of the determinant; that is, the most tedious and primitive approach is the best way for programming code. We rewrite the determinant part in Eq. (114):

$$\left| \begin{pmatrix} (\tilde{\tilde{A}}) & (\tilde{\tilde{B}}') \\ (\tilde{\tilde{B}})^t & (\tilde{\tilde{A}}) \end{pmatrix} \right| = \begin{vmatrix} 1 & 0 & 0 & -x' & 0 & 0 \\ 0 & 1 & 0 & 0 & -x' & 0 \\ 0 & 0 & 1 & 0 & 0 & -x' \\ -x & 0 & 0 & 1 & 0 & 0 \\ 0 & -x & 0 & 0 & 1 & 0 \\ 0 & 0 & -x & 0 & 0 & 1 \end{vmatrix}. \quad (115)$$

At first, we have to divide all the columns into two parts; that is, for instance, the first column is written by

$$\begin{pmatrix} 1 \\ 0 \\ 0 \\ -x \\ 0 \\ 0 \end{pmatrix} = \begin{pmatrix} 1 \\ 0 \\ 0 \\ 0 \\ 0 \\ 0 \end{pmatrix} - x \begin{pmatrix} 0 \\ 0 \\ 0 \\ 1 \\ 0 \\ 0 \end{pmatrix}. \quad (116)$$

As consequences, the determinant can be estimated by the summation of $2^6$ separated determinants which are represented by the polynomial with respect to $x$ and $x'$. Almost all of the separated determinants are exactly 0 except for the power of $xx'$ due to the physical background. Thus, we can imagine that the derivation of the determinant of each term in Eq. (114) is easy task, and all the summation of the terms is obtained with analytical form of the norm kernel. Of course, we have to elegantly arrange them because there are a lot of the exactly same terms. The analytical formula of the norm kernel is written by the following equation:

$$N_{all}(x, x') = f(x, x') \sum_{k=1}^{9} W_k f_k^{-3/2}(xx'), \quad (117)$$

where

$$f(x, x') = \left\{ \left(\frac{2-x}{1-x}\right) \left(\frac{2-x'}{1-x'}\right) \right\}^{-3 \times 3/2},$$

$$f_k(xx') = \sum_{l=0}^{3} c_{kl} \left(-\frac{xx'}{16}\right)^l.$$

The weight $W_k$, which comes from the same exchange character of

$$(G_{1i_1}G_{2j_1}G_{2k_1})_{\uparrow\uparrow}(G_{1i_1}G_{2j_1}G_{2k_1})_{\uparrow\downarrow}$$
$$(G_{1i_1}G_{2j_1}G_{2k_1})_{\downarrow\uparrow}(G_{1i_1}G_{2j_1}G_{2k_1})_{\downarrow\downarrow}, \quad (118)$$

and the coefficients $c_{kl}$ are listed in Table **13**. All the 9 terms with respect to the polynomial have a common factor $(1 - xx')$ because of the inclusion of c.o.m parameter.

| | | $c_{kl}$ | | | | The set of $(i,j,k)$ | | | |
|---|---|---|---|---|---|---|---|---|---|
| $k$ | $W_k$ | 0 | 1 | 2 | 3 | ↑↑ | ↑↓ | ↓↑ | ↓↓ |
| 1 | 1 | 1 | 48 | 768 | 4096 | 123 | 123 | 123 | 123 |
| 2 | -12 | 1 | 36 | 384 | 1024 | 123 | 123 | 123 | 213 |
| 3 | 8 | 1 | 30 | 273 | 784 | 123 | 123 | 123 | 231 |
| 4 | 9 | 1 | 32 | 256 | 0 | 123 | 123 | 213 | 213 |
| 5 | 36 | 1 | 26 | 169 | 144 | 123 | 123 | 213 | 132 |
| 6 | -72 | 1 | 24 | 132 | 64 | 123 | 123 | 213 | 231 |
| 7 | 6 | 1 | 24 | 144 | 256 | 123 | 123 | 231 | 231 |
| 8 | -12 | 1 | 18 | 33 | 16 | 123 | 123 | 231 | 312 |
| 9 | 36 | 1 | 20 | 64 | 0 | 123 | 213 | 132 | 231 |

**Table 13** All the independent terms for the norm kernel. The suffix $l$ for $k$ runs from 0 to 3. The set of $(i,j,k)$ means the representative of $G_{1i}G_{2j}G_{3k}$.

Unfortunately, we cannot find the common factor automatically by the programming code above mentioned. We have already presented the more advanced method in Ref. [144], in which we pointed out that the permutation property of nucleons among the 3 $\alpha$ clusters is naturally found out. Note that the important parameters $x, x'$ relating to the $\alpha$ condensation strength are defined in the range of $0 < x, x' < 1$. In the case of $x, x' \to 0$, the system goes to shell-model state, on the other hand, for $x, x' \to 1$ the system shows a free $\alpha$ gas aggregate.

**(2) The integrand with distance parameter** Let us assume $2\alpha - \alpha$ system. The distance parameter $d$ between $2\alpha$ and $\alpha$ clusters is introduced; namely, In Eq.(110) the matrix elements of overlapping are explicitly written by

$$G_{ij} = \exp\left\{-\frac{1}{4b^2}\left(\bm{R}_i - \bm{R}'_j - \frac{1}{3}\bm{d} + \frac{1}{3}\bm{d}'\right)^2\right\}$$
$$\text{for } i,j \leq 2$$

$$G_{i3} = \exp\left\{-\frac{1}{4b^2}\left(\bm{R}_i - \bm{R}'_3 - \frac{1}{3}\bm{d} - \frac{2}{3}\bm{d}'\right)^2\right\}$$
$$\text{for } i \leq 2$$

$$G_{3j} = \exp\left\{-\frac{1}{4b^2}\left(\bm{R}_3 - \bm{R}'_j + \frac{2}{3}\bm{d} + \frac{1}{3}\bm{d}'\right)^2\right\}$$
$$\text{for } j \leq 2$$

$$G_{33} = \exp\left\{-\frac{1}{4b^2}\left(\bm{R}_3 - \bm{R}'_3 + \frac{2}{3}\bm{d} - \frac{2}{3}\bm{d}'\right)^2\right\} \quad (119)$$

Concerning Eq.(113) related to $(G_{11}G_{22}G_{33})^4$, we write the power part in exponential as

$$\left\{-\left(\tilde{R}_d\right)^t \begin{pmatrix} (\tilde{A}) & (\tilde{B}) & (\tilde{D}_{11}) & (\tilde{D}_{12}) \\ (\tilde{B})^t & (\tilde{A}') & (\tilde{D}_{21}) & (\tilde{D}_{22}) \\ (\tilde{D}_{11})^t & (\tilde{D}_{21})^t & d_{11} & d_{12} \\ (\tilde{D}_{12})^t & (\tilde{D}_{22})^t & d_{21} & d_{22} \end{pmatrix} (\tilde{R}_d)\right\}, \quad (120)$$

where matrices are given by

$$\left(\tilde{R}_d\right)^t = (\bm{R}_1, \bm{R}_2, \bm{R}_3, \bm{R}'_1, \bm{R}'_2, \bm{R}'_3, \bm{d}, \bm{d}')$$

$$\left(\tilde{D}_{11}\right)^t = \frac{1}{b^2}\left(-\frac{1}{3}, -\frac{1}{3}, \frac{2}{3}\right)$$

$$\left(\tilde{D}_{12}\right)^t = \frac{1}{b^2}\left(\frac{1}{3}, \frac{1}{3}, -\frac{2}{3}\right)$$

$$\left(\tilde{D}_{21}\right)^t = \frac{1}{b^2}\left(\frac{1}{3}, \frac{1}{3}, -\frac{2}{3}\right)$$

$$\left(\tilde{D}_{22}\right)^t = \frac{1}{b^2}\left(-\frac{1}{3}, -\frac{1}{3}, \frac{2}{3}\right)$$

and the other scalars are

$$d_{11} = -d_{12} = -d_{21} = d_{22} = \frac{6}{9b^2}. \quad (121)$$

The $D$ matrices and the scalars are independent of the permutation property. The matrices $(\tilde{A}), (\tilde{A}')$ and $(\tilde{B})$ are common to those in Eq. 113.

The multiple integral with respect to the parameters, $\bm{R}_1, \bm{R}_2, \bm{R}_3, \bm{R}'_1, \bm{R}'_2, \bm{R}'_3$ where the parameter $\bm{d}$ and $\bm{d}'$ remain being not carried out the integral is easily obtained because the integrand is written by the summation of the Gaussian form. The result of the first term is as follows:

$$N_1(\beta, \beta') \exp\left\{-(\bm{d}, \bm{d}')\begin{pmatrix} \tilde{\Delta}_{11} & \tilde{\Delta}_{12} \\ \tilde{\Delta}_{21} & \tilde{\Delta}_{22} \end{pmatrix}\begin{pmatrix} \bm{d} \\ \bm{d}' \end{pmatrix}\right\}, \quad (122)$$

where

$$\tilde{\Delta}_{ij} = \frac{\begin{vmatrix} (\tilde{A}) & (\tilde{B}) & (\tilde{D}_{1j}) \\ (\tilde{B})^t & (\tilde{A}') & (\tilde{D}_{2j}) \\ (\tilde{D}_{1i})^t & (\tilde{D}_{2i})^t & d_{ij} \end{vmatrix}}{\begin{vmatrix} (\tilde{A}) & (\tilde{B}) \\ (\tilde{B})^t & (\tilde{A}') \end{vmatrix}}.$$

The analytic form of the numerator should be considered as well as that of the denominator, where the same procedure is applied to how to analytically derive the numerator. Using the representation of double tilde followed by Eq.(114), we rewrite the matrix element $\Delta_{11}$ as

$$\tilde{\tilde{\Delta}}_{11} = \frac{1}{9b^2} \frac{\begin{vmatrix} \left(\tilde{\tilde{A}}\right) & \left(\tilde{\tilde{B}}\right) & \left(\tilde{\tilde{D}}_{11}\right) \\ \left(\tilde{\tilde{B}}\right)^t & \left(\tilde{\tilde{A}}'\right) & \left(\tilde{\tilde{D}}_{21}\right) \\ \left(\tilde{\tilde{D}}'_{11}\right)^t & \left(\tilde{\tilde{D}}'_{21}\right)^t & 6 \end{vmatrix}}{\begin{vmatrix} \left(\tilde{\tilde{A}}\right) & \left(\tilde{\tilde{B}}\right) \\ \left(\tilde{\tilde{B}}\right)^t & \left(\tilde{\tilde{A}}'\right) \end{vmatrix}}, \quad (123)$$

where

$$\left(\tilde{D}_{11}\right)^t = (-1, -1, 2)$$
$$\left(\tilde{D}_{21}\right)^t = (1, 1, -2)$$
$$\left(\tilde{D}'_{11}\right)^t = x(-1, -1, 2)$$
$$\left(\tilde{D}'_{21}\right)^t = x'(1, 1, -2).$$

The explicit form of the numerator in Eq.(123) is

$$\begin{vmatrix} 1 & 0 & 0 & -x' & 0 & 0 & -1 \\ 0 & 1 & 0 & 0 & -x' & 0 & -1 \\ 0 & 0 & 1 & 0 & 0 & -x' & 2 \\ -x & 0 & 0 & 1 & 0 & 0 & 1 \\ 0 & -x & 0 & 0 & 1 & 0 & 1 \\ 0 & 0 & -x & 0 & 0 & 1 & -2 \\ -x & -x & 2x & x' & x' & -2x' & 6 \end{vmatrix}. \quad (124)$$

The same procedure as that of the original THSR derivation is carried out. As a consequence, the analytical form is written by

$$N_{all}(x, x' : \boldsymbol{d}, \boldsymbol{d}') = f(x, x') \sum_{k=1}^{9} f_k^{-3/2}(xx') \sum_n W_{kn}^{(1)} g_{kn}^{(1)}, \quad (125)$$

where

$$g_{kn}^{(1)} = \exp\left\{-\frac{1}{f_k(xx')} (\boldsymbol{d}, \boldsymbol{d}') \begin{pmatrix} \Delta_{11}^{(kn)} & \Delta_{12}^{(kn)} \\ \Delta_{21}^{(kn)} & \Delta_{22}^{(kn)} \end{pmatrix} \begin{pmatrix} \boldsymbol{d} \\ \boldsymbol{d}' \end{pmatrix}\right\}.$$

The coefficients $W_{kn}^{(1)}$ are listed in Table 14, where $W_k = \sum_n W_{kn}^{(1)}$ is satisfied for each type of $k$ in the original THSR ansatz. The property of the function $\Delta_{ij}^{(kn)}$ with

| $k$ | $n$ | $W_{kn}^{(1)}$ | The type of $\Delta_{ij}^{(kn)}$ | $\uparrow\uparrow$ | $\uparrow\downarrow$ | $\downarrow\uparrow$ | $\downarrow\downarrow$ |
|---|---|---|---|---|---|---|---|
| 1 | 1 | 1 | $s$ | 123 | 123 | 123 | 123 |
| 2 | 1 | -4 | $s$ | 123 | 123 | 123 | 213 |
|   | 2 | -6 | $s$ | 123 | 123 | 123 | 132 |
|   | 3 | -2 | $s$ | 123 | 132 | 132 | 132 |
| 3 | 1 | 6 | $s$ | 123 | 123 | 123 | 231 |
|   | 2 | 2 | $s$ | 123 | 231 | 231 | 231 |
| 4 | 1 | 3 | $s$ | 123 | 123 | 213 | 213 |
|   | 2 | 6 | $s$ | 123 | 123 | 132 | 132 |
| 5 | 1 | 18 | $s$ | 123 | 123 | 213 | 132 |
|   | 2 | 6 | $s_0$ | 123 | 123 | 132 | 321 |
|   | 3 | 6 | $ns$ | 123 | 132 | 132 | 231 |
|   | 4 | 6 | $ns$ | 123 | 132 | 132 | 312 |
| 6 | 1 | -18 | $s$ | 123 | 123 | 213 | 231 |
|   | 2 | -12 | $ns_x$ | 123 | 123 | 132 | 231 |
|   | 3 | -12 | $ns_{x'}$ | 123 | 123 | 132 | 312 |
|   | 4 | -12 | $s$ | 123 | 213 | 132 | 132 |
|   | 5 | -6 | $s$ | 123 | 132 | 132 | 321 |
|   | 6 | -6 | $ns$ | 123 | 132 | 231 | 231 |
|   | 7 | -6 | $ns$ | 123 | 132 | 312 | 312 |
| 7 | 1 | 6 | $s_0$ | 123 | 123 | 231 | 231 |
| 8 | 1 | -6 | $s_0$ | 123 | 123 | 231 | 312 |
|   | 2 | -6 | $s$ | 123 | 132 | 231 | 312 |
| 9 | 1 | 12 | $ns_x$ | 123 | 213 | 132 | 231 |
|   | 2 | 12 | $ns_{x'}$ | 123 | 213 | 132 | 312 |
|   | 3 | 12 | $s$ | 123 | 132 | 231 | 321 |

**Table 14** All the independent terms for the norm kernel. The set of $(i, j, k)$ means the representative of $G_{1i}G_{2j}G_{3k}$. The type of $\Delta_{ij}^{(kn)}$ is explained in text.

respect to the polynomial of $x, x'$ is as follows:

$$\text{type } s : \sum_l c_l (xx')^l + (x + x') \sum_l c'_l (xx')^l$$

$$\text{type } s_0 : \sum_l c_l (xx')^l$$

$$\text{type } ns : \sum_l c_l (xx')^l + x \sum_l c'_l (xx')^l + x' \sum_l c''_l (xx')^l$$

$$\text{type } ns_x : \sum_l c_l (xx')^l + x \sum_l c'_l (xx')^l$$

$$\text{type } ns_{x'} : \sum_l c_l (xx')^l + x' \sum_l c'_l (xx')^l .$$

As for this primitive and tedious method, it is difficult to carry out the unique factorization, but easy to do an exact angular-momentum projection because of analytical expression of $(\boldsymbol{d} \cdot \boldsymbol{d}')$.

**(3) The integral operator with the distant parameter** Now we consider the interaction between two aggregates of $\alpha$ clusters. Here, we take interacting system of $2\alpha$ as one aggregate and $\alpha$ as another, which is the simplest aggregate. It is reasonable to introduce two kinds of condensation strengths. Eq.(107) has more complex forms as

$$\hat{C}(\beta_1, \beta_2) = \frac{1}{(\pi\beta_1^2)^{2 \times 3/2}} \frac{1}{(\pi\beta_2^2)^{3/2}}$$
$$\times \int \prod_{i=1}^{2} d\boldsymbol{R}_i \exp\left\{-\frac{(\boldsymbol{R}_i - \frac{1}{3}\boldsymbol{d})^2}{\beta_1^2}\right\}$$
$$\times \int d\boldsymbol{R}_3 \exp\left\{-\frac{(\boldsymbol{R}_3 + \frac{2}{3}\boldsymbol{d})^2}{\beta_2^2}\right\}. \quad (126)$$

In the same way as the previous discussions, the Eq.(123) is

$$\left\{-\left(\tilde{R}_d\right)^t \begin{pmatrix} (\tilde{A}) & (\tilde{B}) & (\tilde{D}_{11}) & (\tilde{D}_{12}) \\ (\tilde{B})^t & (\tilde{A}') & (\tilde{D}_{21}) & (\tilde{D}_{22}) \\ (\tilde{D}_{11})^t & (\tilde{D}_{21})^t & d_{11} & d_{12} \\ (\tilde{D}_{12})^t & (\tilde{D}_{22})^t & d_{21} & d_{22} \end{pmatrix} (\tilde{R}_d)\right\}, \quad (127)$$

where matrices are,

$$\left(\tilde{R}_d\right)^t = (\boldsymbol{R}_1, \boldsymbol{R}_2, \boldsymbol{R}_3, \boldsymbol{R}'_1, \boldsymbol{R}'_2, \boldsymbol{R}'_3, \boldsymbol{d}, \boldsymbol{d}')$$

$$\left(\tilde{D}_{11}\right)^t = \left(-\frac{1}{3\beta_1^2}, -\frac{1}{3\beta_1^2}, \frac{2}{3\beta_2^2}\right)$$

$$\left(\tilde{D}_{12}\right)^t = (0, 0, 0)$$

$$\left(\tilde{D}_{21}\right)^t = (0, 0, 0)$$

$$\left(\tilde{D}_{22}\right)^t = \left(-\frac{1}{3\beta_1'^2}, -\frac{1}{3\beta_1'^2}, \frac{2}{3\beta_2'^2}\right)$$

The scalars $d_{ij}$ are as follows

$$d_{11} = \frac{2}{9\beta_1^2} + \frac{4}{9\beta_2^2}, \; d_{12} = d_{21} = 0, \; d_{22} = \frac{2}{9\beta_1'^2} + \frac{4}{9\beta_2'^2}. \quad (128)$$

First of all, we should scrutinize two-strength THSR ansatz as the case of $\boldsymbol{d} = 0$. Namely, fot the first term of $(G_{11}G_{22}G_{33})^4$ the deteminant is given by

$$\begin{vmatrix} 1 & 0 & 0 & -x'_1 & 0 & 0 \\ 0 & 1 & 0 & 0 & -x'_1 & 0 \\ 0 & 0 & 1 & 0 & 0 & -x'_2 \\ -x_1 & 0 & 0 & 1 & 0 & 0 \\ 0 & -x_1 & 0 & 0 & 1 & 0 \\ 0 & 0 & -x_2 & 0 & 0 & 1 \end{vmatrix}, \quad (129)$$

where

$$x_i = \frac{\frac{\beta_i^2}{b^2}}{1 + \frac{\beta_i^2}{b^2}}, \quad x'_i = \frac{\frac{\beta_i'^2}{b^2}}{1 + \frac{\beta_i'^2}{b^2}} \quad \text{with } i = 1, 2.$$

The function of $f(x, x')$ should be also exchanged by $f(x_1, x_2, x'_1, x'_2)$ as

$$f(x_1, x_2, x'_1, x'_2) = \left\{\left(\frac{2-x_1}{1-x_1}\right)\left(\frac{2-x'_1}{1-x'_1}\right)\right\}^{-3 \times 3/2}$$
$$\times \left\{\left(\frac{2-x_2}{1-x_2}\right)\left(\frac{2-x'_2}{1-x'_2}\right)\right\}^{-3/2}.$$

After arranging all the terms, the number of the independent terms is 25, which is exactly the same as that of the previous section. This is conserved in the case of $\boldsymbol{d} \neq 0$. However, the function of $f_k(xx')$ is written by the combination of 4 kinds of $x$ parameter. As for the case of $\boldsymbol{d} \neq 0$, the function for each term is analytically written by the same form as that of Eq.(122). The numerator $\Delta_{11}$ for the term of $(G_{11}G_{22}G_{33})^4$ is given by dividing the final column as

$$\begin{pmatrix} -\frac{1}{\beta_1^2} \\ -\frac{1}{\beta_1^2} \\ \frac{2}{\beta_2^2} \\ \frac{2}{\beta_2^2} \\ 0 \\ 0 \\ 0 \\ \frac{2}{\beta_1^2} + \frac{4}{\beta_2^2} \end{pmatrix} = \frac{1}{\beta_1^2}\begin{pmatrix} -1 \\ -1 \\ 0 \\ 0 \\ 0 \\ 0 \\ 0 \\ 2 \end{pmatrix} + \frac{1}{\beta_2^2}\begin{pmatrix} 0 \\ 0 \\ 2 \\ 2 \\ 0 \\ 0 \\ 0 \\ 4 \end{pmatrix}. \quad (130)$$

Thus, we can easily write the numerator such as

$$\frac{1}{9\beta_1^2}\begin{vmatrix} 1 & 0 & 0 & -x_1' & 0 & 0 & -1 \\ 0 & 1 & 0 & 0 & -x_1' & 0 & -1 \\ 0 & 0 & 1 & 0 & 0 & -x_2' & 0 \\ -x_1 & 0 & 0 & 1 & 0 & 0 & 0 \\ 0 & -x_1 & 0 & 0 & 1 & 0 & 0 \\ 0 & 0 & -x_2 & 0 & 0 & 1 & 0 \\ -1+x_1 & -1+x_1 & 2+2x_2 & 0 & 0 & 0 & 2 \end{vmatrix} +$$

$$\frac{1}{9\beta_2^2}\begin{vmatrix} 1 & 0 & 0 & -x_1' & 0 & 0 & 0 \\ 0 & 1 & 0 & 0 & x_1' & 0 & 0 \\ 0 & 0 & 1 & 0 & 0 & -x_2' & 2 \\ -x_1 & 0 & 0 & 1 & 0 & 0 & 0 \\ 0 & -x_1 & 0 & 0 & 1 & 0 & 0 \\ 0 & 0 & -x_2 & 0 & 0 & 1 & 0 \\ -1+x_1 & -1+x_1 & 2+2x_2 & 0 & 0 & 0 & 4 \end{vmatrix}, (131)$$

where we use the relation of

$$\frac{1}{1+\frac{\beta_i^2}{b^2}} = 1 - x_i, \quad \text{with } i = 1, 2$$

for the final row. Thus, we can give the analytical expression of the case of two-center THSR ansatz after obtaining the independent terms. The property is exactly same as that of the introduction of distant parameter in the single THSR ansatz. This is because the special role is given for two $\alpha$-cluster aggregates. Nevertheless, the Pauli principle through all the nucleons is completely carried out.

**(4) Partial integral for the THSR ansatz** According to the previous discussions, the partial integral can be easily done in the THSR framework. Fixing $\boldsymbol{R}_3, \boldsymbol{R}_3'$ to the origin ($=0$), we carry out the integral with respect to $\boldsymbol{R}_1, \boldsymbol{R}_2, \boldsymbol{R}_1', \boldsymbol{R}_2'$. For the first term of $(G_{11}G_{22}G_{33})^4$, the determinant is written by

$$\begin{vmatrix} 1 & 0 & -x' & 0 \\ 0 & 1 & 0 & -x' \\ -x & 0 & 1 & 0 \\ 0 & -x & 0 & 1 \end{vmatrix}. \quad (132)$$

The resultant norm kernel is as follows:

$$N_{all}^{(p)}(x,x') = f^{(p)}(x,x')\sum_{k=1}^{20} W_k^{(p)}\left\{f_k^{(p)}(xx')\right\}^{-3/2}, \quad (133)$$

| $k$ | $n$ | $W_{kn}^{(p)}$ | $c_{kl}^{(p)}$ 0 | 1 | 2 | $\uparrow\uparrow$ | $\uparrow\downarrow$ | $\downarrow\uparrow$ | $\downarrow\downarrow$ |
|---|---|---|---|---|---|---|---|---|---|
| 1 | 1 | 1 | 1 | 32 | 256 | 123 | 123 | 123 | 123 |
| 2 | 1 | 2 | 1 | 16 | 64 | 123 | 123 | 123 | 132 |
|   | 2 | -6 | 1 | 25 | 144 | 123 | 123 | 123 | 132 |
|   | 3 | -2 | 1 | 17 | 16 | 123 | 132 | 132 | 132 |
| 3 | 1 | 6 | 1 | 19 | 81 | 123 | 123 | 123 | 231 |
|   | 2 | 2 | 1 | 11 | 1 | 123 | 231 | 231 | 231 |
| 4 | 1 | 3 | 1 | 16 | 0 | 123 | 123 | 213 | 213 |
| 5 | 1 | 18 | 1 | 15 | 25 | 123 | 123 | 213 | 132 |
|   | 2 | 6 | 1 | 18 | 81 | 123 | 123 | 132 | 321 |
|   | 3 | 12 | 1 | 11 | 9 | 123 | 132 | 132 | 231 |
|   | 4 | -6 | 1 | 13 | 36 | 123 | 132 | 132 | 321 |
| 6 | 1 | -18 | 1 | 13 | 4 | 123 | 123 | 213 | 231 |
|   | 2 | -24 | 1 | 14 | 36 | 123 | 123 | 132 | 231 |
|   | 3 | -12 | 1 | 12 | 4 | 123 | 213 | 132 | 132 |
|   | 4 | -12 | 1 | 9 | 4 | 123 | 132 | 231 | 231 |
| 7 | 1 | 6 | 1 | 12 | 16 | 123 | 123 | 231 | 231 |
| 8 | 1 | -6 | 1 | 10 | 9 | 123 | 123 | 231 | 312 |
|   | 2 | -6 | 1 | 7 | 1 | 123 | 132 | 231 | 312 |
| 9 | 1 | 24 | 1 | 10 | 0 | 123 | 213 | 132 | 231 |
|   | 2 | 12 | 1 | 11 | 1 | 123 | 132 | 231 | 321 |

**Table 15** All the independent terms for the norm kernel. The suffix $l$ for $k$ runs from 0 to 2. The set of $(i,j,k)$ means the representative of $G_{1i}G_{2j}G_{3k}$.

where

$$f^{(p)}(x,x') = \left\{\left(\frac{2-x}{1-x}\right)\left(\frac{2-x'}{1-x'}\right)\right\}^{-2\times 3/2},$$

$$f_k^{(p)}(xx') = \sum_{l=0}^{2} c_{kl}^{(p)}\left(-\frac{xx'}{16}\right)^l.$$

The coefficients in this ansatz are listed in Table 15. The number of the independent terms is 20, which is larger than that of the original THSR ansatz because of the break of the symmetry. However, comparing the case of the introduction of the distant parameter $\boldsymbol{d}$, the number of the independent terms is degenerated.

In summary, on the basis of the complete consideration of the Pauli principle, the THSR ansatz has been successfully applied to various cluster systems. The positions of clusters are not fixed, then they can freely move in the container. The introduction of the dimensionless parameters $x, x'$ in the THSR ansatz plays an decisive role in clarifying the property of the cluster aggregates.

Here, we mainly present two kinds of extensions including distant parameter $\boldsymbol{d}$. The first one is related to the inner part in the single container. Its introduction enables us to find out the even-odd difference of two-cluster system in the container. The other is introduced to consider the interaction between two containers. Analytical derivation of the exchange kernel needs to arrange a tremendous number of the permutation coming prom the Pauli principle. In addition, the multi-integral concerning the number of clusters looks difficult task. In order to avoid such a cumbersomeness, numerical integral has been proposed by some authors. We cannot deny such efforts because ,for instance, even 10 $\alpha$ system has $(10!)^4 \approx 1.73 \times 10^{26}$ terms. Nevertheless, the analytical derivation of the exchange kernel makes an important base to scrutinize the physical property of nuclear cluster system.

Naturally, as the next step, we should present a new method to treat resonant states or scattering states within the framework of the THSR ansatz. One of the key notes is to express explicitly the real variables, $\boldsymbol{r}_1, \cdots, \boldsymbol{r}_n$ with nucleon number $n$. We are now going to prepare a report on this subject.

# 6 Evolution of cluster structure

## 6.1 Duality property of mean-field states and formation of cluster states

### 6.1.1 Duality property of shell-model states with good SU(3) symmetry

As we discussed in Sec. 3.1, the wave functions of the ground states of the self-conjugate $4n$ nuclei ("$N=Z=$ even" nuclei) have duality property. The duality property means that the wave function has both the character of mean-field-type structure and that of cluster structure. The ground-state wave functions of these nuclei up to at least around 40 have dominant component of good SU(3) symmetry. According to the Bayman-Bohr theorem [145], the wave function with good SU(3) symmetry has dual character of mean-field-type structure and cluster structure. A typical example which we referred to in previous sections is the duality property of the double closed-shell wave function $\det|(0s)^4(0p)^{12}|$ of $^{16}$O. This wave function with $(\lambda,\mu)=(0,0)$ symmetry of SU(3) can be rewritten in the form of cluster-model wave function of $^{12}$C + $\alpha$ configuration,

$$\frac{1}{\sqrt{16!}}\det|(0s)^4(0p)^{12}|$$
$$= N_g \frac{1}{\sqrt{C_{16}^4}} \mathcal{A}\{[R_4(\boldsymbol{r},3\nu_N)\phi_{(0,4)}(^{12}\mathrm{C})]_{(0,0)}\phi(\alpha)\} \quad (134)$$

$$g(\boldsymbol{r}(^{16}\mathrm{O}),16\nu_N)[R_4(\boldsymbol{r},3\nu_N)\phi_{(0,4)}(^{12}\mathrm{C})]_{(0,0)}$$
$$= \sum_{L=}^{0,2,4} C_L [R_{4,L}(\boldsymbol{r},3\nu_N)\phi_{(0,4)L}(^{12}\mathrm{C})]_{J=0}. \quad (135)$$

$$g(\boldsymbol{X},\gamma) = (\frac{2\gamma}{\pi})^{3/4}\exp(-\gamma \boldsymbol{X}^2), \quad \boldsymbol{r}(^{16}\mathrm{O}) = \frac{1}{16}\sum_{i=1}^{16}\boldsymbol{r}_i,$$

$$C_{16}^4 = \frac{16!}{12!4!}, C_L = \langle (4,0)L,(0,4)L||(0,0)0\rangle. \quad (136)$$

Here $\boldsymbol{r}$ is the relative coordinate between $^{12}$C and $\alpha$ clusters and $R_{4,L}(\boldsymbol{r},3\nu_N)$ stands for the harmonic oscillator function $R_{4,LM}(\boldsymbol{r},3\nu_N) = R_{4,L}(r,3\nu_N)Y_{LM}(\widehat{r})$ with the oscillator quanta $N=4$, namely the nodal number $n=(4-L)/2$, and with the size parameter $3\nu_N$. $\phi_{(0,4)L}(^{12}\mathrm{C})$ is the internal wave function of $^{12}$C with SU(3) symmetry (0,4) and is given by the 0$p$-shell shell-model wave function as

$$\phi_{(0,4)L}(^{12}\mathrm{C})g(\boldsymbol{r}(^{12}\mathrm{C}),12\nu_N) = |(0s)^4(0p)^8;(0,4)L\rangle, \quad (137)$$

$$\boldsymbol{r}(^{12}\mathrm{C}) = \frac{1}{12}\sum_{i=1}^{12}\boldsymbol{r}_i. \quad (138)$$

The relative wave function $R_{4,LM}(\boldsymbol{r},3\nu_N)$ has SU(3) symmetry (4,0). $[R_4(\boldsymbol{r},3\nu_N)\phi_{(0,4)}(^{12}\mathrm{C})]_{(0,0)}$ stands for the SU(3) vector-coupling of $R_{4,L}(\boldsymbol{r},3\nu_N)$ with (4,0) symmetry and $\phi_{(0,4)L}(^{12}\mathrm{C})$ with (0,4) symmetry to the SU(3) symmetry (0,0) by the use of the SU(3) reduced Clebsch-Gordan coefficient $\langle(4,0)L,(0,4)L||(0,0)0\rangle; (4,0)\times(0,4)\to(0,0)$.

The angular-momentum-coupled wave function of $^{12}$C + $\alpha$ configuration can be expressed by the linear combination of the SU(3)-coupled wave function of $^{12}$C + $\alpha$ configuration as

$$\mathcal{A}\{[R_{4,L}(\boldsymbol{r},3\nu_N)\phi_{(0,4)L}(^{12}\mathrm{C})]_{J=0}\phi(\alpha)\} =$$
$$\sum_{(\lambda,\mu)}\langle(4,0)L,(0,4)L||(\lambda,\mu)0\rangle \mathcal{A}\{[R_4(\boldsymbol{r},3\nu_N)\phi_{(0,4)}(^{12}\mathrm{C})]_{(\lambda,\mu)}\phi(\alpha)\}. \quad (139)$$

The number of total oscillator quanta, $Q$, of the double-closed-shell wave function $\det|(0s)^4(0p)^{12}|$ is $Q=8$. Since the $^{16}$O wave function having $Q=8$ is restricted only to this double-closed-shell wave function having (0,0) SU(3) symmetry, we know that there hold

$$\mathcal{A}\{[R_4(\boldsymbol{r},3\nu_N)\phi_{(0,4)}(^{12}\mathrm{C})]_{(\lambda,\mu)}\phi(\alpha)\} = 0$$
$$\text{for } (\lambda,\mu) \neq (0,0). \quad (140)$$

Therefore Eq.(139) can be written as

$$\mathcal{A}\{[R_{4,L}(\boldsymbol{r}, 3\nu_N)\phi_{(0,4)L}(^{12}\text{C})]_{J=0}\phi(\alpha)\} =$$
$$\langle (4,0)L, (0,4)L || (0,0)0 \rangle \mathcal{A}\{[R_4(\boldsymbol{r},3\nu_N)\phi_{(0,4)}(^{12}\text{C})]_{(0,0)}\phi(\alpha)\} \quad (141)$$

Thus the double-closed-shell wave function can be equivalently expressed in three ways by the use of $^{12}\text{C} + \alpha$ cluster-model wave functions, $^{12}\text{C}(L=0) + \alpha(S\text{-wave})$, $^{12}\text{C}(L=2) + \alpha(D\text{-wave})$, and $^{12}\text{C}(L=4) + \alpha(G\text{-wave})$.

Besides the case of the double closed-shell wave function of $^{16}\text{O}$, many other cases of duality property have been discussed rather in detail in many review papers of nuclear clustering [6, 146]. They include the ground-state wave functions of $^{12}\text{C}$, $^{20}\text{Ne}$, $^{28}\text{Si}$, $^{40}\text{Ca}$, and $^{44}\text{Ti}$. The shell-model wave function with duality property is not restricted to the ground state. A typical example is the $4\hbar\omega$ excited state of $^{32}\text{S}$ which is the main component of the superdeformed state and has the configuration with 4 particles raised from $1s0d$-shell to $1p0f$-shell. The intrinsic wave function of this state has the duality character that it is equivalent to the $^{16}\text{O}$-$^{16}\text{O}$ cluster model wave function as shown below,

$$(0,0,0)^4 (1,0,0)^4 (0,1,0)^4 (0,0,1)^4 (1,0,1)^4 \times$$
$$(0,1,1)^4 (0,0,2)^4 (0,0,3)^4$$
$$= n_0 \mathcal{A}\{X_{(0,0,24)}(\boldsymbol{r}_{\text{O-O}}, 8\nu_N)\phi^2(^{16}\text{O})\}\, g(\boldsymbol{r}(^{32}\text{S}), 32\nu_N), \quad (142)$$

$$\boldsymbol{r}(^{32}\text{S}) = \frac{1}{32}\sum_{i=1}^{32} \boldsymbol{r}_i. \quad (143)$$

Here $(n_x, n_y, n_z)$ expresses the Cartesian harmonic oscillator single-particle function with $n_k$ being the number of oscillator quanta along $k$ direction ($k = x, y,$ or $z$). The coordinate $\boldsymbol{r}_{\text{O-O}}$ stands for the relative coordinate between two $^{16}\text{O}$ clusters and $X_{(n_x,n_y,n_z)}(\boldsymbol{r},\gamma)$ stands for the Cartesian harmonic oscillator function of coordinate $\boldsymbol{r}$ with size parameter $\gamma$. This intrinsic wave function belongs to the SU(3) symmetry $(\lambda,\mu) = (24,0)$.

### 6.1.2 Formation of cluster states from mean-field state with duality property

The duality property of the shell-model wave function of the ground state was already known in 1950's among people who studied microscopic cluster model. For example, Perring and Skyrme wrote at the beginning of the abstract of their paper [147] "It is shown that it is possible to write down $\alpha$-particle wave functions for the ground states of $^8\text{Be}$, $^{12}\text{C}$, and $^{16}\text{O}$, which become, when antisymmetrized, identical with shell-model wave functions." The microscopic cluster model study by Wildermuth which he started with his paper at 1958 [148, 149] was on the basis of the duality property of the shell-model wave function of the ground state. The year 1958 was about 10 years after the proposal of the shell-model by Mayer and Jensen at 1949 and therefore the shell-model had already been well established [150]. Wildermuth aimed to discuss the cluster physics which coexists with mean-field physics. Wildermuth considered that the formation of cluster states is due to the activation of cluster degrees of freedom embedded in the ground state due to its duality property. Namely he considered that, since the ground state wave function possesses the duality property, there occurs two kinds of excitation of the ground state, one is the activation of the mean-field degrees of freedom of the ground state yielding the mea-field-type excited states and the other is the activation of the cluster degrees of freedom of the ground state yielding the cluster states. His philosophy can be shown as in Fig. 52.

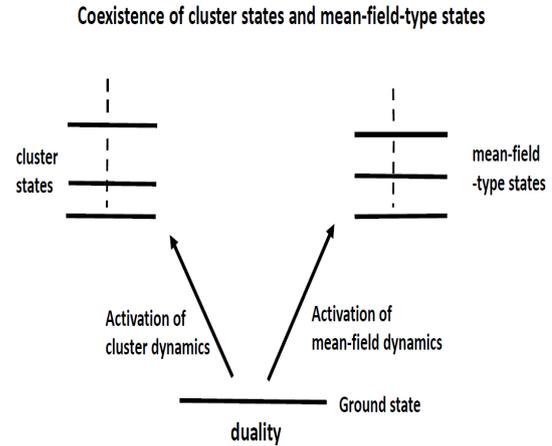

**Fig. 52** Coexistence of cluster states and mean-field-type states due to the duality of the ground state. Ground state is a compact shell-model state but it has degrees of freedom of clustering dynamics together with that of mean-field dynamics. Activation of cluster degrees of freedom yields cluster states while that of mean-field degrees of freedom yields mean-field-type excited states.

Wildermuth assigned cluster structures to many excited states in light nuclei [151]. For example, in $^{16}\text{O}$ he assigned $^{12}\text{C}(0^+) + \alpha$ structure to $0^+(6.06 \text{ MeV})$, $2^+(6.91 \text{ MeV})$, $4^+(10.36 \text{ MeV})$, $1^-(9.58 \text{ MeV})$, and $3^-(11.62 \text{ MeV})$. He also assigned $^{12}\text{C}(2^+) + \alpha$ structure to $2^+(9.84 \text{ MeV})$. Wildermuth's assignments are mostly supported by the present nuclear structure study.

In Refs. [6, 146], detailed discussions are given that, in many self-conjugate $4n$ nuclei from $^{12}\text{C}$ up to $^{44}\text{Ti}$, many cluster states are really existent as excited states in accordance with the prediction by the duality property of the ground state. We can say that the existence of cluster states in addition to mean-field-type states is an inevitable consequence of the duality of the ground state.

We discussed the duality property of the excited state by using the example in $^{32}$S given in Eq. (142). The AMD study of Ref. [152] showed that the excitation of the $^{16}$O-$^{16}$O clustering degree of freedom embedded in the $4\hbar\omega$ excited state of $^{32}$S expressed in Eq.(142) yields the molecular resonance states of $^{16}$O-$^{16}$O. This result means that the formation of the $^{16}$O-$^{16}$O molecular resonance states by the evolution of nuclear structure from the $^{32}$S ground state is through two steps, one is the formation of the $4\hbar\omega$ excited state from the ground state by activating mean-field degree of freedom and the other is the formation of the $^{16}$O-$^{16}$O molecular resonance states from the $4\hbar\omega$ excited state by activating the cluster degree of freedom.

Mean-field state can have the duality property when the state has a dominant component of good SU(3) symmetry. Therefore the ground state of heavy nucleus is considered to have no duality property since the SU(3) symmetry of Elliott is considered to be nonexistent in such ground state. The cluster states we know until now except for lighter nuclei than $^{12}$C can be regarded as being formed by the excitation of cluster degree of freedom from some respective mean-field states, but it does not necessarily mean the nonexistence of cluster state which are not related to any mean-field state through the duality nature. In lighter nuclei than $^{12}$C, the formation of the mean field of nucleons in the ground state is not so firm as is seen in the ground state of $^8$Be which has $\alpha + \alpha$ cluster structure.

### 6.1.3 Observed large monopole transition strengths between ground state and cluster states

Cluster states have very different structure from the ground state with mean-field-type structure. For example, the Hoyle state of $^{12}$C is a $3\alpha$ gas-like state and its density is about $1/3$ of the ground state. Also, the 6th $0^+$ state of $^{16}$O is predicted to be a $4\alpha$ gas-like state and its density is calculated to be more dilute [19, 20]. However, the observed strengths of the monopole ($E0$) transitions between cluster states and the ground state are large and comparable with the single-nucleon strength in its order of magnitude, in spite of large difference of structures of initial and final states of the transition [153].

The $E0$ single-nucleon strength [72] is roughly given by $(3/5)R^2$ with $R$ standing for the nuclear radius. It is obtained by calculating $\langle u_f|r^2|u_i\rangle$ by using uniform-density approximation for $u_f(r)$ and $u_i(r)$, $u(r) \approx (3/R^3)^{1/2}$ for $0 \leq r \leq R$. If we adopt $R \approx 3$ fm for $R$ for light nuclei like $^{12}$C and $^{16}$O, we have $(3/5)R^2 \approx 5.4$ fm$^2$. The observed $E0$ value between the Hoyle state and the ground state in $^{12}$C is $5.4 \pm 0.2$ fm$^2$, and that between the first-excited $0^+$ state and the ground state in $^{16}$O is $3.55 \pm 0.21$ fm$^2$. The first-excited $0^+$ state in $^{16}$O is a well-known cluster state which has the structure of $^{12}$C$(0_1^+) + \alpha(S)$ with $\alpha(S)$ standing for an $\alpha$ cluster moving around $^{12}$C$(0_1^+)$ with $S$-wave ($L = 0$).

It looks very contradictory that we have, on one hand, the large difference of structures between cluster states and the ground state with mean-field-type structure, but, on the other hand, we have the large magnitude of the monopole ($E0$) transitions between cluster states and the ground state. The reason why it looks contradictory comes from our way of understanding of cluster states as being described by superpositions of many particle-hole excited configurations which are very much complicated compared with the ground state. The matrix element of the monopole transition operator between the ground state and the cluster state expressed by the superposition of many particle-hole configurations looks to be much smaller than the $E0$ single-nucleon strength. Therefore in order to resolve this seeming contradiction we have to abondon above-described way of understanding of cluster states in the language of mean-field theory.

It was shown in Ref. [72] that the resolution of this seeming contradiction is very easy if we rewrite the ground-state wave function in the form of cluster-model wave function according to the duality property of the ground state. When we express the ground-state wave function in the form of cluster-model wave function, it is immediately clear that the monopole transition between the ground state and a cluster state is nothing but the monopole transition of the inter-cluster relative motion of the ground state to that of the cluster state. For example, we consider the monopole transition in $^{16}$O between the first-excited $0^+$ ($0_2^+$) state and the ground state ($0_1^+$). Because of the duality property the ground state is expressed as $\mathcal{A}\{R_{4,0}(r, 3\nu_N)h_0\}$ with $h_0 = Y_{0,0}(\hat{r})\phi_{(0,4),L=0}(^{12}\text{C})\phi(\alpha)$. On the other hand, since the first-excited $0^+$ state has the $^{12}$C$(0_1^+) + \alpha(S)$ structure, it is expressed as $\mathcal{A}\{\chi_0(r)h_0\}$. Thus the $E0$ transition between two states is the transition between $R_{4,0}(r, 3\nu_N)$ and $\chi_0(r)$. Since the $E0$ transition operator $O(E0,^{16}\text{O})$ can be written as

$$O(E0,^{16}\text{O}) = \frac{1}{2}\sum_{i=1}^{16}(\mathbf{r}_i - \mathbf{r}(^{16}\text{O}))^2$$
$$= O(E0,^{12}\text{C}) + O(E0,\alpha) + \frac{1}{2}\frac{12\times 4}{16}r^2, \quad (144)$$

we see that only the relative-motion part, $(1/2)((12 \times 4)/16)r^2$, contributes to the $E0$ transition. We can easily show that both the operators $O(E0,^{12}\text{C})$ and $O(E0,\alpha)$ do not contribute [72]. The fact that the $E0$ transition reduces to the $E0$ transition of the inter-cluster relative motion means that the $E0$ transition is due to only single degree of freedom, inter-cluster relative motion. Therefore it is very natural that the observed strength of $E0$ transition is comparable to the single-nucleon strength. One may think that the effect of the antisymmetrization operator $\mathcal{A}$ invalidates the above argument. But in Ref. [72], it is shown that the effect of the antisymmetrization operation is small. Actually the result of the analytical calculation of the $E0$ matrix element is given as

$$M(E0, 0_2^+ \leftrightarrow 0_1^+) = \langle ((0s)^4 (0p)^{12}, J=0|O(E0, ^{16}\text{O})| \\ \times \mathcal{A}\{\chi_0(r)[Y_0(\hat{r})\phi_0(^{12}\text{C})]_{J=0}\ \phi(\alpha)\}g(\mathbf{r}(^{16}\text{O}), 16\nu)\rangle \tag{145}$$

$$= \frac{1}{2}\sqrt{\frac{\tau_{0,4}}{\tau_{0,6}}}\eta_6 \langle R_{40}(r,\nu)|r^2|R_{60}(r,\nu)\rangle, \tag{146}$$

$$\tau_{L,N} = \langle \Psi_{L,N}|\mathcal{A}\{\Psi_{L,N}\}\rangle, \tag{147}$$

$$\Psi_{L,N} = R_{N,L}(r_{\text{C}-\alpha}, 3\nu)[Y_L(\hat{r}_{\text{C}-\alpha})\phi_L(^{12}\text{C})]_{J=0}\phi(\alpha), \tag{148}$$

$$\mathcal{A}\{\chi_0(r)[Y_0(\hat{r})\phi_0(^{12}\text{C})]_{J=0}\ \phi(\alpha)\} = \sum_{N=6}^{\infty} \eta_N (C_N \mathcal{A}\{\Psi_{0,N}\}), \tag{149}$$

$$||C_N \mathcal{A}\{\Psi_{0,N}\}|| = 1. \tag{150}$$

The quantity $\tau_{L,N}$ represents the effect of the antisymmetrization and actually is fairly smaller than unity in general for non-large $N$. However, in the above analytical formula, quantities $\tau_{0,N}$ appear in the form of ratio, $\tau_{0,4}/\tau_{0,6}$, and the magnitude of the ratio is close to unity, which implies that the effect of antisymmetrization has only little influence on the $M(E0)$ value. The quantity $\eta_6$ is the coefficient of the $2\hbar\omega$-jump component contained in $\mathcal{A}\{\chi_0(r)[Y_0(\hat{r})\phi_0(^{12}\text{C})]_{J=0}\ \phi(\alpha)\}$, and its magnitude is around 0.4. Note that $\eta_6$ is not percentage quantity, $(\eta_6)^2$. The $E0$ matrix element of the relative motion, $\langle R_{40}(r,\nu)|r^2|R_{60}(r,\nu)\rangle$ is larger than the corresponding $E0$ matrix elements of the single-nucleon motion, $\langle R_{00}(r,\nu)|r^2|R_{20}(r,\nu)\rangle$ and $\langle R_{11}(r,\nu)|r^2|R_{31}(r,\nu)\rangle$, by about 50%. We thus see that the order of magnitude of $M(E0, 0_2^+ \to 0_1^+)$ is the same as the single-nucleon strength.

In $^{16}$O, the $E0$ transition from the second-excited $0^+$ ($0_3^+$) state to the ground state has been also observed and its observed strength is large as 4.03 fm$^2$. The dominant configuration of this excited state is the cluster structure of $^{12}\text{C}(2_1^+) + \alpha(D)$ and the double-closed-shell wave function of the ground state can be expressed, because of its duality property, by the cluster-model wave function of the configuration of $^{12}\text{C}(2_1^+) + \alpha(D)$ as we explained in Sec. 6.1.1. Therefore the $E0$ transition in this case can also be regarded as being the $E0$ transition of the relative motion between $^{12}\text{C}(2_1^+)$ cluster and $\alpha$ cluster. Here also the $E0$ transition is due to only single degree of freedom, inter-cluster relative motion and hence it is very natural that the observed strength of $E0$ transition is comparable to the single-nucleon strength. In Ref. [72], there is shown the analytical formula of the $E0$ matrix element between $0_3^+$ and $0_1^+$ states which is quite similar to that between $0_2^+$ and $0_1^+$ states.

The numerical values of $M(E0, 0_2^+ \to 0_1^+)$ and $M(E0, 0_3^+ \to 0_1^+)$ calculated by the analytical formulae mentioned above with suitable parameter values are 1.97 fm$^2$ and 3.89 fm$^2$, respectively. These values are somewhat smaller than the observed values although the order of magnitudes are well reproduced. In Ref. [72] it was shown that the inclusion of the ground-state correlation into the ground-state wave function makes the reproduction of the observed values by theory fairly satisfactory. The ground-state correlation adopted is that of the inter-cluster relative wave function; namely the $^{12}$C-$\alpha$ relative wave function of the ground state is not simply $R_{4,L}(r, 3\nu)$ but has contributions from higher $N$ components, $R_{N,L}(r, 3\nu)$ with $N > 4$. The important point of this result is that the cluster degree-of-freedom due to the duality property induces the ground-state correlation.

There have been many cluster-model calculations of $E0$ transition strengths by many authors. As far as the microscopic wave functions like those of RGM, Brink-GCM, AMD, and THSR are used, and also so far as the OCM is adopted, the duality property of the ground state is automatically taken into account, and hence there have been no difficulty in obtaining the $E0$ transition strengths between the ground state and cluster states with reasonably good agreement with the observed data. What was done in Ref. [72] was the clarification of the reason why the $E0$ transition strengths between the ground state and cluster states have the magnitude comparable with the single-nucleon strength. The most important point is the fact that the $E0$ transition is due to few degrees of freedom, namely the inter-cluster relative motion. Especially, for cluster states composed of two clusters, the number of degrees of freedom is only one. Another important point is the fact that the antisymmetrization operation by $\mathcal{A}$ has only small effect. This point was proved by obtaining the explicit analytical formulas of the $E0$ matrix elements between the ground state and cluster states. An example in the two-cluster system is given in Eq.(146).

In the above we explained that large $E0$ transition strengths comparable with the single-nucleon strength between the ground state and cluster states can be considered as being quite natural because of the duality property of the ground state. Conversely if some excited state has a large $E0$ transition strength to the ground state which is comparable to the single-nucleon strength, we can say that it is quite probable that this state is a cluster state if this state can be judged not to be a mean-field state having one-particle one-hole excited configuration from the ground state. Therefore large $E0$ transition strength is a good probe for identifying cluster states just like large $E2$ transition strength which is a good probe for identifying states with large quadrupole deformation.

### 6.2 Pauli-allowed states of many-cluster systems

#### 6.2.1 Two-cluster Pauli-forbidden states and many-cluster Pauli-allowed states

In the container model, clusters make nonlocalized motion in containers under inter-cluster Pauli repulsion. The inter-cluster Pauli repulsion between clusters $C_p$ and $C_q$ originates from the Pauli-forbidden states $\chi_u^{F,pq}(\mathbf{r}_{pq})$ defined by

$$\mathcal{A}\{\chi_u^{F,pq}(\mathbf{r}_{pq})\phi(C_p)\phi(C_q)\} = 0, \tag{151}$$

where $\phi(C_j)$ is the internal wave functions of cluster $C_j$ ($j = p, q$) and $\mathbf{r}_{pq}$ is the relative coordinate between $C_p$ and $C_q$. In the case of $\alpha + \alpha$ system, the Pauli-forbidden states are harmonic oscillator wave functions $R_{N,\ell}(r_{\alpha\alpha}, \nu_N/2)Y_{\ell,m}(\widehat{r}_{\alpha\alpha})$ with $N = 2n + \ell < 4$ [33]. The wave function of the relative motion between clusters $C_p$ and $C_q$ should be orthogonal to the Pauli-forbidden states. This orthogonality makes the relative wave function to avoid the spatial region where the Pauli-forbidden wave functions are existent. This is the effect of the repulsive force by the Pauli-forbidden states. A little more detailed explanation of the repulsive force effect of the Pauli-forbidden states is as follows. The orthogonality to the Pauli-forbidden states creates the oscillatory behavior of the relative wave function in the inner spatial region where the Pauli-forbidden wave functions are existent. The oscillatory behavior of the wave function means the large value of the kinetic energy. Therefore in order to avoid the large kinetic energy, the amplitude of the wave function becomes small in the inner spatial region. The small amplitude of the relative wave function in the inner spatial region is nothing but the repulsion between two clusters avoiding to come close to each other. The effective repulsive force coming from the orthogonality to the Pauli-forbidden states is called structural repulsive core [154]. The structural repulsive core does not kill the inner-part wave function and so, if the attractive force is strong, the inner-part wave function can have large amplitude in order to gain the potential energy. Its difference from the rigid repulsive core is in many ways. For example at high scattering energy of clusters, high frequency oscillatory behavior of the relative wave function satisfies automatically the orthogonality to the Pauli-forbidden states, which means the fadeaway of the repulsive force effect. Another example which is very important is the weakening of the repulsive effect in three or many cluster systems where inter-cluster attractive force is strengthened by the Borromean effect. In the system of many clusters composed of $C_j$ ($j = 1 \sim n$), every pair of clusters of $C_p$ and $C_q$ ($p, q = 1 \sim n$) should have their relative wave function orthogonal to their Pauli-forbidden states $\chi_u^{F,pq}(\mathbf{r}_{pq})$. It means that the relative wave function $\chi(\boldsymbol{\xi}_1, \cdots, \boldsymbol{\xi}_{n-1})$ of the many-cluster system should be orthogonal to all the Pauli-forbidden states of all pairs of clusters. Here $\boldsymbol{\xi}_k$ ($k = 1 \sim (n-1)$) are relative coordinates among clusters like Jacobi coordinates. In the case of Jacobi coordinates, we have $\boldsymbol{\xi}_k = \mathbf{X}_{k+1} - (1/\sum_{j=1}^k A_j)\sum_{j=1}^k A_j \mathbf{X}_j$, where $\mathbf{X}_j$ and $A_j$ are center-of-mass coordinate and mass number of cluster $C_j$, respectively.

In discussing the effect of the Pauli principle in the many-cluster system, the RGM norm kernel $\mathcal{N}$ [37] is useful which is defined as

$$\begin{aligned}&\mathcal{N}(\mathbf{a}_1, \cdots, \mathbf{a}_{n-1}; \mathbf{a}_1', \cdots, \mathbf{a}_{n-1}') \\ &= \langle \delta(\boldsymbol{\xi}_1 - \mathbf{a}_1) \cdots \delta(\boldsymbol{\xi}_{n-1} - \mathbf{a}_{n-1})\phi(C_1) \cdots \phi(C_n)| \\ &\mathcal{A}\{\delta(\boldsymbol{\xi}_1 - \mathbf{a}_1') \cdots \delta(\boldsymbol{\xi}_{n-1} - \mathbf{a}_{n-1}')\phi(C_1) \cdots \phi(C_n)\}\rangle\end{aligned} \tag{152}$$

The eigen functions and eigen values of $\mathcal{N}$ are denoted as $\chi_i(\boldsymbol{\xi}_1, \cdots, \boldsymbol{\xi}_{n-1})$ and $\mu_i$, respectively, which satisfy [37, 155]

$$\int d\mathbf{a}_1' \cdots d\mathbf{a}_{n-1}' \mathcal{N}(\mathbf{a}_1, \cdots, \mathbf{a}_{n-1}; \mathbf{a}_1', \cdots, \mathbf{a}_{n-1}')$$
$$\chi_i(\mathbf{a}_1', \cdots, \mathbf{a}_{n-1}') = \mu_i \chi_i(\mathbf{a}_1, \cdots, \mathbf{a}_{n-1}), \tag{153}$$
$$\langle \phi(C_1) \cdots \phi(C_n)|\mathcal{A}\{\chi_i(\boldsymbol{\xi}_1, \cdots, \boldsymbol{\xi}_{n-1})\phi(C_1) \cdots \phi(C_n)\}\rangle$$
$$= \mu_i \chi_i(\boldsymbol{\xi}_1, \cdots, \boldsymbol{\xi}_{n-1}). \tag{154}$$

The eigen functions $\{\chi_i(\boldsymbol{\xi}_1, \cdots, \boldsymbol{\xi}_{n-1})\}$ constitute an orthonormal complete set and any wave function $\chi(\boldsymbol{\xi}_1, \cdots, \boldsymbol{\xi}_{n-1})$ can be expanded by this complete set. The eigen functions with zero eigen value $\mu_i = 0$ are Pauli forbidden because $\mathcal{A}\{\chi_i(\boldsymbol{\xi}_1, \cdots, \boldsymbol{\xi}_{n-1})\phi(C_1) \cdots \phi(C_n)\}$ =0 for $\mu_i = 0$, while the eigen functions with non-zero eigen value $\mu_i \neq 0$ are Pauli allowed because $\mathcal{A}\{\chi_i(\boldsymbol{\xi}_1, \cdots, \boldsymbol{\xi}_{n-1})\phi(C_1) \cdots \phi(C_n)\} \neq 0$ for $\mu_i \neq 0$. The functional space spanned by all the Pauli-allowed eigen functions is designated as $S_n$.

Two-cluster Pauli-forbidden state $\chi_u^{F,pq}(\mathbf{r}_{pq})$ is also a many-cluster Pauli-forbidden state because there holds $\mathcal{A}\{\chi_u^{F,pq}(\mathbf{r}_{pq})\phi(C_1) \cdots \phi(C_n)\} = 0$. Then we have

$$0 = \langle \chi_u^{F,pq}(\mathbf{r}_{pq})\phi(C_1) \cdots \phi(C_n)\}|\mathcal{A}\{\chi_i(\boldsymbol{\xi}_1, \cdots, \boldsymbol{\xi}_{n-1}) \tag{155}$$
$$\phi(C_1) \cdots \phi(C_n)\}\rangle = \mu_i \langle \chi_u^{F,pq}(\mathbf{r}_{pq})|\chi_i(\boldsymbol{\xi}_1, \cdots, \boldsymbol{\xi}_{n-1})\rangle. \tag{156}$$

From this relation we know that every Pauli-allowed eigen function $\chi_i(\boldsymbol{\xi}_1, \cdots, \boldsymbol{\xi}_{n-1})$ with $\mu_i \neq 0$ is orthogonal to all the two-cluster Pauli-forbidden state $\{\chi_u^{F,pq}(\mathbf{r}_{pq})\}$. Therefore any Pauli-allowed state which is expressed by a linear combination of Pauli-allowed eigenfunctions $\{\chi_i\}$ is orthogonal to all the two-cluster Pauli-forbidden states $\{\chi_u^{F,pq}(\mathbf{r}_{pq})\}$. Now we consider the functional space $S_n'$ which is spanned by many-cluster wave functions that are orthogonal to all the two-cluster Pauli-forbidden state $\{\chi_u^{F,pq}(\mathbf{r}_{pq})\}$. Clearly the space $S_n'$ contains in it the space $S_n$. However explicit comparisons of $S'_n$ and $S_n$ in many kinds of many-cluster systems have shown that $S'_n$ and $S_n$ are equal to each other ($S'_n = S_n$) [37, 155, 156]. In the OCM for many-cluster system, the total wave function is required to be orthogonal to all the two-cluster Pauli-forbidden state $\{\chi_u^{F,pq}(\boldsymbol{r}_{pq})\}$. Thus the functional space of many-cluster OCM is $S'_n$. The fact that $S'_n = S_n$ ensures that the functional space of many-cluster OCM is the correct Pauli-allowed space $S_n$ spanned by all the Pauli-allowed eigenstates of the norm Kernel $\mathcal{N}$.

**Table 16** Pauli-allowed eigen states of the norm kernel of $4\alpha$ system. Their eigen values are given below their $(\lambda, \mu)$.

| N | $(\lambda, \mu)$ / $\mu_i$ | | | | | | | |
|---|---|---|---|---|---|---|---|---|
| 12 | (0,0) 0.6592 | | | | | | | |
| 13 | (2,1) 0.4614 | | | | | | | |
| 14 | (3,1) 0.3955 | (2,0) 0.6152 | (4,2) 0.3811 | (0,4) 0.3365 | | | | |
| 15 | (6,0) 0.3399 | (4,1) 0.4944 | (2,2) 0.4944 | (0,3) 0.4614 | (3,0) 0.5933 | (5,2) 0.3626 | (3,3) 0.3376 | (6,3) 0.3677 | (2,5) 0.2848 |
| 16 | (5,1) 0.5324 0.4237 | (3,2) 0.4635 | (1,3) 0.5191 | (4,0) 0.5982 0.5628 | (0,2) 0.6633 | (8,1) 0.3208 | (6,2) 0.4363 0.3252 | (4,3) 0.4366 | (2,4) 0.4390 0.3856 |
| | (7,3) 0.3845 | (5,4) 0.3450 | (3,5) 0.2957 | (8,4) 0.4082 | (4,6) 0.2765 | (0,8) 0.2295 | | |
| 17 | (8,0) 0.3816 | (6,1) 0.5434 0.5192 0.4434 | (4,2) 0.5410 0.4979 | (2,3) 0.5730 0.4944 | (5,0) 0.5933 | (3,1) 0.6406 | (1,2) 0.6691 | ······ | ······ |

### 6.2.2 Many-cluster Pauli-allowed state and evolution of cluster structure

In Table 16 we show the Pauli-allowed eigen states of the norm kernel of $4\alpha$ system. Since the eigen states of the norm kernel have, as their quantum numbers, the total number of harmonic oscillator quanta $N$ and SU(3) label $(\lambda, \mu)$, the allowed states are indicated by using $N$ and $(\lambda, \mu)$. Their eigen values $\mu_i$ are given below their $(\lambda, \mu)$. The Pauli-allowed eigen state with lowest harmonic oscillator quanta has $N = 12$. This eigen state which we denote as $\chi_{12,(0,0)}(\boldsymbol{\xi}_1, \boldsymbol{\xi}_2, \boldsymbol{\xi}_3,)$ has the following property,

$$\mathcal{A}\{\chi_{12,(0,0)}(\boldsymbol{\xi}_1, \boldsymbol{\xi}_2, \boldsymbol{\xi}_3)\phi(\alpha_1)\cdots\phi(\alpha_4)\} \propto \det|(0s)^4(0p)^{12}|_{^{16}O}$$

This is of course the duality property of the double closed-shell wave function. When we denote the Pauli-allowed eigen state with $N = 13$ as $\chi_{13,(2,1)}(\boldsymbol{\xi}_1, \boldsymbol{\xi}_2, \boldsymbol{\xi}_3,)$, we have the equality that $\mathcal{A}\{\chi_{13,(2,1)}(\boldsymbol{\xi}_1, \boldsymbol{\xi}_2, \boldsymbol{\xi}_3)\phi(\alpha_1)\cdots\phi(\alpha_4)\}$ is the same as the 1p-1h shell-model wave function $C|(0p)^{-1}(1s0d)^1;(\lambda,\mu)=(2,1)\rangle$. This is the duality property of the 1p-1h shell-model wave function. The formation of this state by the excitation of inter-cluster wave function from the ground state is equivalent to the formation of this state by the excitation of mean-field degree of freedom from the ground state.

As we discussed in Sec. 4, in two-cluster system, the inter-cluster Pauli repulsion works so as to separate two clusters, which means that the inter-cluster Pauli repulsion works so as to separate spatial localization of clusters. The spatial localization-cluster Pauli repulsion does not necessarily generate spatial localization of clusters. A typical example is the Hoyle state in $^{12}$C which has the condensate-like gas structure of three $\alpha$ clusters with no deformation. We also know that it is highly probable that the sixth $0^+$ state at 15.1 MeV in $^{16}$O has condensate-like $4\alpha$ gas structure. Moreover in general there can be variety of deformations in many-cluster systems. We can see this point in Table 16. Except the compact shell-model states for $N = 12$ and 13, there are variety of Pauli-allowed states with different deformations ranging from prolate to oblate. For example, we see many Pauli-allowed states with SU(3) label $(\lambda \neq 0, \mu = 0)$ corresponding to prolate deformation and also many Pauli-allowed states with SU(3) label $(\lambda = 0, \mu \neq 0)$ corresponding to oblate deformation.

The most compact Pauli-allowed state has the lowest number of oscillator quanta $N = N_{\text{lowest}}$ and expresses the shell-model wave function which has the dual-ity property and is a dominant component of the ground state. The Pauli-allowed state with $N = N_{\text{lowest}} + 1$ often expresses a dominant component of a shell-model state with $1p-1h$ character. According to the Ikeda diagram the cluster state which appears first as the excitation energy rises is two-cluster state. In general the cluster state composed of two or more clusters has the wave function which is a linear combination of Pauli-allowed state with many different $N$. The two-cluster state has spatially localized clusters because of inter-cluster Pauli repulsion and its deformation is prolate. As the excitation energy goes up further higher, there appears three-cluster state. Three-cluster states do not necessarily have spatial localization of clusters as is typically seen in the case of the Hoyle state which has $3\alpha$ condensate-like structure. When a three-cluster state has a spatial localization of clusters, its shape or deformation can be both prolate and oblate. The shape or deformation of a three-cluster state is largely influenced by the orthogonality requirement to the states lying in the energy region lower than the three-cluster state. A recent study of this point is given in Ref. [17] in the case of the three-$\alpha$ states above the Hoyle state in $^{12}$C.

### 6.3 Container model and evolution of cluster structure

The so-called Ikeda diagram [23] discusses the necessary condition of the excitation energy for the formation of the cluster state. It says that a cluster state appears around or above the threshold energy of the breakup into constituent clusters. The argument is that if the excitation energy is fairly below the threshold energy, attractive interaction between clusters is no more weak and clusters will lose their identity and dissolve into mean-field-like structure. Fig. 53 displays the Ikeda diagram given in Ref. [23].

This diagram shows what kind of cluster structures are formed along the increase of excitation energy. This diagram may be understood such that they express the evolution path of cluster structure along the increase of excitation energy. In Fig. 54 a new diagram of the cluster formation is also shown, which is along the container picture discussed throughout this review paper. The path of cluster formation along the increase of excitation energy can be described by dilatation of the container, from spatially compact ground states to more dilute cluster structures. This point of view of cluster evolution, which was referred to as the container evolution, was demonstrated in the case of $^{16}$O, in Sec. 5.4. The concept of the container evolution was here generalized in other cluster formations appearing in the Ikeda diagram.

The evolution path of cluster structure along the increase of excitation energy, which can be identified with the container evolution process, is closely associated by the duality property discussed in the previous section. For example, in $^{12}$C, because of the duality property of the ground state, the activation of cluster degree-of-freedom leads to dilatation of container for $3\alpha$ clusters and generates the $3\alpha$ cluster states around or above the $3\alpha$ breakup threshold energy including the Hoyle state. In $^{16}$O, because of the duality property of the ground state, the activation of $^{12}$C + $\alpha$ clustering degree-of-freedom leads to dilatation of container for $^{12}$C + $\alpha$ clusters and generates the $^{12}$C + $\alpha$ cluster states around or above the $^{12}$C + $\alpha$ breakup threshold energy. Then the $^{12}$C cluster of the $^{12}$C + $\alpha$ cluster state evolves into $3\alpha$ cluster structure due to the $^{12}$C duality property forming the $4\alpha$ cluster states around or above the $4\alpha$ breakup threshold energy. We see that each step of the evolution process of the cluster structure, firstly from the ground state to the $^{12}$C + $\alpha$ cluster states and next from the $^{12}$C + $\alpha$ cluster states to the $4\alpha$ cluster states, can be understood according to the duality property of $^{16}$O and $^{12}$C. We here say that, in these nuclei, the evolution of cluster structure is governed by the duality path along the evolution path.

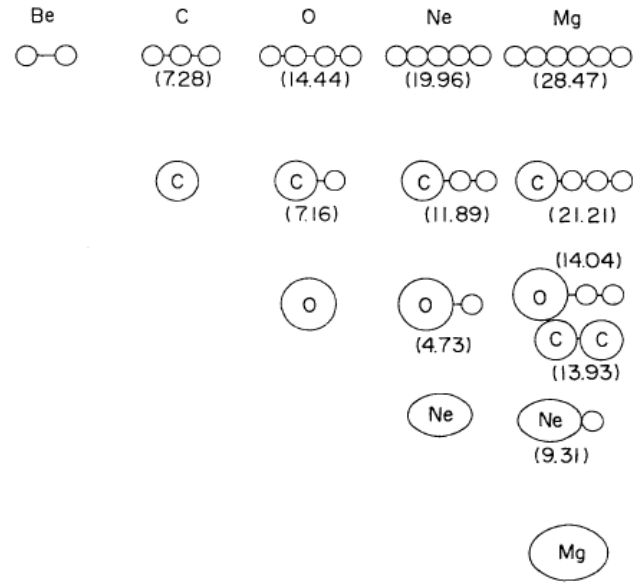

**Fig. 53** Threshold energies of breakup into constituent clusters in light $4n$ self-conjugate nuclei. Threshold energy value in the unit of MeV is given in parenthesis. Figure was reproduced from Ref. [23]. Small circle stands for an $\alpha$ cluster. Linear alignment of three or more clusters does not imply that the cluster structure near this threshold energy has a linear-chain cluster structure.

However, there exist evolution paths of cluster structure which contain non-duality path. A typical example is the formation of the $^{16}$O + $^{16}$O molecular resonance states. The dominant component of the ground state of $^{32}$S has the SU(3) symmetry $(\lambda, \mu) = (0, 12)$ and this component does not have the clustering degree of freedom of $^{16}$O + $^{16}$O configuration whose SU(3) symmetry is $(N, 0)$ with $N \geq 24$. Thus the formation of the $^{16}$O + $^{16}$O molecular resonance states is not made only by the activation of the cluster degree of freedom from the ground state. As we already discussed in the previous sec. 6.1.1, the $^{16}$O + $^{16}$O molecular resonance states are formed by the activation of the $^{16}$O + $^{16}$O cluster degree of freedom from the superdeformed mean-field state which are formed from the ground state by the activation of mean-field degree of freedom. Therefore the Ikeda diagram, as well as the container evolution diagram, does not describe the evolution path of cluster structure which involves non-duality path and does not show the mean-field-type excited state like the superdeformed mean-field state in $^{32}$S.

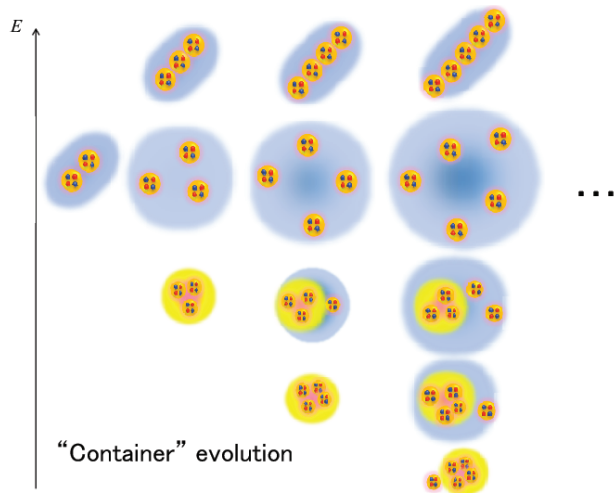

**Fig. 54** New diagram of container evolution along the THSR wave function.

Special attention needs to be paid for the ground states of $^8$Be, $^{20}$Ne, and $^{44}$Ti which are the nuclei having two protons and two neutrons outside the double-closed shell nuclei $^4$He, $^{16}$O, and $^{40}$Ca, respectively. In the case of $^8$Be, as we already pointed out, the ground state is not a mean-field-type state but a cluster state. We can conjecture that the shell-model wave function $|(0s)^4(0p)^4, [44]J = 0\rangle$ is spread into ground state and also into many excited states, and there is no excited state to which this shell-model wave function concentrates as a dominant component. $^8$Be in the Ikeda diagram indicates the existence of a very important change of nuclear structure, namely the formation of the nucleon mean field in the ground state is not energetically favourable in lighter nuclei than $^{12}$C. In the case of $^{20}$Ne, the ground state rotational band constitutes an inversion-doublet together with the negative-parity rotational band with $K^\pi = 0^-$ built upon the $J^\pi = 1^-$ at 5.80 MeV. The intrinsic state of the inversion-doublet bands is the $^{16}$O + $\alpha$ cluster state. Thus it looks similar to the $^8$Be case. However in Refs. [21, 23] the authors argue that the clustering feature is much weaker in the ground band than in the $K^\pi = 0^-$ band. The AMD calculation of $^{20}$Ne of Ref. [157] supports the argument of Refs. [21, 23]. For example the intrinsic wave function of the ground state obtained by AMD does not have density distribution of $^{16}$O + $\alpha$ clustering while that of the $J^\pi = 1^-$ state of the $K^\pi = 0^-$ band shows clearly density distribution of $^{16}$O + $\alpha$ clustering. Although the density distribution of the ground state does not show $^{16}$O + $\alpha$ clustering, the component of $^{16}$O + $\alpha$ cluster wave function contained in the ground state wave function has the squared amplitude of 70%, which is due to the duality property of the shell-model wave function, $|(0s)^4(0p)^{12}(1s0d)^4; [44444](\lambda, \mu) = (8, 0), J = 0\rangle$. In Table 17 we show the properties of the levels of the inversion-doublet bands of $^{20}$Ne obtained by the AMD study of Ref. [157]. We see that the squared amplitude of the $K^\pi = 0^-$ band component $W^J$ in the ground band decreases rather rapidly as the spin value $J$ increases from 70% of the $0^+$ state to 28% of the $8^+$ state. This decrease of the cluster character is due to the increase of the nucleon-spin alignment for larger $J$ [158]. It can be seen in the increase of the expectation value of the two-body spin-orbit interaction $\langle \widehat{V}_{ls} \rangle$ for larger $J$. This feature of the spin-alignment well reflects the mean-field-type character of the ground bend of $^{20}$Ne. On the other hand, in the $K^\pi = 0^-$ band, the $W^J$ value of the $1^-$ state is as large as 95% and the smallest value of $W^J$ for the $9^-$ state is still as large as 70%. This result means that the $K^\pi = 0^-$ band has good clustering character of $^{16}$O + $\alpha$. It is supported by the small expectation values $\langle \widehat{V}_{ls} \rangle$ for all the states of the $K^\pi = 0^-$ band.

In the case of $^{44}$Ti, just like the $^{20}$Ne case, there exists inversion-doublet bands composed of the ground rotational band and the $K^\pi = 0^-$ band built upon the $J^\pi = 1^-$ at 7.0 MeV. The negative-parity band levels were found by the $\alpha$-transfer reaction [94]. In Table 18 we show the squared amplitude of the $^{40}$Ca + $\alpha$ component $W^J$ of each level of the inversion-doublet bands of $^{44}$Ti obtained by the AMD study of Ref. [159]. We see that the $^{40}$Ca + $\alpha$ clustering component is very small in the ground band. Even the largest clustering percentage for the ground $0^+$ state is 40%, which means that non-clustering component is 60% for the ground state. The $10^+$ and $12^+$ states have almost no clustering component. As for the $K^\pi = 0^-$ band, the largest clustering percentage is 56% for the $1^-$ state, which is fairly smaller than 95% of the $1^-$ state of $^{20}$Ne. The small clustering component in the inversion doublet bands is considered to be due to the strong spin-orbit potential and also to the strong attraction from the $^{40}$Ca core nucleus. Although the $^{40}$Ca + $\alpha$ percentage of the ground band is not larger than 40%, the AMD calculation does not show the existence of $J^\pi = 0^+$ excited state whose $^{40}$Ca + $\alpha$ component is larger than 40% and has relative-motion nodal number $n_r$ =6. Here the relative-motion nodal number $n_r$ means the nodal number of the reduced width amplitude $\mathcal{Y}(r)$ defined as

$$\mathcal{Y}(r) = r \langle \frac{\delta(r_{\text{rel}} - r)}{r^2} Y_{0.0}(\widehat{r}_{rel}) \phi(^{40}\text{Ca}) \phi(\alpha) | \Phi^{\text{AMD}} \rangle \quad (158)$$

**Table 17** Squared amplitude of the $^{16}$O + $\alpha$ component $W^J$ and the expectation value of the two-body spin-orbit interaction $\langle \widehat{V}_{ls} \rangle$ (in MeV) for each state of the inversion-doublet rotational bands of $^{20}$Ne obtained by AMD calculation [157].

| $K^\pi$ | $J^\pi$ | $W^J$ | $\langle \widehat{V}_{ls} \rangle$ | $K^\pi$ | $J^\pi$ | $W^J$ | $\langle \widehat{V}_{ls} \rangle$ |
|---|---|---|---|---|---|---|---|
| $0^+$ | $0^+$ | 0.70 | -5.2 | $0^-$ | $1^-$ | 0.95 | -0.8 |
|  | $2^+$ | 0.68 | -5.3 |  | $3^-$ | 0.93 | -0.8 |
|  | $4^+$ | 0.54 | -5.9 |  | $5^-$ | 0.88 | -0.7 |
|  | $6^+$ | 0.34 | -8.4 |  | $7^-$ | 0.71 | -0.9 |
|  | $8^+$ | 0.28 | -10.9 |  | $9^-$ | 0.70 | -1.3 |

**Table 18** Squared amplitude of the $^{40}$Ca + $\alpha$ component $W^J$ for each state of the inversion-doublet rotational bands of $^{44}$Ti obtained by AMD calculation [159].

| $K^\pi = 0^+$ | $J^\pi =$ | $0^+$ | $2^+$ | $4^+$ | $6^+$ | $8^+$ | $10^+$ | $12^+$ |
|---|---|---|---|---|---|---|---|---|
|  | $W^J =$ | 0.39 | 0.34 | 0.32 | 0.25 | 0.21 | 0.06 | 0.06 |
| $K^\pi = 0^-$ | $J^\pi =$ | $1^-$ | $3^-$ | $5^-$ | $7^-$ | $9^-$ |  |  |
|  | $W^J =$ | 0.56 | 0.50 | 0.43 | 0.38 | 0.32 |  |  |

The relative-motion nodal number $n_r = 6$ is the smallest nodal number of $\mathcal{Y}(r)$ which is allowed by the intercluster Pauli principle. The value of $n_r$ of the $J^\pi = 1^-$ of the $K^\pi = 0^-$ band is also $n_r$ =6. The fact that the percentage of $^{40}$Ca + $\alpha$ component of the ground state is 40% means that the percentage of the SU(3) shell-model component $\Phi^{(12,0)} = |^{40}\text{Ca-core},(1p0f)^4;[4^{10}](\lambda,\mu) = (12,0)\rangle$ is less than 40% because $\Phi^{(12,0)}$ has the duality property that it is equivalent the cluster-model wave function of $^{40}$Ca + $\alpha$. The decrease of the shell-model component with good spatial and SU(3) symmetry seen in the $^{44}$Ti ground state is considered to be more pronounced in the ground states of heavier nuclei.

## 7 Summary

In this report, we mainly focus on the new concept of nonlocalized clustering and the container picture for the evolution of cluster structure in nuclei, which can be another important fruition from the creative THSR idea in addition to the first important achievement of $\alpha$ condensation. In the recent five years, various cluster systems from n$\alpha$ nuclei, neutron-rich nuclei, to $\Lambda$−hypernuclei, are investigated with different THSR-type wave functions, which all provide the concept of nonlocalized clustering and the container picture. It was surprising that, even for the very old and historical subjects, like the $\alpha$ +$^{16}$O clustering, $\alpha$-linear-chain states, etc., there is still a large room for deepening the basic understanding of them.

We start with the famous Hoyle state of $^{12}$C. This novel 3$\alpha$ gas-like state triggers great interests in a wide field in physics and it has been studied for half a century by numerous models in nuclear physics. It came as a surprise that the THSR wave function with a simple structure gives a quite accurate description of this state and this Hoyle state is finally proved to be the $\alpha$ condensation state. Now, the search for the $\alpha$ condensation in $^{16}$O and $^{20}$Ne nuclei is becoming a very important subject both from theory and experiment. The great successful description of the gas-like states by the THSR wave function encourages us to extend this idea to the general cluster systems. By the study of the inversion doublet bands of $^{20}$Ne one significance underlying the THSR wave function is clarified, i.e., the clusters make a nonlocalized motion occupying the lowest orbit of the cluster mean-field potential instead of the localized motion from the traditional understanding. Subsequent calculations show that, this kind of nonlocalized motion, characterized by the size parameter in the THSR wave function, is not only the essential feature of the Hoyle state and typical $\alpha$+$^{16}$O cluster states but also of the *alpha*-linear-chain structures, molecular structures in neutron-rich nuclei like $^{9,10}$Be, and even cluster structures in hypernuclei.

The container picture was developed further for expressing this essential concept of nonlocalized clustering. In the container picture, the clusters are making almost independent motion in a nonlocalized way from a dynamical aspect, while the Pauli principle plays a key role for the formation of clusters from a kinematical aspect. The evolution of cluster structure along the increase of excitation energy can be naturally described by solving the Hill-Wheeler equation with the size parameter being the generator coordinate. This description can be identified with the container evolution, in which the size parameter characterizing the container expands along the path of the cluster evolution. We discussed the diverse clusters $0^+$ states in $^{12}$C and $^{16}$O. These evolution of cluster structures are well described and explained in our container picture in a unified way.

To investigate various complex nuclear systems in the container model, some new generations of the THSR wave functions and also new techniques are developed. For instance, the different widths of clusters can be treated exactly in an extended THSR wave function. The Monte-Carlo arithmetic has been applied to deal with the complicated integration problem for multi-cluster systems. These extensions show the original THSR wave function has a very flexible character and it is becoming a powerful tool in nuclear cluster physics. In the future, we are extending this container idea to study the cluster structure of nuclei in an *ab initio* way.

## 8 acknowledgements

The authors thanks for discussions with Drs. Peter Schuck, Gerd Röpke, Zhongzhou Ren, Chang Xu, Mengjiao Lyu, and Qing Zhao. This work was supported by JSPS KAKENHI Grant Numbers 17K14262, JP16K05351, 17K05440, and 18K03658. One of the authors (A.T.) is supported by the Ministry of Education and Science of the Republic of Kazakhstan, and the research grand number is IRN: AP05132476. Numerical computation in this work was carried out at the Yukawa Institute Computer Facility in Kyoto University.